\documentclass{aa}  
\usepackage{graphicx}
\usepackage{txfonts}
\usepackage{natbib}
\bibpunct{(}{)}{;}{a}{}{,}
\usepackage{color}
\usepackage{longtable}
\usepackage{epstopdf}

\begin{document}
   \title{Ammonia from cold high-mass clumps discovered in the inner Galactic disk by the ATLASGAL survey}


   \author{M. Wienen\inst{1}\thanks{Member of the International Max Planck Research School (IMPRS) for Astronomy and Astrophysics at the Universities of Bonn and Cologne.}, F. Wyrowski\inst{1}, F. Schuller\inst{1,}\inst{2}, K. M. Menten\inst{1}, C. M. Walmsley\inst{3,}\inst{4}
          \and L. Bronfman\inst{5}
          \and F. Motte\inst{6}
         }

    \offprints{M. Wienen \\ \email{mwienen@mpifr-bonn.mpg.de}}
  
   \institute{\inst{1}Max-Planck-Institut f\"ur Radioastronomie, Auf dem H\"ugel 69, 53121 Bonn, Germany\\
              \inst{2}Alonso de Cordova 3107, Casilla 19001, Santiago 19, Chile\\
              \inst{3}Osservatorio Astrofisico di Arcetri, Largo E. Fermi, 5, I-50125 Firenze, Italy\\
              \inst{4}Dublin Institute of Advanced Studies, Fitzwilliam Place 31, Dublin 2, Ireland\\
              \inst{5}Departamento de Astronom\'{\i}a, Universidad de Chile, Casilla 36-D, Santiago, Chile\\
              \inst{6}Laboratoire AIM, CEA/IRFU - CNRS/INSU - Universit\'e Paris Diderot, CEA-Saclay, 91191 Gif-sur-Yvette Cedex, France
                           }

   \date{Received 16 September 2011/ Accepted 26 May 2012}
                  
\abstract
 {The APEX Telescope Large Area Survey: The Galaxy (ATLASGAL) is an unbiased continuum survey of the inner Galactic disk at 870 $\mu$m. It covers $\pm 60^{\circ}$ in Galactic longitude and aims to find all massive clumps at various stages of high-mass star formation in the inner Galaxy, particularly the earliest evolutionary phases.}
{We aim to determine properties such as the gas kinetic temperature and dynamics of new massive cold clumps found by ATLASGAL. Most importantly, we derived their kinematical distances from the measured line velocities.} 
{We observed the ammonia ($J$,$K$) = (1,1) to (3,3) inversion transitions toward 862 clumps of a flux-limited sample of submm clumps detected by ATLASGAL and extracted $^{13}$CO (1-0) spectra from the Galactic Ring Survey (GRS). We determined distances for a subsample located at the tangential points (71 sources) and for 277 clumps whose near/far distance ambiguity is resolved.}
{Most ATLASGAL clumps are cold with rotational temperatures from 10-30 K with a median of 17 K. They have a wide range of NH$_3$ linewidths (1-7 km~s$^{-1}$) with 1.9 km~s$^{-1}$ as median, which by far exceeds the thermal linewidth, as well as a broad distribution of high column densities from $10^{14}$ to $10^{16}$ cm$^{-2}$ (median of $2 \times 10^{15}$ cm$^{-2}$) with an NH$_3$ abundance in the range of 5 to $30 \times 10^{-8}$. ATLASGAL sources are massive, $\gtrsim$ 100 M$_{\odot}$, and a fraction of clumps with a broad linewidth is in virial equilibrium. We found an enhancement of clumps at Galactocentric radii of 4.5 and 6 kpc. The comparison of the NH$_3$ lines as high-density probes with the GRS $^{13}$CO emission as low-density envelope tracer yields broader linewidths for $^{13}$CO than for NH$_3$. The small differences in derived clump velocities between NH$_3$ (representing dense core material) and $^{13}$CO (representing more diffuse molecular cloud gas) suggests that the cores are essentially at rest relative to the surrounding giant molecular cloud.}
{The high detection rate (87\%) confirms ammonia as an excellent probe of the molecular content of the massive, cold clumps revealed by ATLASGAL. A clear trend of increasing rotational temperatures and linewidths with evolutionary stage is seen for source samples ranging from 24 $\mu$m dark clumps to clumps with embedded HII regions. The survey provides the largest ammonia sample of high-mass star forming clumps and thus presents an important repository for the characterization of statistical properties of the clumps and the selection of subsamples for detailed, high-resolution follow-up studies.}

\titlerunning{Ammonia in massive clumps discovered with ATLASGAL}
\authorrunning{M. Wienen et al.}
\maketitle
%
\section{Introduction}
High-mass stars are essential for the evolution of galaxies because they energize the interstellar medium and release heavy elements, which are determining Galactic cooling mechanisms. Although the understanding of high-mass star formation is consequently of great importance, only little is known about the early stages of the formation of massive stars in contrast to the well-established evolutionary sequence of isolated low-mass stars \citep{2000prpl.conf...59A}. 
This is partly because of the difficult observational conditions that one encounters. High-mass stars form in clusters and are rarer according to the initial mass function \citep{2011arXiv1112.3340K}, therefore they are usually found at larger distances than molecular clouds, in which low-mass stars form. Moreover, high-mass stars evolve on shorter time scales and within dynamic and complex environments and are still deeply embedded in their earliest phases.\\ 
It is well known that high-mass star formation starts in giant molecular clouds \citep{2001ApJ...547..792D}. In analogy to low-mass star formation, one expects a cold and massive starless clump of $\gtrsim$ 500 M$_{\odot}$ consisting of molecular gas for the initial conditions of massive star/cluster formation \citep{2007prpl.conf..165B}. Young massive (proto) stars evolving in dense cores within the clumps emit ultraviolet radiation that leads to the formation of ultracompact HII regions (UCHIIRs). Many of them were found in radio continuum surveys with the VLA \citep[e.g.][]{1989ApJS...69..831W}, which revealed their distribution and number. Those regions were very helpful in tracing a more evolved stage of massive star formation in the Galaxy. Often UCHIIRs are associated with so-called hot molecular cores \citep{1992A&A...256..618C}, which are possible precursors of UCHIIRs, also called high-mass protostellar objects (HMPOs) or massive young stellar objects (MYSOs), which have also been discovered recently \citep{2002ApJ...566..945B}. Still earlier phases are deeply embedded in a massive envelope and are so cold that mostly they cannot be probed by means of mid-infrared emission \citep[e.g.][]{2005ApJ...634L..57S, 2010ApJ...723..555P,2007A&A...476.1243M}. Under certain circumstances, these clumps can be seen in absorption against a bright background provided mostly by emission from polycyclic aromatic hydrocarbon molecules (PAHs) as infrared dark clouds \citep[IRDCs,][]{1996A&A...315L.165P, 1998ApJ...494L.199E}. \cite{2006ApJ...653.1325S} and \cite{2010ApJ...723..555P} have presented large catalogues of these IRDCs. But to date, only a small fraction of them has been investigated in detail \citep{1998ApJ...508..721C,2000ApJ...543L.157C,2006A&A...450..569P,2006ApJ...641..389R}. Previous studies revealed that IRDCs are cold ($<$ 25 K), dense ($>$ $10^5$ cm$^{-3}$) and have high column densities \citep[$\sim$ 10$^{23}$-10$^{25}$ cm$^{-2}$,][]{2006ApJ...641..389R,2006ApJ...639..227S}. Their sizes ($\sim$ 2 pc) and masses ($\sim$ 10$^2$-10$^4$ M$_{\odot}$) are similar to those of cluster-forming molecular clumps. This, together with the cold temperatures and fragmented substructure of IRDCs, suggests that they are protoclusters \citep{2006ApJ...641..389R,2005ApJ...630L.181R}. Kinematic distances to IRDC samples in the first and fourth Galactic quadrants have been determined using $^{13}$CO (1-0) and CS (2-1) emission \citep{2006ApJS..163..145J,2008ApJ...680..349J}. However, surveys targeted at the early phases of high-mass star formation \citep[e.g.][]{2002ApJ...566..945B,1989ApJ...340..265W} usually only trace one of those stages, e.g. IRDCs, HMPOs or UCHIIRs. In contrast, a first attempt to obtain an unbiased sample using a complete dust continuum imaging at the scale of the molecular complex Cygnus X \citep{2007A&A...476.1243M} was able to derive some more systematic results on the existence of a cold phase (IR-quiet massive dense cores and Class 0-like massive protostars) for high-mass star formation \citep{2010A&A...524A..18B,2011IAUS..270...53C}. Another survey of the molecular complex NGC 6334/NGC 6357 was conducted by \cite{2010A&A...515A..55R}, who identified IR-quiet massive dense cores and estimated high-mass protostellar lifetimes. The IR-quiet massive dense cores are suggested to host massive class 0 protostars as seen by the Herschel Space Telescope \citep{2011A&A...535A..76N}.\\
Only a few hundred high-mass proto- or young stellar objects have been studied up to now. Hence, these and especially objects in still earlier phases of massive star formation need to be investigated in more detail. To achieve significant progress in that, a large-area survey of the Galactic plane was conducted, which now provides a global view of star formation at submillimeter wavelengths. ATLASGAL (\textit{The APEX Telescope Large Area Survey of the Galaxy at 870 $\mu$m}) is the first unbiased continuum survey of the whole inner Galactic disk at 870 $\mu$m \citep{2009arXiv0903.1369S}. It aims to find all massive clumps that form high-mass stars in the inner Galaxy by using the Large APEX Bolometer Camera LABOCA, which is an array with 295 bolometer pixels operated at the APEX telescope \citep{2007Msngr.129....2S, 2008SPIE.7020E...2S} with a field of view of 11$\arcmin$ and a beamwidth of 19.2$\arcsec$ FWHM at the wavelength of 870 $\mu$m.\\
ATLASGAL, which reaches a Galactic longitude of $\pm 60^{\circ}$ and latitude of $\pm 1.5^{\circ}$, is able to detect objects associated with massive star formation at various stages and to compare them \citep{2009arXiv0903.1369S}. Other Galactic plane surveys, conducted over a similar range, are the MSX survey covering 8 to 21 $\mu$m \citep{2001AJ....121.2819P}, 2MASS around 2 $\mu$m \citep{2006AJ....131.1163S}, GLIMPSE from 3 to 8 $\mu$m \citep{2003PASP..115..953B}, MIPSGAL at 24 and 70 $\mu$m \citep{2009PASP..121...76C} and HiGAL from 70 to 500 $\mu$m \citep{2010PASP..122..314M}.  While submillimeter dust continuum surveys are essential for identifying high-mass clumps, they lack information on important parameters, especially the distances to the newly found sources. These are needed to determine other properties such as their masses and luminosities. But all of this can be addressed by molecular line observations of the high-mass star forming regions. Because their density is at about the critical NH$_3$ density, this molecule is appropriate for determining properties of the clumps without contamination from large-scale cloud structures with lower density. Since the molecular gas of the cores is very dense \citep[$\sim$ 10$^5$ cm$^{-3}$][]{2002ApJ...566..945B} and cold, many molecules such as CS and CO are partly frozen onto dust grains. In contrast to those, the fractional gas abundance of ammonia remains constant in prestellar cores \citep{2002ApJ...569..815T}. Hence, we carried out follow-up observations of northern ATLASGAL sources in the lowest NH$_3$ inversion transitions.\\ 
NH$_3$ is known as a reliable temperature probe of interstellar clouds \citep{2004A&A...416..191T,1983A&A...122..164W}. The rotational energy levels are given by the total angular momentum, $J$, and its projection along the molecular axis, $K$. Radiative transitions between different $K$-ladders are forbidden and the lowest metastable energy levels, for which $J = K$, are thus collisionally excited. The intensity ratio of their inversion transitions therefore provides the rotational temperature of the gas  \citep{1983ARA&A..21..239H}, which can be used to estimate the kinetic temperature of the cores. This is needed for proper mass estimates from the submm data. Independently, the linewidth can be used to derive virial masses. Moreover, the inversion transitions are split into distinct hyperfine components and their ratio provides a measure of the optical depth, knowledge of which leads to reliable column density and rotational temperature determinations.\\
As this article was ready for submission, we became aware of the recent study by \cite{2011ApJ...741..110D}, who observed NH$_3$ lines of sources in the first Galactic quadrant as well, identified by the Bolocam Galactic Plane Survey (BGPS), which measures the 1.1 mm continuum emission. In contrast to our sample, the observations of \cite{2011ApJ...741..110D} do not cover the whole northern Galactic longitude range up to 60$^{\circ}$, but are conducted only within four different ranges in longitude, and within a smaller Galactic latitude range of $\pm 0.5^{\circ}$ compared to our NH$_3$ survey. Our analysis is thus complementary to the results of \cite{2011ApJ...741..110D}.\\
In Section \ref{s:obs}, we describe which ATLASGAL sources were selected for our NH$_3$ follow-up observations and provide details of the measurements. In Section \ref{data reduction}, we present how we reduced the data and derived different clump properties from fits to the spectra of NH$_3$ inversion transitions. In Section \ref{results}, different correlations of the NH$_3$ line parameters such as the velocity distribution, linewidth, rotational temperature and column density are shown. Then, we compare these with submillimeter dust continuum properties of the clumps such as the H$_2$ column density in Section \ref{dust continuum} and estimate gas and virial masses. We analyse additional line parameters derived from $^{13}$CO (1-0) emission \citep{2006ApJS..163..145J} in Section \ref{13CO} to investigate clump-to-cloud motions etc. In Section \ref{discussion}, the determined gas properties are compared with those of other high-mass dust-selected star forming regions. A summary and conclusions of this NH$_3$ investigation are given in Section \ref{conclusion}.\\
This paper analyses statistics of the NH$_3$ data obtained for the northern ATLASGAL sources. We provide their near and far distance, but do not distinguish between both, although some of the molecular clouds have known distances from a previous study \citep{2009ApJ...699.1153R}. In a second paper (Wienen et al. in prep.) ammonia lines of ATLASGAL sources in the southern hemisphere will be studied and distances to southern molecular clouds will be derived.

\section{Observations}\label{s:obs}
\subsection{Source selection}
\label{select}
\begin{figure}[h]
\centering
\includegraphics[angle=-90,width=8.0cm]{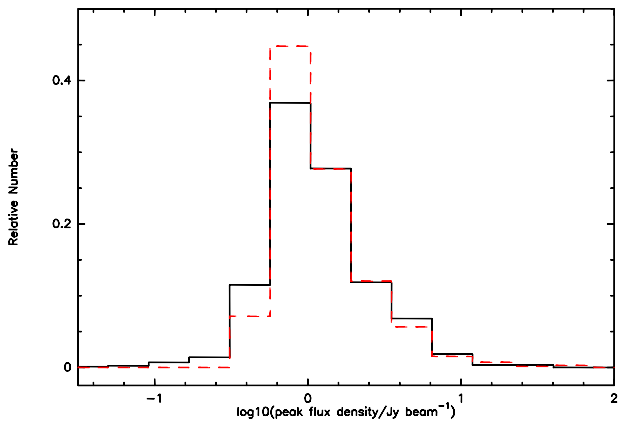}
\caption[correlation between the (1,1) linewidth and the rotational temperature]{Relative number distribution of NH$_3$ observations as solid black curve and of the whole northern ATLASGAL sample as dashed red curve vs. the logarithm of the 870 $\mu$m peak flux density.}\label{flux-histo}
\end{figure}

From a preliminary ATLASGAL point source catalogue, we selected a flux-limited subsample of compact (smaller than 50$\arcsec$) clumps down to about 0.4 Jy/beam peak flux density, whose positions were selected using the Miriad task sfind \citep{1995ASPC...77..433S}, and observed $\sim$ 63\% of these 1361 sources. From 2008 to 2010 we observed the ammonia emission of a total of 862 dust clumps in a Galactic longitude range from $5^\circ$ to $60^\circ$ and latitude within $\pm$1.5$^\circ$. The 870 $\mu$m peak flux density distribution of observed sources is plotted as a solid black curve in Fig. \ref{flux-histo} and that of the whole ATLASGAL sample of 1361 sources as a dashed red curve. Their comparison reveals that the 870 $\mu$m peak flux distributions of both samples are similar and therefore we expect the NH$_3$ subsample to be representative of the ATLASGAL sample as a whole. Most NH$_3$ observations as well as most ATLASGAL sources from the whole sample have a peak flux density of $\sim$ 0.8 Jy/beam, the distributions above those values can be described by a power law.\\ 
We estimate the ATLASGAL sample to be complete above 1 Jy/beam peak flux density, which results in 644 observed clumps. Of those the NH$_3$ sample contains 406 sources (63\%) within this flux limit. Seventy-three clumps (11\%) were observed in previous ammonia surveys \citep{2002ApJ...566..931S,1990A&AS...83..119C,1996A&A...308..573M,2006A&A...450..569P}. 

\subsection{Observational setup}
Observations of the NH$_3$ (1,1), (2,2) and (3,3) inversion transitions were made from 2008 to 2010 with the Effelsberg 100-m telescope. The frontend was a 1.3 cm cooled HEMT receiver, which is used in a frequency range between 18 GHz and 26 GHz and mounted at the primary focus. In 2008 January, two different spectrometers were used, the 8192 channel autocorrelator (AK 90) and the Fast Fourier Transform Spectrometer (FFTS). In May 2008, the AK90 was replaced by the FFTS, which possesses a greater bandwidth. The autocorrelator contains eight individual modules. A bandwidth of 20 MHz was chosen for each of them, which results in a spectral resolution of about 0.5 km~s$^{-1}$, while the FFTS consists of two modules with a chosen bandwidth of 500 MHz each and a spectral resolution of $\sim 0.7$ km~s$^{-1}$. Both spectrometers are able to measure two polarizations of the three observed NH$_3$ lines simultaneously. The scans observed in 2008 January with the two spectrometers were summed to obtain a better signal-to-noise ratio. The beamwidth (FWHM) at the frequencies of the (1,1) to (3,3) ammonia inversion transitions at about 24 GHz is 40$\arcsec$. Pointed observations towards the dust emission peaks were conducted in frequency-switching mode with a frequency throw of 7.5 MHz. 
The median system temperature was about 70 K, with a zenith optical depth that usually varied between 0.02 and 0.09 in the beginning of the year in about January and was more constant, about 0.07 some months later, in about May and October. The total integration time for each source was about 5 minutes. For some clumps at low declination, which could only be observed at low elevations and consequently have high system temperatures of $\sim 150$ K, longer integration times of up to 25 minutes were chosen.

\section{Data reduction and analysis}
\label{data reduction}
To reduce the NH$_3$ spectra we used the CLASS software\footnote{available at http://www.iram.fr/IRAMFR/GILDAS}. Because the ammonia lines from each source are detected in two polarizations, all scans that belong to one polarization of an inversion transition were summed and then the spectra of both polarizations were averaged. The frequency switching results in two lines, which were inverted and separated in frequency by 7.5 MHz. This procedure was reversed by a folding, which shifts the two lines by 3.75 MHz and subtracts one of them from the other. The large frequency throw led to fluctuations of the baseline, which needed to be corrected. For this we subtracted a polynomial baseline of the order of 3 to 7, which produces inaccuracies in the subsequent analysis of the spectra. The NH$_3$(1,1) hyperfine structure of very many ammonia sources shows different line profiles with various linewidths, which had to be reduced in the same way. Hence, we developed a strategy to set windows, spaces around NH$_3$ lines, that were excluded from the baseline fitting: The width of the window around each hyperfine component of the (1,1) line was chosen according to the width of the main line; the order of the baseline depends on the linewidth as well. A polynomial of high order, up to 7, was subtracted for lines with narrow width, $<$ 2 km~s$^{-1}$, while a low order, mostly 3, was chosen for lines with a broad width, greater than 4 km~s$^{-1}$.
To test if this introduced any systematic errors, we compared the linewidths of simulated NH$_3$ (1,1) lines with the linewidths resulting from a fit of the same spectra added on observed baselines. For linewidths more narrow than 1.5 km~s$^{-1}$ the model and the fitted observation agreed, the deviation of the two becomes sligthly higher with increasing linewidths. However, the contribution of systematic errors for lines with broad linewidths is only small, $\sim$ 6\% on average. Some examples for reduced and calibrated spectra of observed (1,1) to (3,3) inversion transitions are given in Fig. \ref{nh3lines}.\\
The hyperfine structure of the NH$_3$(1,1) line was fitted with CLASS. Assuming that all hyperfine components have the same excitation temperature, a nonlinear least-squares fit was made to the observed spectra with the CLASS software, which yielded as independent parameters radial velocity, $v_{\mbox{\tiny LSR}}$, the linewidth, $\Delta v$, at the full width at half maximum of a Gaussian profile and the optical depth of the main line, $\tau_{\mbox{\tiny m}}$, with their errors.\\
Because the hyperfine satellite lines of the (2,2) and (3,3) transitions are mostly too weak to be detected, their optical depth could not be determined. A single Gaussian was fitted to the main line of the (2,2) and (3,3) transitions with CLASS. The three independent fit parameters were the integrated area, $A$, the radial velocity of the main line, $v_{\mbox{\tiny LSR}}$, and the FWHM linewidth, $\Delta v$. The temperature of the NH$_3$ (1,1) line was derived from the peak intensity of the Gaussian fit to the main line. For the (2,2) and (3,3) transitions the temperature were calculated with 
\begin{eqnarray}\label{Temperatur-cygx}
T_{\mbox{\tiny MB}}=\frac{A}{\Delta v \times 1.064}
\end{eqnarray}
with the error given by the rms value. If lines were not detected, the noise level was used as an upper limit. We list the positions of the sources, the optical depths of the NH$_3$ (1,1) lines, $\tau$(1,1), their LSR velocities, $v$(1,1), linewidths, $\Delta v$(1,1), and main beam brightness temperatures, $T_{\mbox{\tiny MB}}$(1,1), together with their formal fit errors in Table \ref{parline11-atlasgal}. The LSR velocities, linewidths, and main beam brightness temperatures of the (2,2) and (3,3) lines together with their formal fit errors are given in Table \ref{parline22_line33-atlasgal}. The other physical parameters such as the rotational temperature ($T_{\mbox{\tiny rot}}$), the kinetic temperature ($T_{\mbox{\tiny kin}}$) and the ammonia column density ($N_{\mbox{NH}_3}$) were derived using the standard formulation for NH$_3$ spectra \citep{1983ARA&A..21..239H,1986A&A...157..207U}, see Sect. \ref{rotational temperature}. They are given with the errors (1$\sigma$) calculated from Gaussian error propagation in Table \ref{parabgel-atlasgal}.

\subsection{Line parameters derived from ammonia}
\begin{figure*}
\centering
\includegraphics[angle=-90,width=15.0cm]{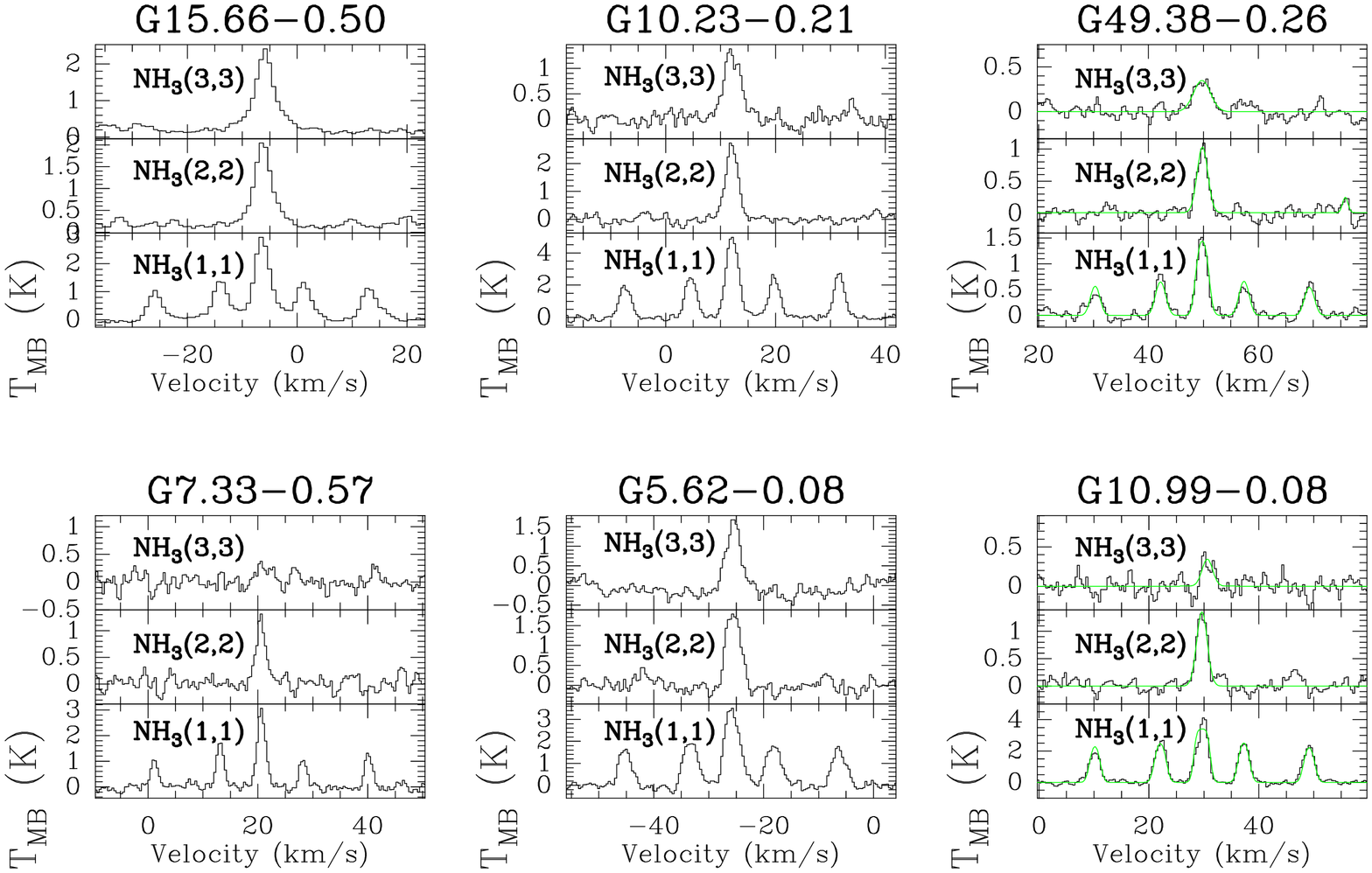}\vspace*{0.5cm}
\includegraphics[angle=-90,width=15.0cm]{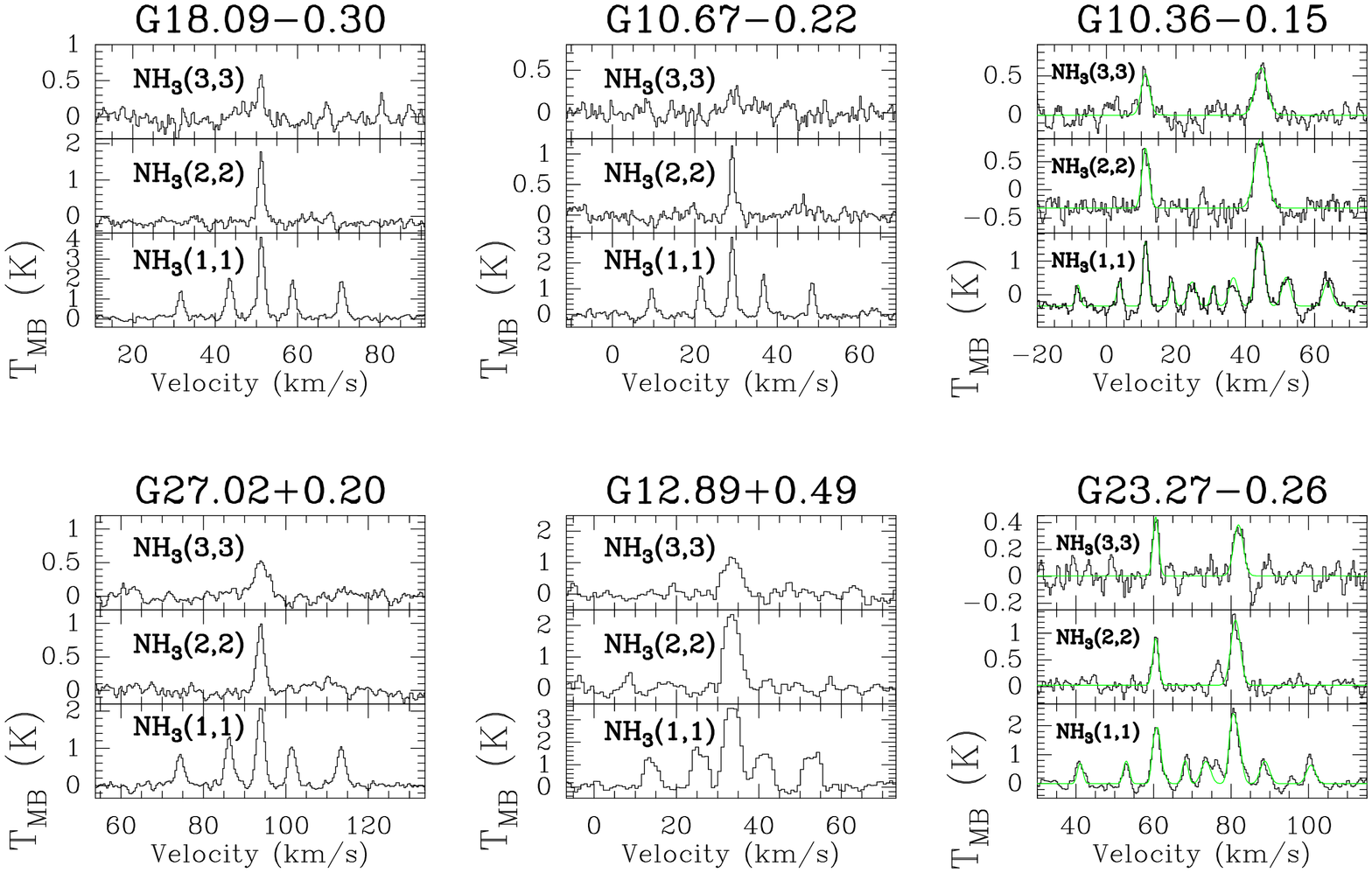}
\caption[spectra of observed sources]{Reduced and calibrated spectra of observed NH$_3$(1,1), (2,2) and (3,3) inversion transitions, the fit is shown in green for some sources. Towards G10.36$-$0.15 and G23.27$-$0.26 two lines are seen at different LSR velocities.}\label{nh3lines}
\end{figure*}

Of the 862 observed sources, 752 were detected in NH$_3$ (1,1) (87\%), 710 clumps (82\%) in the (2,2) line and 415 sources (48\%) in NH$_3$ (3,3) with a signal-to-noise (S/N) ratio $>$ 3. The hyperfine structure of most (1,1) lines is clearly detected, while that of the (2,2) and (3,3) lines is mostly too weak to be visible. The sample contains sources of different intensities; examples of them are shown in Fig. \ref{nh3lines}. There are many strong sources such as G5.62$-$0.08 with an S/N ratio of 30. An example for a weak line is the (3,3) line in G10.99$-$0.08. Its signal-to-noise ratio is 3, its fit is shown together with that of some other sources in Fig. \ref{nh3lines}. The comparison of peak intensities using Tables \ref{parline11-atlasgal} and \ref{parline22_line33-atlasgal} yields an average peak intensity of the (2,2) line of 53\% of the average (1,1) peak line intensity and an average (3,3) main beam brightness temperature of 40\% of the average (1,1) peak intensity.\\
Moreover, a few sources like G10.36$-$0.15 and G23.27$-$0.26 (see Fig. \ref{nh3lines}) exhibit two different velocity components. Those are probably two NH$_3$ clumps, that lie on the same line of sight to the observer with different velocities, e.g. 11 km~s$^{-1}$ and 44 km~s$^{-1}$ for G10.36$-$0.15, at different distances. The two main components of the inversion transitions can be separated, while the satellite lines partly interfere with each other. These sources were fitted with a model with two velocity components.

\section{Results and analysis}
\label{results}
We first plot different correlations of line parameters, which are derived from the NH$_3$ spectra observed with the Effelsberg 100 m telescope (FWHM beamwidth = $40''$) (Fig. \ref{coords-v11-atlasgal} - \ref{dv-t33-hotcores}). An ammonia (3,3) maser line together with the thermal emission is displayed in Fig. \ref{33maser}. Then, ammonia results are compared with the 870 $\mu$m flux integrated over the Effelsberg beam, extracted from submillimeter dust continuum maps of the ATLASGAL project \citep{2009arXiv0903.1369S} (Figs. \ref{nh3-mm-atlasgal} and \ref{mass-virialmass}), and with line parameters from the $^{13}$CO (1-0) emission detected by the Boston University-Five College Radio Astronomy Observatory Galactic Ring Survey \citep[GRS][]{2006ApJS..163..145J}. Fig. \ref{co-nh3sources} compares NH$_3$ (1,1) and $^{13}$CO (1-0) lines. Fig. \ref{diff-vnh3vco-atlasgal} shows our analysis of $^{13}$CO and NH$_3$ line-centre velocities, both molecular lines for extreme sources of this plot are displayed in Fig. \ref{two-velocity-spectra}. NH$_3$ (1,1) and $^{13}$CO (1-0) linewidths are compared in Fig. \ref{dv11-dvco-atlasgal}.  

\subsection{Ammonia velocities}
\begin{figure}[h]
\centering
\includegraphics[angle=0,width=9.0cm]{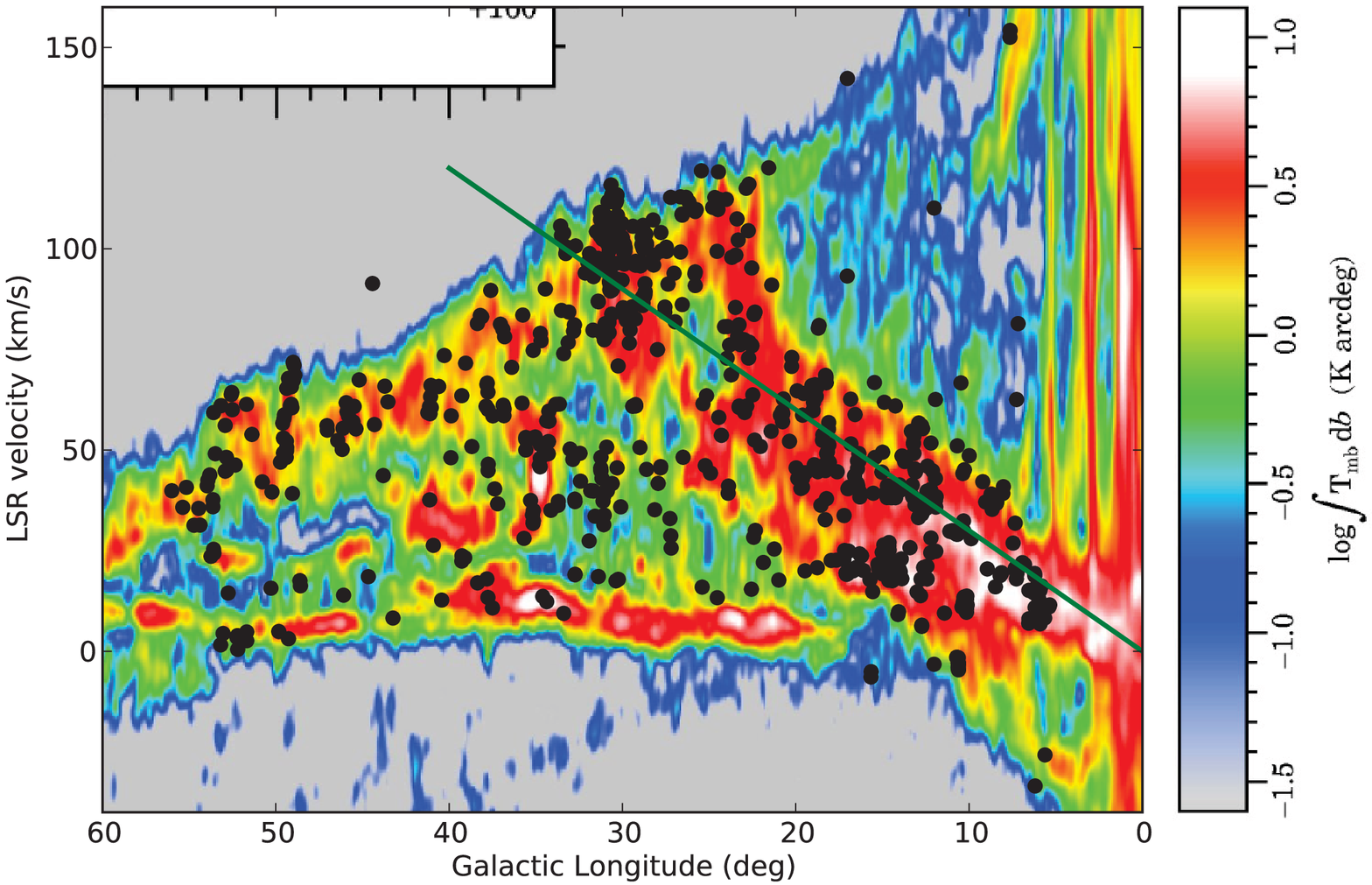}
\caption[dependence of the (1,1) LSR velocity from the Galactic longitude]{LSR velocities of detected sources are plotted against the Galactic longitude with CO (1-0) emission \citep{2001ApJ...547..792D}, which is shown in the background. The straight line denotes the 5 kpc molecular ring \citep{2006ApJ...653.1325S}.}\label{coords-v11-atlasgal}
\end{figure}

\begin{figure}[h]
\centering
\includegraphics[angle=-90,width=9.0cm]{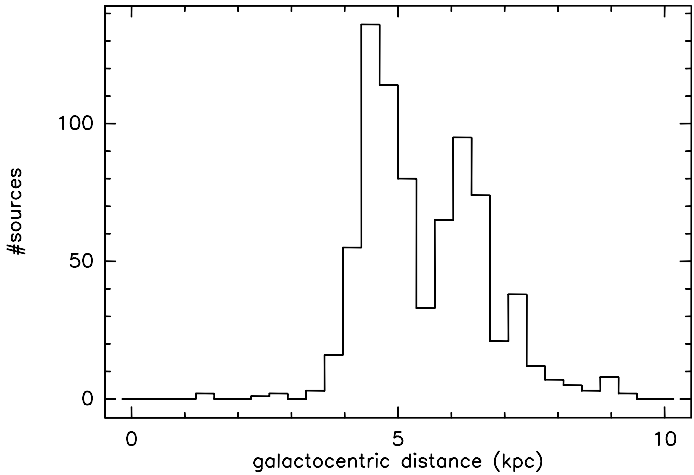}
\caption[dependence of the (1,1) radial velocity from the Galactic longitude]{Number distribution of northern ATLASGAL sources with Galactocentric distance.}\label{galactocentric}
\end{figure}

The distribution of NH$_3$ velocities of the ATLASGAL clumps with Galactic longitude is shown in Fig. \ref{coords-v11-atlasgal}. For most clumps, the velocities of the ammonia (1,1) line lie between -10 km~s$^{-1}$ and 120 km~s$^{-1}$, only few exhibit extreme velocities of between -20 km~s$^{-1}$ and -30 km~s$^{-1}$ and between 120 km~s$^{-1}$ to 150 km~s$^{-1}$. They are compared with CO (1-0) emission, observed with the CfA 1.2 m telescope \citep{2001ApJ...547..792D}, which is shown in the background. Most clumps are strongly correlated with CO emission, which probes the larger giant molecular clouds, but some with extreme velocities seem to be related to only weak CO emission. These probably consist of compact clumps, which are detected with the Effelsberg beamwidth of 40$\arcsec$, but are not visible in the larger CfA telescope beam of 8.4$\arcmin$. The straight line indicates the 5 kpc molecular ring that represents the most massive concentration of molecular gas and star formation activity in the Milky Way \citep{2006ApJ...653.1325S}. Its high CO intensity results from integration over many molecular clouds, which are identified by $^{13}$CO observations of the Galactic Ring Survey \citep[GRS,][]{2006ApJ...653.1325S}. For the ATLASGAL sample the NH$_3$ velocities are crucial to obtain kinematical distances to the clumps, which were calculated using the revised rotation parameters of the Milky Way presented by \cite{2009ApJ...700..137R}. We have derived near and far distances (cf. Table \ref{par870mikrom-atlasgal}) and we will distinguish between them by comparing ATLASGAL with extinction maps and absorption in HI lines (Wienen et al. in prep.). Sources associated with infrared extinction peaks are most likely at the near distance. Hence, the addition of the ammonia measurements to the ATLASGAL data reveals the three-dimensional distribution of massive star forming clouds in the first quadrant of our Galaxy. Some clumps are located within GRS molecular clouds, whose kinematic distances have already been determined by using absorption features in the HI 21 cm line against continuum emission \citep{2009ApJ...699.1153R}. These known distances are marked by a star in Table \ref{par870mikrom-atlasgal}. In addition, sources located near the tangential points have similar near and far distances, which are marked by a star in Table \ref{par870mikrom-atlasgal}.\\
The distribution of Galactocentric distances can be studied independently of near/far kinematic distance ambiguity. Fig. \ref{galactocentric} shows an enhancement of the number of ATLASGAL sources at Galactocentric radii of 4.5 and 6 kpc, where the Scutum arm and the Sagittarius arm are located. This agrees with previous results. \cite{2011A&A...529A..41N} used ATLASGAL and complimentary data to show that the region around Galactic longitude of 30$^\circ$, where the W43 Molecular Complex is located, has a very high-mass concentration of sources and star formation activity at a Galactocentric radius of 4.5 kpc. The two peaks in Fig. \ref{galactocentric} are also detected in \cite{2000A&A...358..521B}, based on a CS(2-1) survey of IRAS point-like sources with FIR colours of UCHII regions (see Fig. 3 and Table 2). Moreover, the same features of Galactic structure were discovered already by the first H II region radio recombination line survey \citep{1970A&A.....4..357R} and more recently in \cite{2009ApJ...690..706A} in a sample of H II regions over the extent of the GRS (see Fig. 9). 

\subsection{Linewidth}
\label{linewidth}
\begin{figure}[h]
\centering
\includegraphics[angle=-90,width=9.0cm]{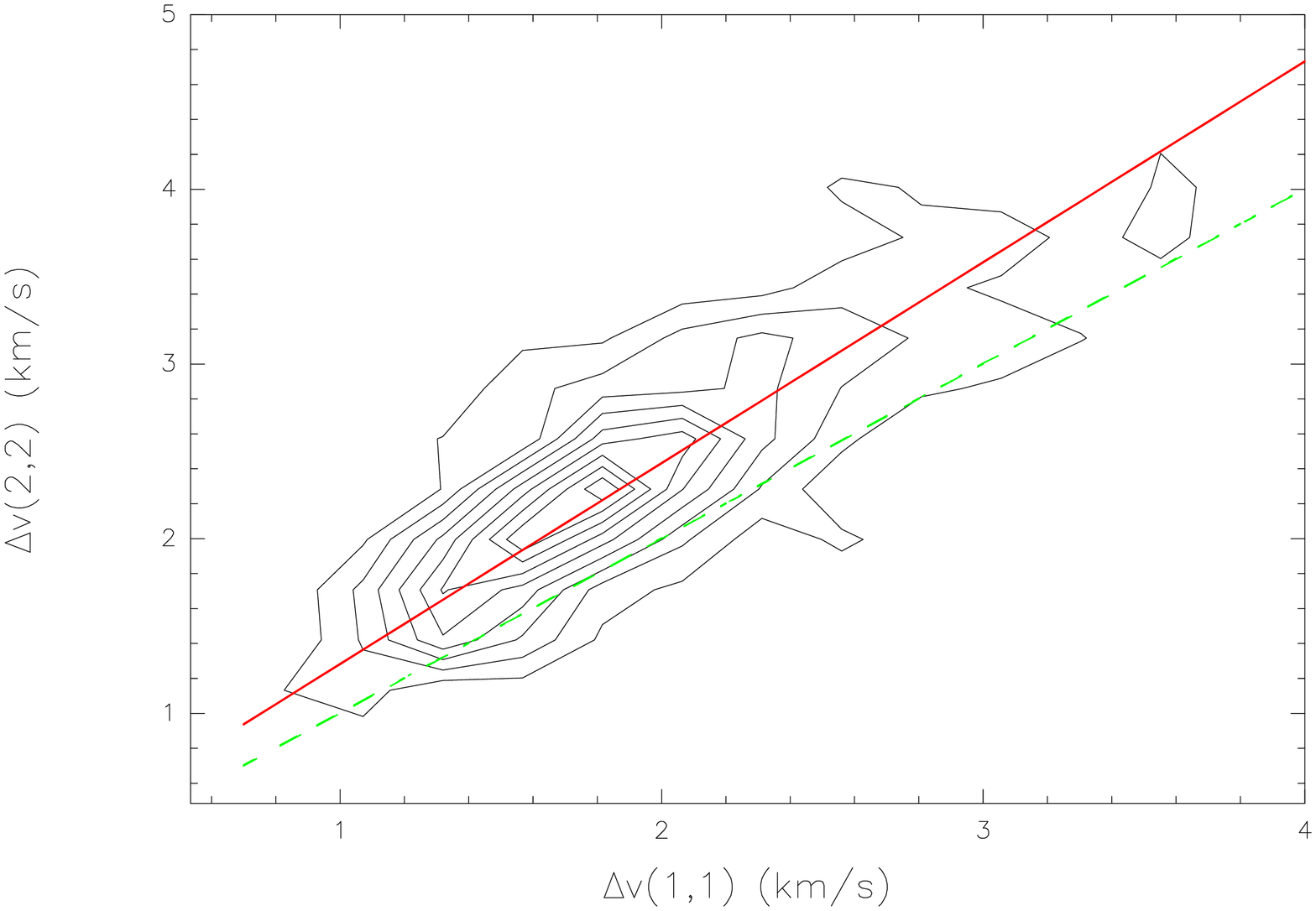}\vspace*{0.5cm}
\includegraphics[angle=-90,width=9.0cm]{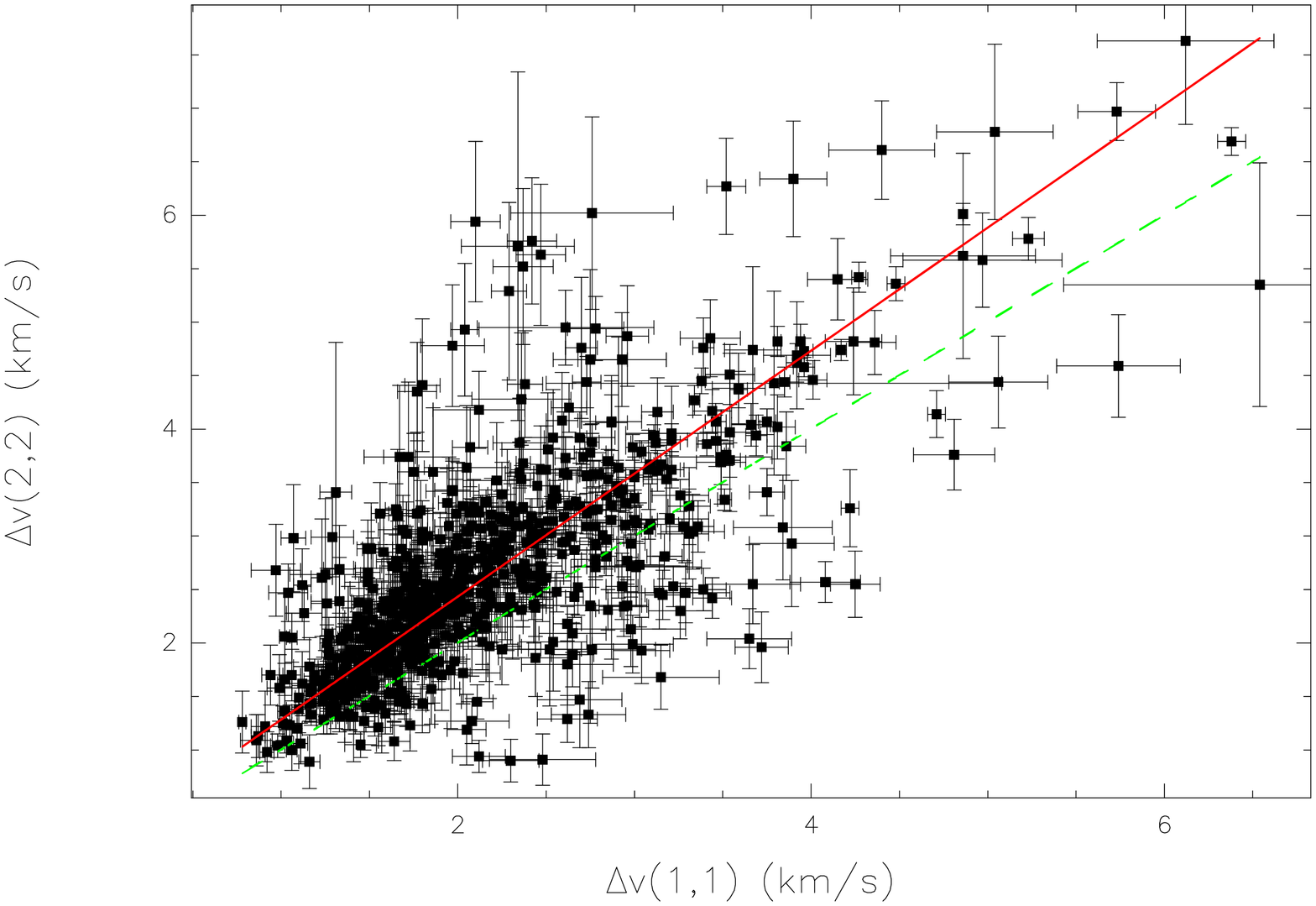}
\caption[comparison of (1,1) and (2,2) linewidth]{Correlation plot of the NH$_3$ (1,1) and (2,2) linewidths. The straight green dashed line corresponds to equal widths, the red solid line includes the correction for the hyperfine structure of the (2,2) line $\Delta v_{\mbox{Gauss}}(2,2) = 1.15 \cdot \Delta v_{\mbox{HFS}}(2,2) + 0.13$. The (1,1) and (2,2) linewidths are mostly equally distributed around the red solid line, as shown by the contour plot in the top panel. For this upper plot we counted the number of sources in each (1,1) and (2,2) linewidth bin of 0.1 km~s$^{-1}$. The contours give 10 to 90\% in steps of 10\% of the peak source number per bin, these levels are used for all contour plots in this article.}\label{dv11-dv22-atlasgal}
\end{figure}

\begin{figure}[h]
\centering
\includegraphics[angle=-90,width=9.0cm]{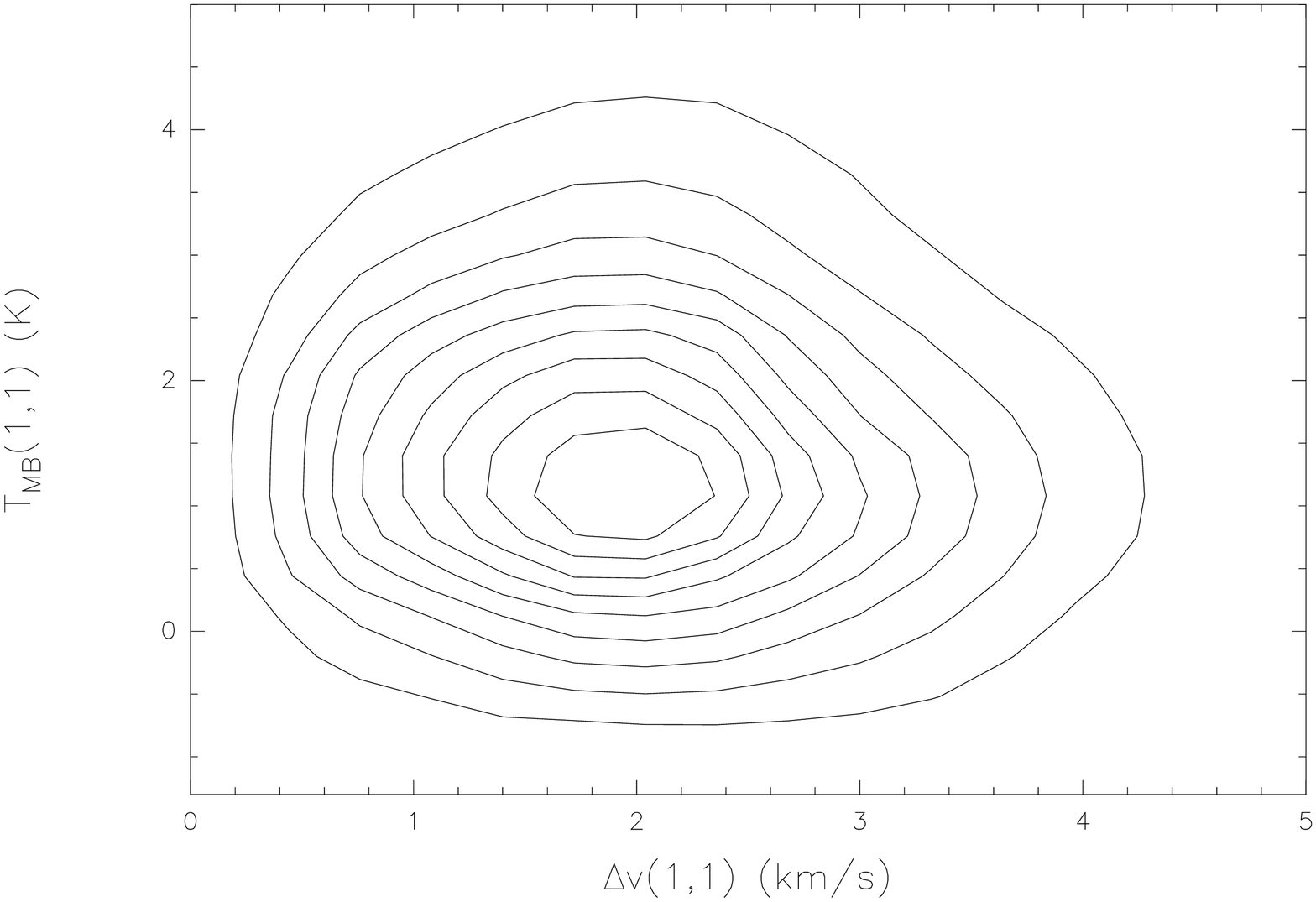}\vspace*{0.5cm}
\includegraphics[angle=-90,width=9.0cm]{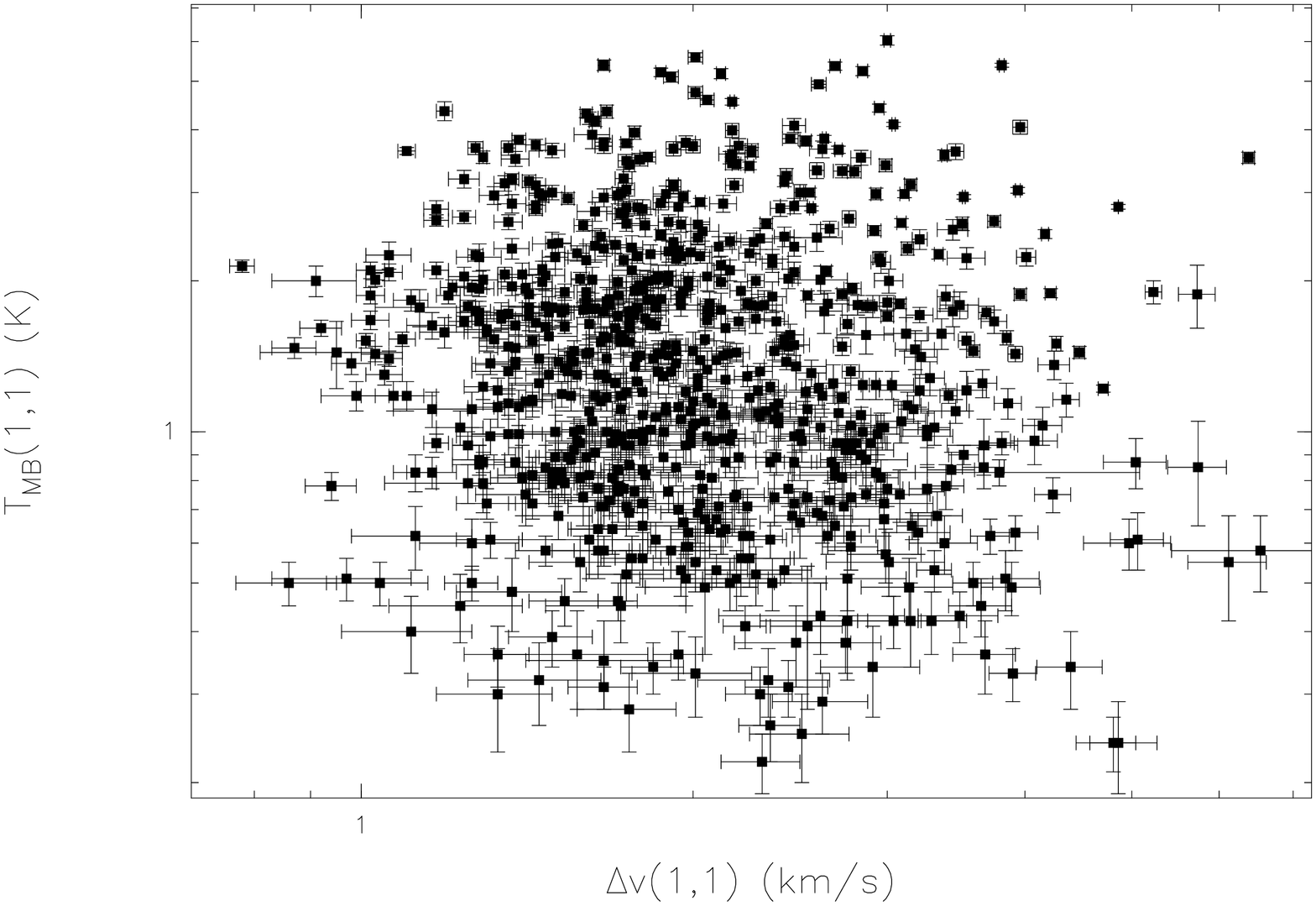}
\caption[comparison of (1,1) linewidth and (1,1) main beam brightness temperature]{NH$_3$ (1,1) main beam brightness temperature plotted against the (1,1) linewidth. The upper panel displays the contour plot with a binning of 1 km/s for the (1,1) linewidth and of 1 K for the (1,1) temperature, while the lower panel shows the scatter plot.}\label{dv11-t11-atlasgal}
\end{figure}

\begin{figure}[h]
\centering
\includegraphics[angle=-90,width=9.0cm]{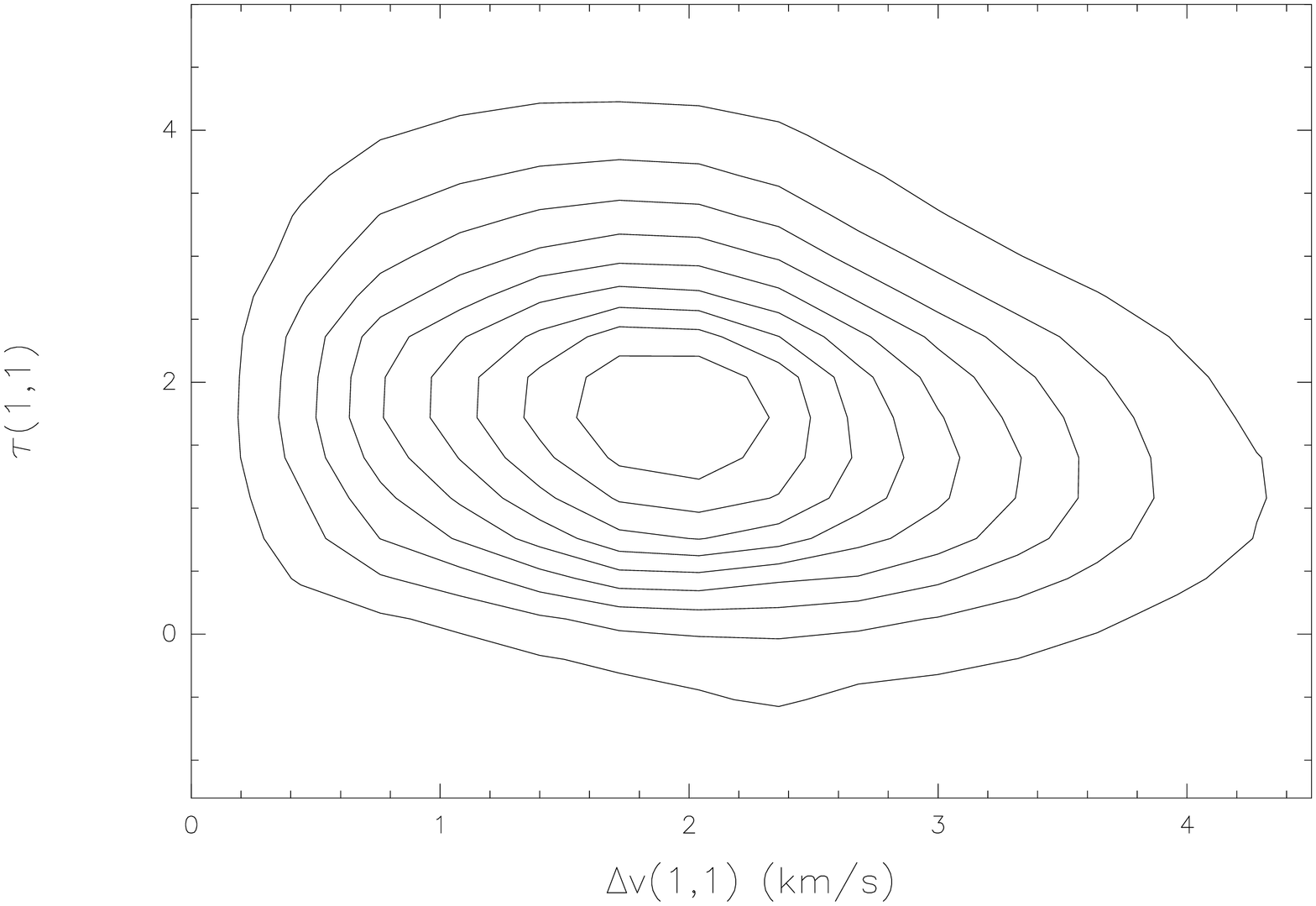}\vspace*{0.5cm}
\includegraphics[angle=-90,width=9.0cm]{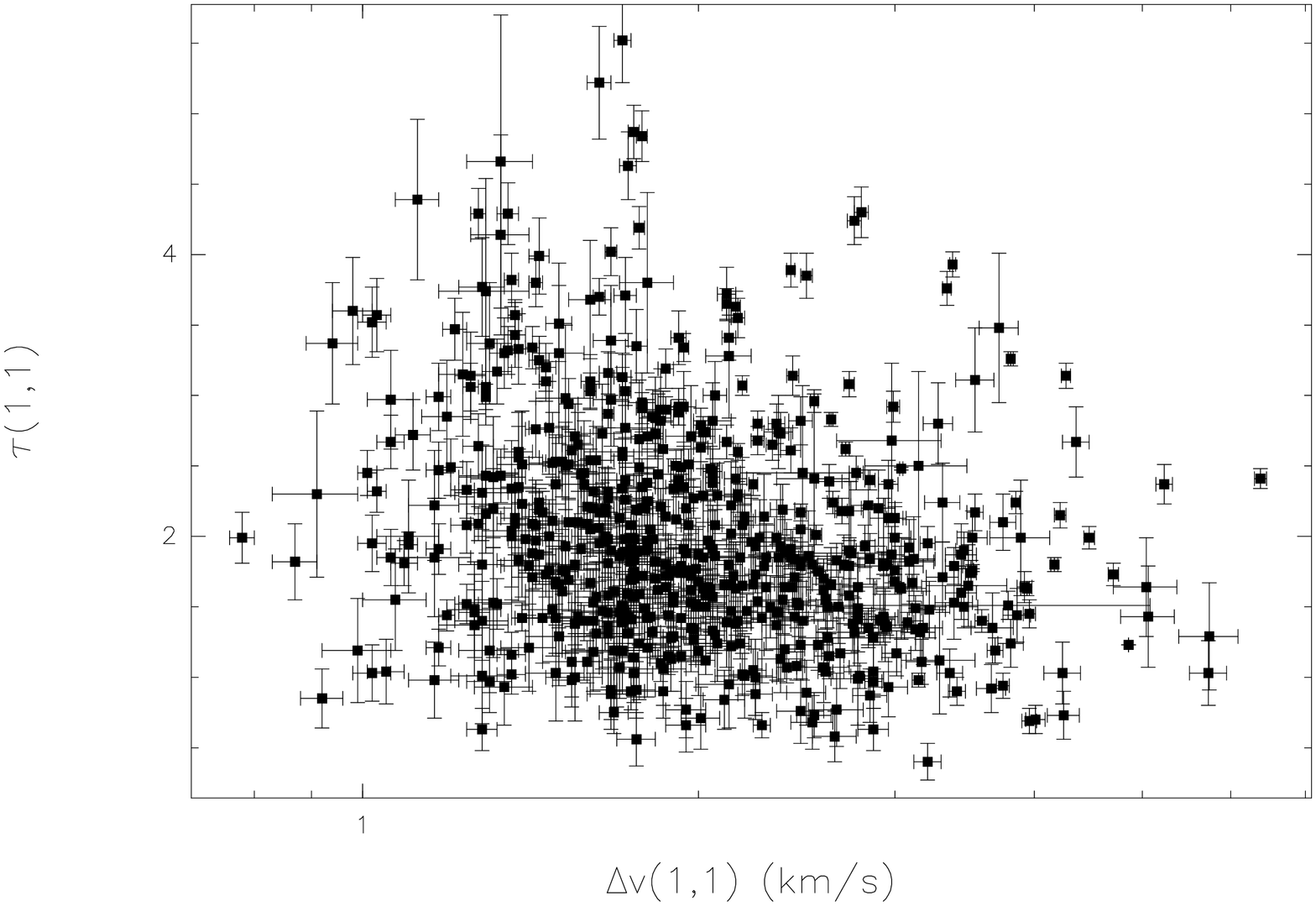}
\caption[comparison of (1,1) linewidth and (1,1) optical depth]{NH$_3$ (1,1) optical depth compared with the (1,1) linewidth. Contour lines of both parameters are plotted in the upper panel with the range of (1,1) linewidths divided into bins of 1 km/s and that of (1,1) optical depths into bins of 1.}\label{dv11-tau11-atlasgal}
\end{figure}

The NH$_3$ (1,1) inversion transitions exhibit widths from 0.7 km~s$^{-1}$ to 6.5 km~s$^{-1}$ (cf. Fig. \ref{dv11-dv22-atlasgal} upper panel), while those of the (2,2) lines range up to 7.5 km~s$^{-1}$. Linewidths narrower than 0.7 km~s$^{-1}$ cannot be resolved with the backend spectral resolution. One of the clumps with broad linewidths is G10.47+0.03 with $\Delta v$(1,1) = 6.4 km~s$^{-1}$ and $\Delta v$(2,2) = 6.7 km~s$^{-1}$, which is a bright, well-known hot core \citep{1992A&A...256..618C}. Because the hyperfine structure of the (2,2) line of most clumps could not be detected, we determined the linewidth by a Gaussian fit. The NH$_3$(1,1) and (2,2) linewidths are compared in the lower panel of Fig. \ref{dv11-dv22-atlasgal}, the straight dashed line in green shows equal values. In addition, we created a contour plot by dividing the (1,1) and (2,2) linewidth ranges into bins of 0.1 km~s$^{-1}$ and counted the number of sources in each bin (upper panel of Fig. \ref{dv11-dv22-atlasgal}). The widths of the (2,2) transition, fitted by a Gaussian, are on average broader than those of the (1,1) line. The fitting of the (2,2) transition by a Gaussian does not take optical depths effects and the underlying hyperfine structure splitting into account. For the (2,2) line the Gaussian fits can therefore result in broader linewidths than for the (1,1) transition, for which we used a hyperfine structure fit. To investigate this difference, we used a subsample of southern ATLASGAL sources with higher S/N data (Wienen et al. in prep.), for which we were able to detect the hyperfine structure of the (2,2) transition, and compared their linewidths derived from Gaussian and hyperfine structure fits. This resulted in the correlation $\Delta v_{\mbox{\tiny Gauss}}(2,2) = 1.15 \cdot \Delta v_{\mbox{\tiny HFS}}(2,2) + 0.13$, which is a better fit to the data as shown by the solid red line in Fig. \ref{dv11-dv22-atlasgal}. The contribution to the line broadening from the optical depth of the line is 15\% with a (2,2) optical depth of 1.24, an additional broadening results from the magnetic hyperfine structure of the (2,2) line, which is only of about 0.1 km~s$^{-1}$ and therefore cannot be resolved. The contour plot in the top panel of Fig. \ref{dv11-dv22-atlasgal} shows that the contour lines are distributed equally around the red line. However, the lower panel reveals that some sources still lie above the red line, which indicates broader (2,2) than (1,1) linewidths. Because we assumed that the same volume of gas emits the (1,1) and (2,2) inversion lines, the same beam filling factor was used for all transitions in calculating the temperature. The difference in the (1,1) and (2,2) linewidths now shows that some measured NH$_3$ lines do not exactly trace the same gas and using equal beam filling factors for those is therefore only an approximation.\\
The widths of the NH$_3$ inversion lines consist of a thermal and a non-thermal contribution. The thermal linewidth is given by 
\begin{eqnarray}
\Delta v_{\mbox{\tiny therm}}=\sqrt{\frac{8ln(2)kT_{\mbox{\tiny kin}}}{m_{\mbox{\tiny NH}_3}}}
\end{eqnarray}
with the kinetic temperature $T_{\mbox{\tiny kin}}$, the Boltzmann constant $k$ and the mass of the ammonia molecule $m_{\mbox{\tiny NH}_3}$. Assuming a mean $T_{\mbox{\tiny kin}}$ of 20 K, the thermal linewidth is about 0.22 km~s$^{-1}$. On average measured NH$_3$(1,1) linewidths are $\sim 2$ km~s$^{-1}$, i.e. they are dominated by non-thermal contributions.\\
In comparison to \cite{1999ApJS..125..161J}, who measured NH$_3$ lines of low-mass cores and derived a mean linewidth of about 0.74 km~s$^{-1}$, the linewidths of the ATLASGAL sources are broader on average. The beamwidth of the Effelsberg 100 m telescope is 40$\arcsec$ at frequencies of the NH$_3$ lines, corresponding to a linear scale of about 0.8 pc at a typical near distance of our sources of $\sim$ 4 kpc. Hence, several cores might fill the Effelsberg beam (see discussion of beam filling factors in Sect. \ref{beamfilling}). Assuming that each core possesses a narrow linewidth, they may add up to the observed linewidths due to a velocity dispersion of the cores within the telescope beam. High-resolution measurements using  interferometers are important for investigating individual cores, because they might reveal possible non-thermal contributions to the linewidth of each of them.\\
To characterize the ammonia sample, Figs. \ref{dv11-t11-atlasgal} and \ref{dv11-tau11-atlasgal} illustrate the main beam brightness temperature and the optical depth of the NH$_3$ (1,1) lines plotted against the (1,1) linewidth. The binning used for the (1,1) linewidth is 1 km/s, the (1,1) peak intensity values are divided into bins of 1 K and the (1,1) optical depth into bins of 1 as well. The (1,1) peak intensity ranges between 0.2 and 6 K with a peak at 1 K, as displayed by the contour plot. Values of the (1,1) optical depth lie between 0.5 and 5.5 with an average error of 14\% and a peak at 1.8. There are no clear trends seen in the (1,1) main beam brightness temperature and the optical depth.

\subsection{Rotational temperature}
\label{rotational temperature}
\begin{figure}[h]
\centering
\includegraphics[angle=-90,width=9.0cm]{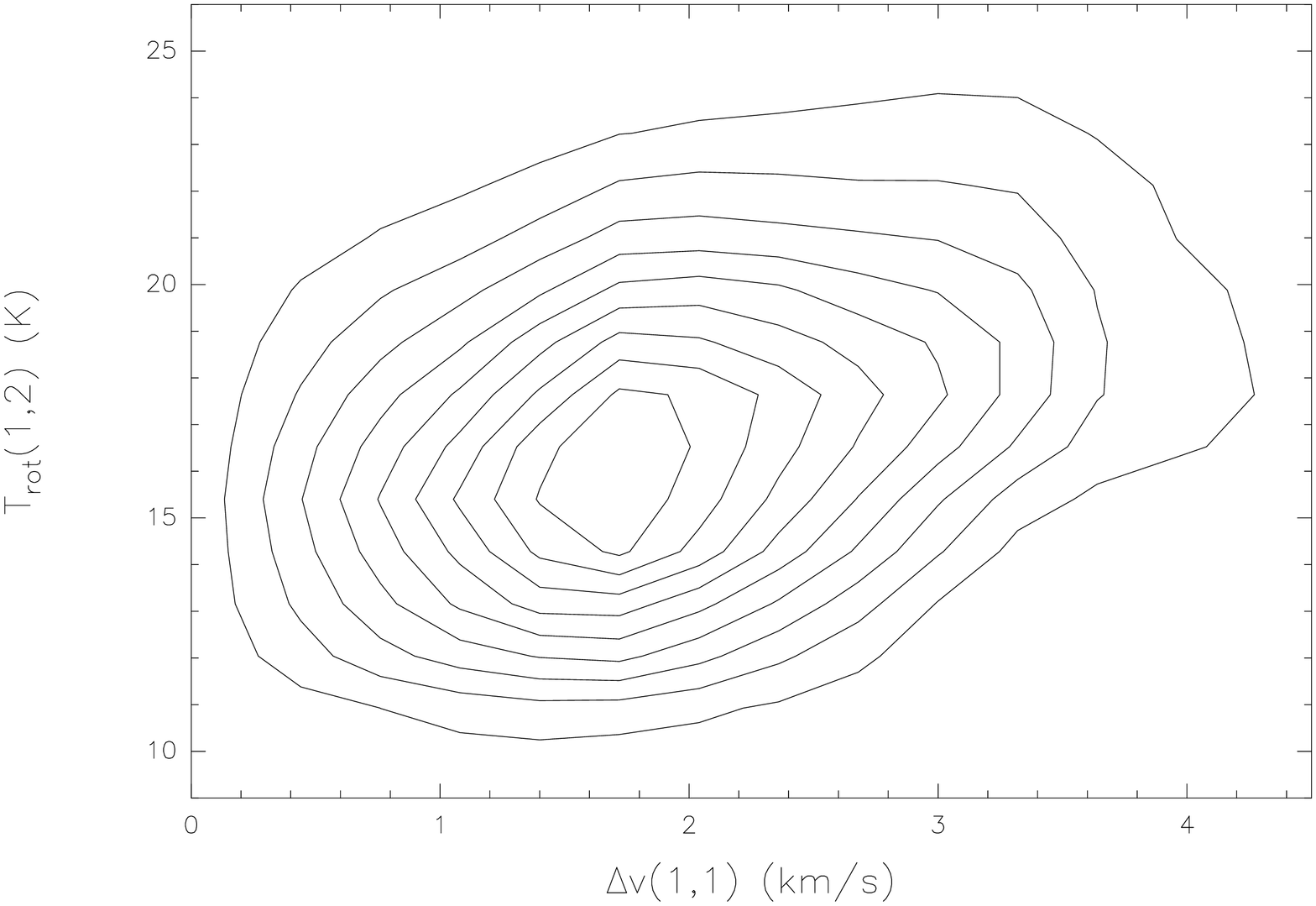}\vspace*{0.5cm}
\includegraphics[angle=-90,width=9.0cm]{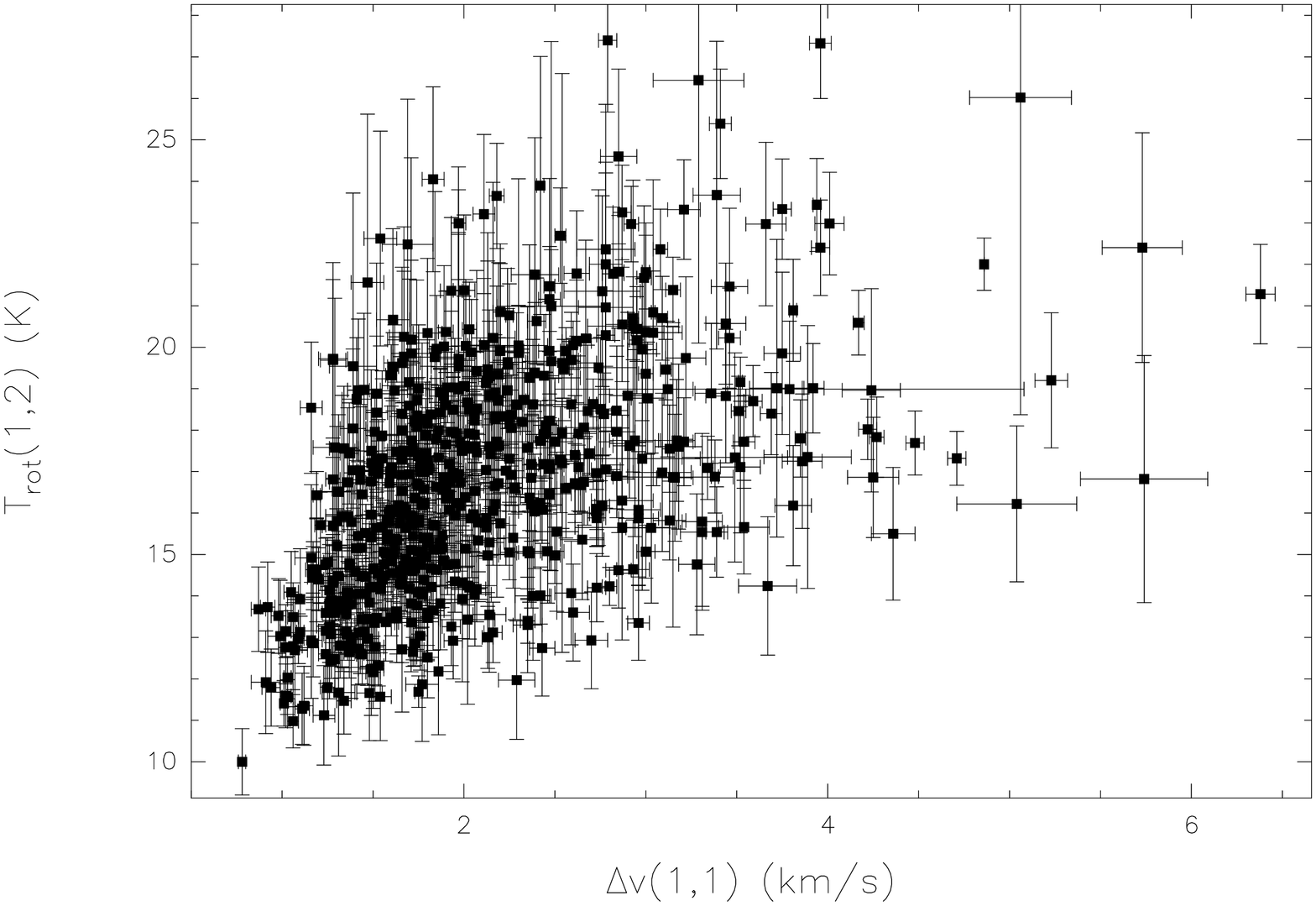}
\caption[correlation between the (1,1) linewidth and the rotational temperature]{Dependence of the rotational temperature between the (1,1) and (2,2) inversion transition on the NH$_3$(1,1) linewidth is shown as contour plot in the top panel and as a scatter plot in the lower panel. The binning of the (1,1) linewidth in the contour plot is 1 km/s and that of the rotational temperature is 2 K.}\label{dv11-trot-atlasgal}
\end{figure}

The rotational temperature between the (1,1) and (2,2) inversion transition can be determined via the relation \citep{1983ARA&A..21..239H}
\begin{eqnarray}\label{Trot}
T_{\mbox{\tiny rot}}=\frac{-41.5}{\ln\left( \frac{-0.282}{\tau_m (1,1)}\ln \left( 1-\frac{T_{\mbox{\tiny MB}}(2,2)}{T_{\mbox{ \tiny MB}}(1,1)}\left( 1- \rm exp(-\tau_m (1,1))\right) \right) \right)}
\end{eqnarray}
with the optical depth of the (1,1) main line, $\tau_m (1,1)$, and the main beam brightness temperatures, $T_{\mbox{\tiny MB}}$, of the (1,1) and (2,2) inversion transitions. The NH$_3$ lines of most sources are well fitted, although the maximum intensity of some lines is slightly underestimated by the fit. The rotational temperature lies between 10 K and 28 K with an average error of 9\% (Table \ref{parabgel-atlasgal}) and a peak at $\sim$ 16 K; it is plotted against the (1,1) linewidth in Fig. \ref{dv11-trot-atlasgal}. We chose only sources with reasonably well-determined rotational temperatures (errors smaller than $\sim 50$\%) for the correlation plot. There is a trend of increasing rotational temperature with broader width of the (1,1) lines. This, together with the broadening of the (2,2) lines, which probe higher temperature regions compared to the (1,1) lines, shows an increase of turbulence with temperature and is an indication for star formation feedback such as stellar winds or outflows. In contrast, some clumps exhibit similar (2,2) and (1,1) linewidths and low rotational temperatures. Those sources could be prestellar clumps without any star formation yet and therefore low temperatures. There are also some clumps with broad (1,1) lines and low rotational temperatures that have large virial masses. The virial mass is a measure of the kinetic energy of a cloud (cf. Sect. \ref{virial mass}). Some examples of the most extreme sources in Fig. \ref{dv11-trot-atlasgal} are, e.g. the clump with an extremely narrow (1,1) linewidth and/or low rotational temperature, G34.37$-$0.66 with $\Delta v$(1,1) = 0.78 km~s$^{-1}$ and $T_{\mbox{\tiny rot}}$ = 10 K, a known IRDC \citep{2009A&A...505..405P}. Among the clumps with broad linewidth are well-known sources, such as the UCHIIR/hot core G10.47+0.03 \citep{1992A&A...256..618C} (see Sect. \ref{linewidth}) with $\Delta v$(1,1) = 6.38 km~s$^{-1}$ and $T_{\mbox{\tiny rot}}$ = 21 K. Another source with a high $T_{\mbox{\tiny rot}}$ of 26 K and $\Delta v$(1,1) of 5.06 km~s$^{-1}$ is G43.80$-$0.13, well-known for harbouring OH \citep{1983A&AS...54..167B} and H$_2$O maser emission \citep{2000A&AS..141..185L}. G12.21$-$0.10 has already been observed with the VLA at 5 GHz by \cite{1994ApJS...91..347B}, who identified it as a possible UCHII region. Hence, this clump is probably in a more evolved phase of high-mass star formation and therefore has a high $\Delta v$(1,1) of 5.73 km~s$^{-1}$ as well and $T_{\mbox{\tiny rot}}$ of 22 K, it is also associated with an IRAS point source \citep{1990ApJS...74..181Z}.

\subsection{Source-averaged NH$_3$ column density and kinetic temperature}
\label{column density}
\begin{figure}[h]
\centering
\includegraphics[angle=-90,width=9.0cm]{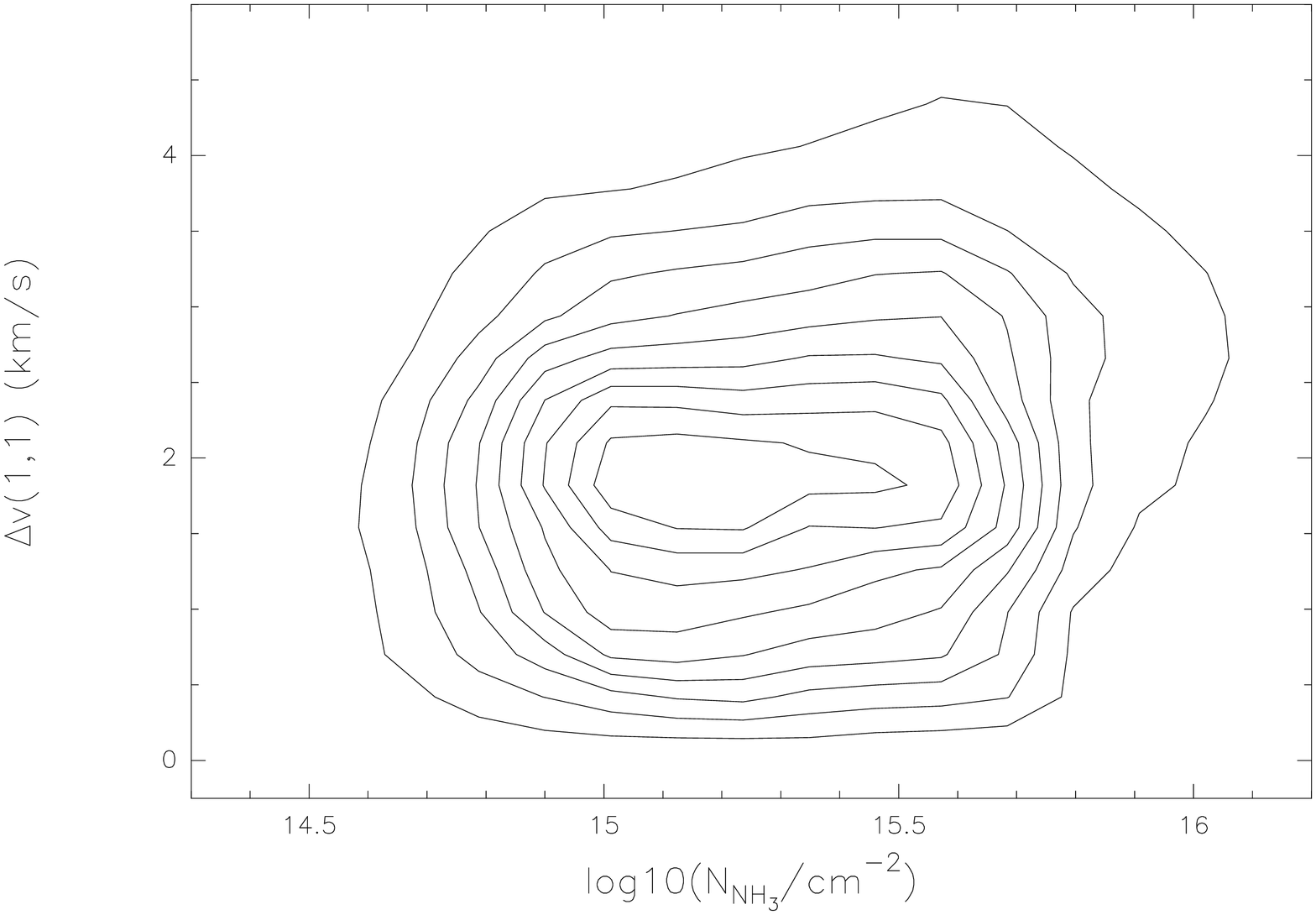}\vspace*{0.5cm}
\includegraphics[angle=-90,width=9.0cm]{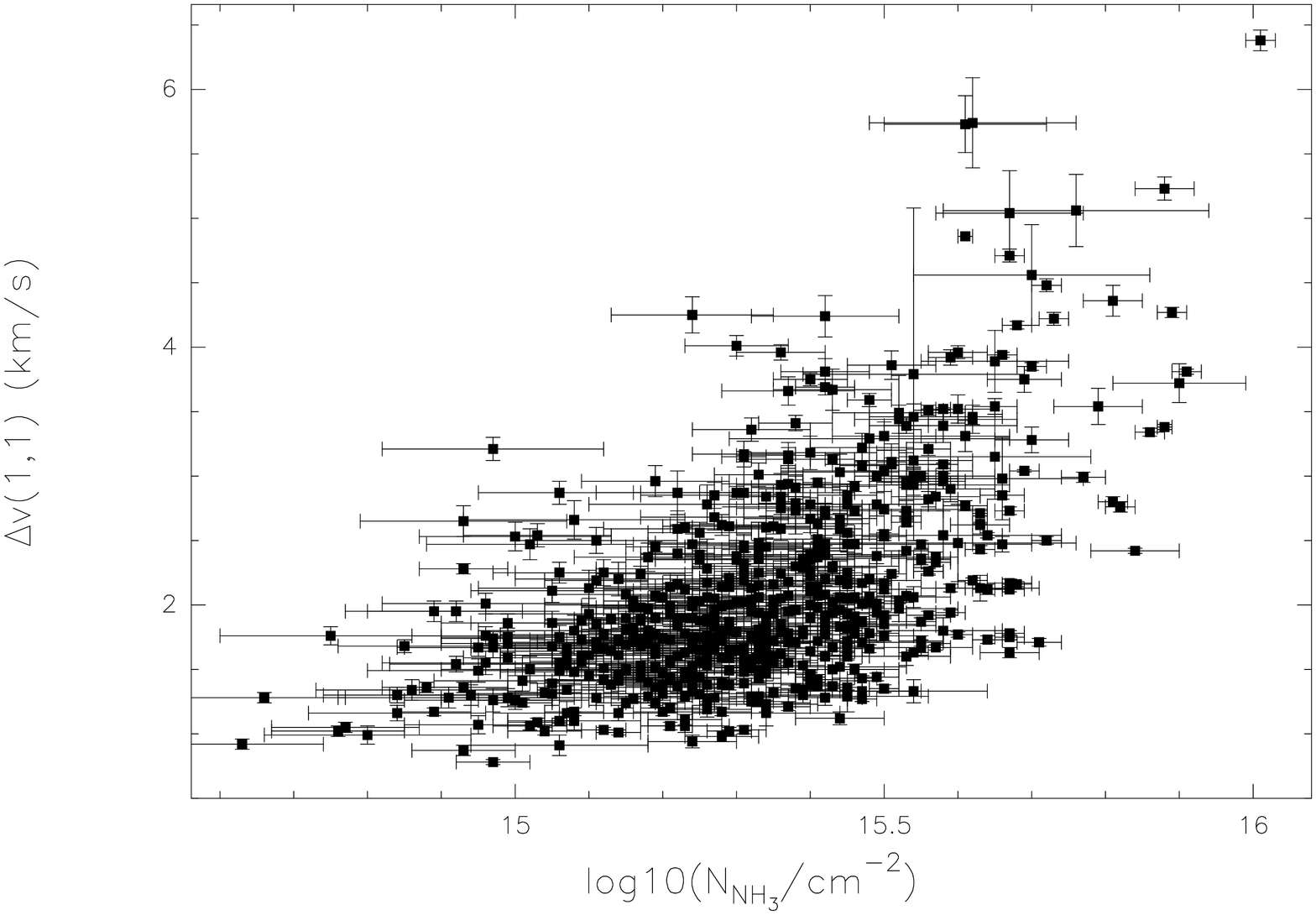}
\caption[dependence of the (1,1) linewidth and the kinetic temperature on the logarithm of the column density]{Correlation plot of the logarithm of the column density and the (1,1) linewidth. The contour plot is illustrated in the upper panel. For the logarithm of the column density bins of 0.4 cm$^{-2}$ and for the (1,1) linewidth bins of 1 km/s are chosen.}\label{nh3-dv11-atlasgal}
\end{figure}

\begin{figure}[h]
\centering
\includegraphics[angle=-90,width=9.0cm]{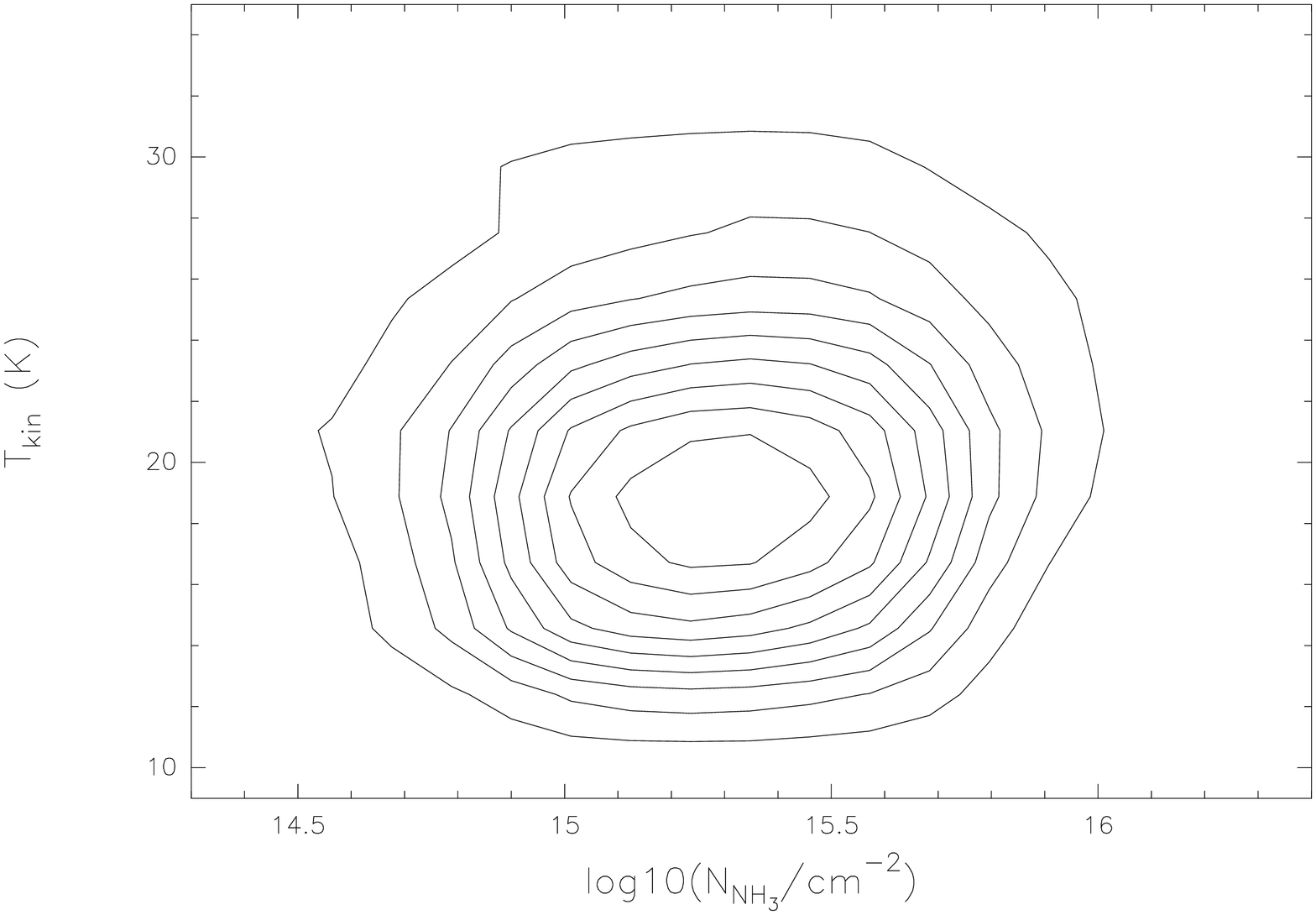}\vspace*{0.5cm}
\includegraphics[angle=-90,width=9.0cm]{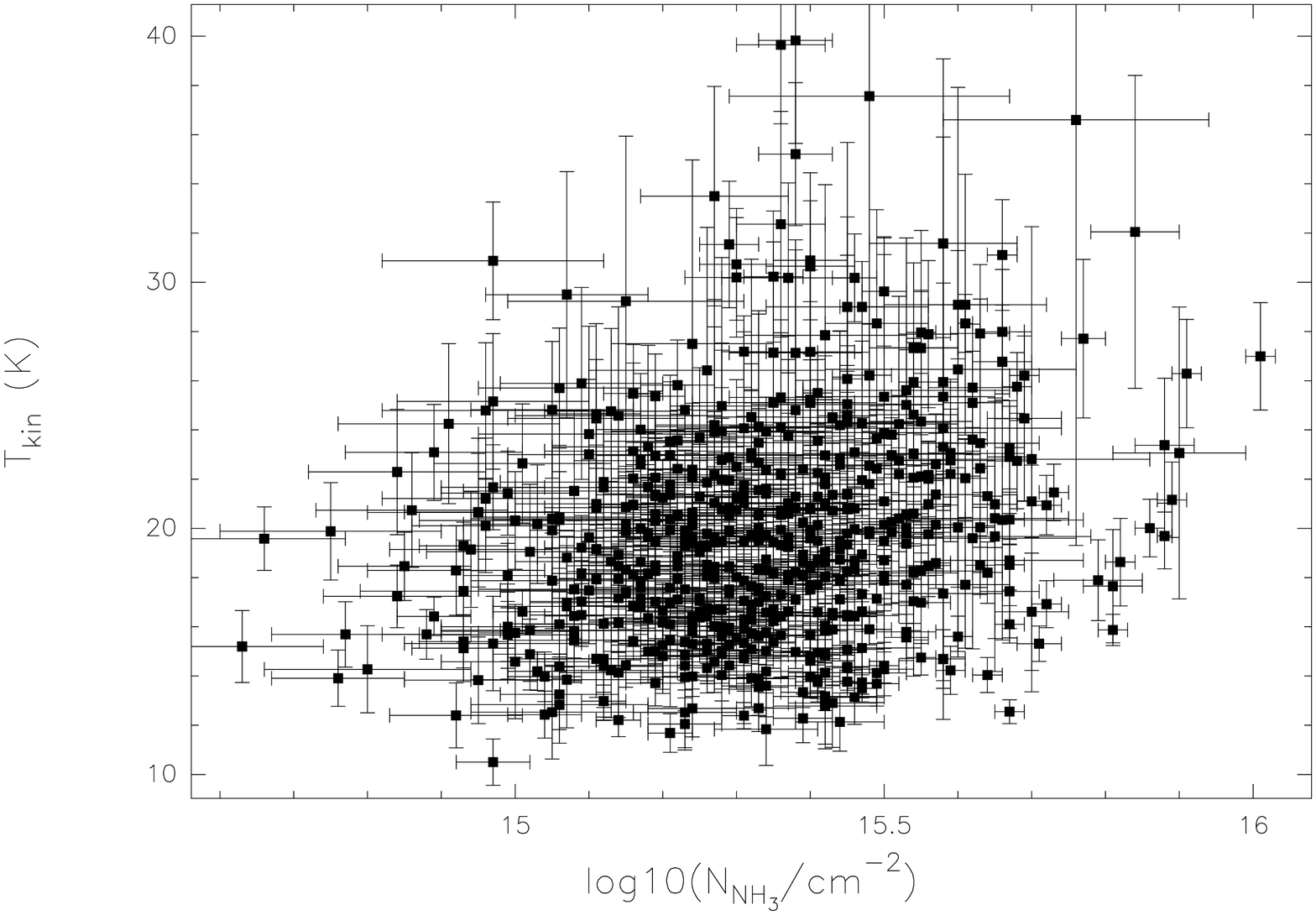}
\caption[dependence of the (1,1) linewidth and the kinetic temperature on the logarithm of the column density]{Logarithm of the column density compared with the kinetic temperature. There is no correlation between both. Contour lines of the NH$_3$ parameters are shown in the upper plot. Bins of 0.4 cm$^{-2}$ are used for the logarithm of the column density and bins of 2 K for the kinetic temperature.}\label{nh3-tkin-atlasgal}
\end{figure}

To calculate the source-averaged ammonia column density, the optical depth and linewidth of the (1,1) inversion transition are needed as well as the rotational temperature, which is derived from equation \ref{Trot}. Because the optical depth and the rotational temperature depend only on line ratios, the resulting column density is a source-averaged quantity. The values of the rotational temperature, kinetic temperature, and the logarithm of the NH$_3$ column density are given in Table \ref{parabgel-atlasgal}. The total column density was derived from the column density of the (1,1) level ($N(1,1)$) assuming that the energy levels are populated via a Boltzmann distribution, defined by the rotational temperature. Another assumption is that only the four lowest metastable levels are populated because of the low rotational temperatures of the cloud clump, which is ranging from 10 to 25 K, and H$_2$ densities of 10$^5$ cm$^{-3}$. Hence, the total column density is given by \cite{2004tra..book.....R}
\begin{eqnarray}
N_{\mbox{\tiny tot}} \approx N(1,1) \Bigg( \frac{1}{3} \, {\rm exp} \left(\frac{23.1}{T_{\mbox{\tiny rot}}(1,2)}\right)+1+\frac{5}{3} \, {\rm exp} \left(-\frac{41.2}{T_{\mbox{\tiny rot}}(1,2)}\right)  \\ \nonumber
+\frac{14}{3} \, {\rm exp}\left( -\frac{99.4}{T_{\mbox{\tiny rot}}(1,2)}\right)\Bigg) 
\end{eqnarray}
with
\begin{eqnarray}
N(1,1) = 2.07\cdot 10^3\frac{g_l \ \nu^2 \ T_{\mbox{\tiny rot}}}{g_uA_{ul}}\tau \Delta v,
\end{eqnarray}
where $g_l$ and $g_u$ are the statistical weights of the lower and upper inversion level, $A_{ul}$ the Einstein $A$ coefficient, $\nu$ the inversion frequency in GHz and $\Delta v$ the linewidth in km~s$^{-1}$. Weak sources, whose errors of the column density are greater than 50\%, were excluded from correlation plots (cf. Fig. \ref{nh3-dv11-atlasgal}).\\
The kinetic temperature is obtained from the rotational temperature \citep{2004A&A...416..191T}:
\begin{eqnarray}\label{Tkin}
T_{\mbox{\tiny kin}}=\frac{T_{\mbox{\tiny rot}}(1,2)}{1-\frac{T_{\mbox{\tiny rot}}(1,2)}{42 {\rm K}}\ln\left( 1+1.1 \, {\rm exp}\left(\frac{-16 {\rm K}}{T_{\mbox{\tiny rot}}(1,2)}\right)\right) },
\end{eqnarray}
where the energy difference between the NH$_3$ (1,1) and (2,2) inversion levels is 42 K. \cite{2004A&A...416..191T} ran different Monte Carlo models to compare their derived rotational temperatures with the kinetic temperatures used in the models. Equation \ref{Tkin} is derived from their fit of the kinetic and rotational temperatures. These authors predicted kinetic temperatures in the range of 5 K to 20 K, in which most of our sources fall with a peak of $T_{\mbox{\tiny kin}}$ at 19 K, to better than 5\%. The logarithm of the averaged NH$_3$ column density is compared with the (1,1) linewidth in Fig. \ref{nh3-dv11-atlasgal} and with the kinetic temperature in Fig. \ref{nh3-tkin-atlasgal}. The column density ranges between $4 \times 10^{14}$ cm$^{-2}$ and $10^{16}$ cm$^{-2}$ with an average error of 17\% (cf. Table \ref{parabgel-atlasgal}) and a peak at $2\times 10^{15}$ cm$^{-2}$. The source with the highest column density of $10^{16}$ cm$^{-2}$ is G10.47+0.03 \citep{1992A&A...256..618C}. The range of column densities of ATLASGAL sources compares well with values found for UCHIIRs by \cite{1989ApJS...69..831W} and high-mass protostellar objects (HMPOs), that are considered to be in a pre-UCHII region phase \citep{2002ApJ...570..758M, 2002ApJ...566..945B}. It is somewhat above that of low-mass, dense cores from a NH$_3$ database \citep{1999ApJS..125..161J}, whose column densities lie between 10$^{14}$ cm$^{-2}$ and $3\times 10^{15}$ cm$^{-2}$.

\subsection{Beam filling factor}
\label{beamfilling}
\begin{figure}[h]
\centering
\includegraphics[angle=-90,width=9.0cm]{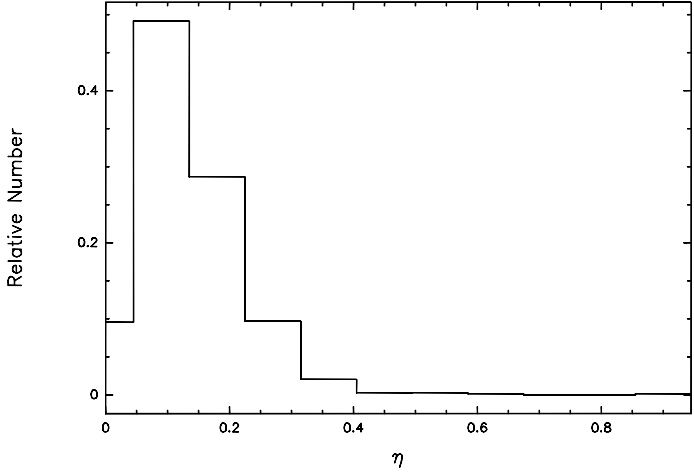}
\caption[beam filling factor]{Histogram of beam filling factor computed according to equation \ref{beamfactor}. Most sources have a beam filling factor of 0.1.}\label{beamfactor-histo}
\end{figure}

The beam filling factor, $\eta$, which gives the fraction of the beam filled by the observed source, is derived from equations of radiative transfer
\begin{eqnarray}\label{beamfactor}
 \eta = \frac{T_{\mbox{\tiny MB}} (1,1)}{(T_{\mbox{\tiny ex}}-T_{\mbox{\tiny bg}})\cdot (1- {\rm exp}( -\tau (1,1)))},
\end{eqnarray}
where $T_{\mbox{\tiny MB}} (1,1)$ is the the main beam brightness temperature of the (1,1) inversion transition, $\tau (1,1)$ the optical depth of the (1,1) main line and $T_{\mbox{\tiny bg}}$=2.73 K. Because the H$_2$ density of the ATLASGAL clumps is $\sim 10^5$ cm$^{-3}$ \citep{2002ApJ...566..945B, 2003ApJ...582..277M}, we assumed local thermodynamic equilibrium (LTE) and accordingly used the kinetic temperature as excitation temperature in Eq. \ref{beamfactor}. 
We found filling factors mostly in the range between 0.02 and 0.6 with one higher value of $\sim 0.9$ (cf. Table \ref{par870mikrom-atlasgal}). Their distribution is shown as a histogram in Fig. \ref{beamfactor-histo}. Equation \ref{beamfactor} can lead to underestimation of the filling factor in cases of sub-thermal excitation. To investigate the deviation of the excitation temperature from the kinetic temperature, we used the non LTE molecular radiative transfer program RADEX \citep{2007A&A...468..627V}. For a background temperature of 2.73 K, an average kinetic temperature of our sample of 19 K, an H$_2$ density of 10$^5$cm$^{-3}$, a column density of 10$^{15}$ cm$^{-2}$ and a linewidth of 1 km~s$^{-1}$ we obtained for para-NH$_3$ an excitation temperature between the lower and upper (1,1) inversion level of $\sim 18$ K. Hence, the excitation and kinetic temperature agree within 94\%, which does not hint at sub-thermal excitation. Since typical ATLASGAL source sizes, obtained from Gaussian fits of the 870 $\mu$m dust continuum flux, are $\sim$ 40$\arcsec$ (cf. Sect. \ref{virial mass}), low beam filling factors indicate clumpiness on smaller scales, as is also evident from interferometer NH$_3$ observations \citep[e.g.][]{2011ApJ...733...44D, 2011ApJ...736..163R, 2010ApJ...715.1132O}.

\subsection{NH$_3$ (3,3) line parameters}
\begin{figure}[h]
\centering
\includegraphics[angle=-90,width=9.0cm]{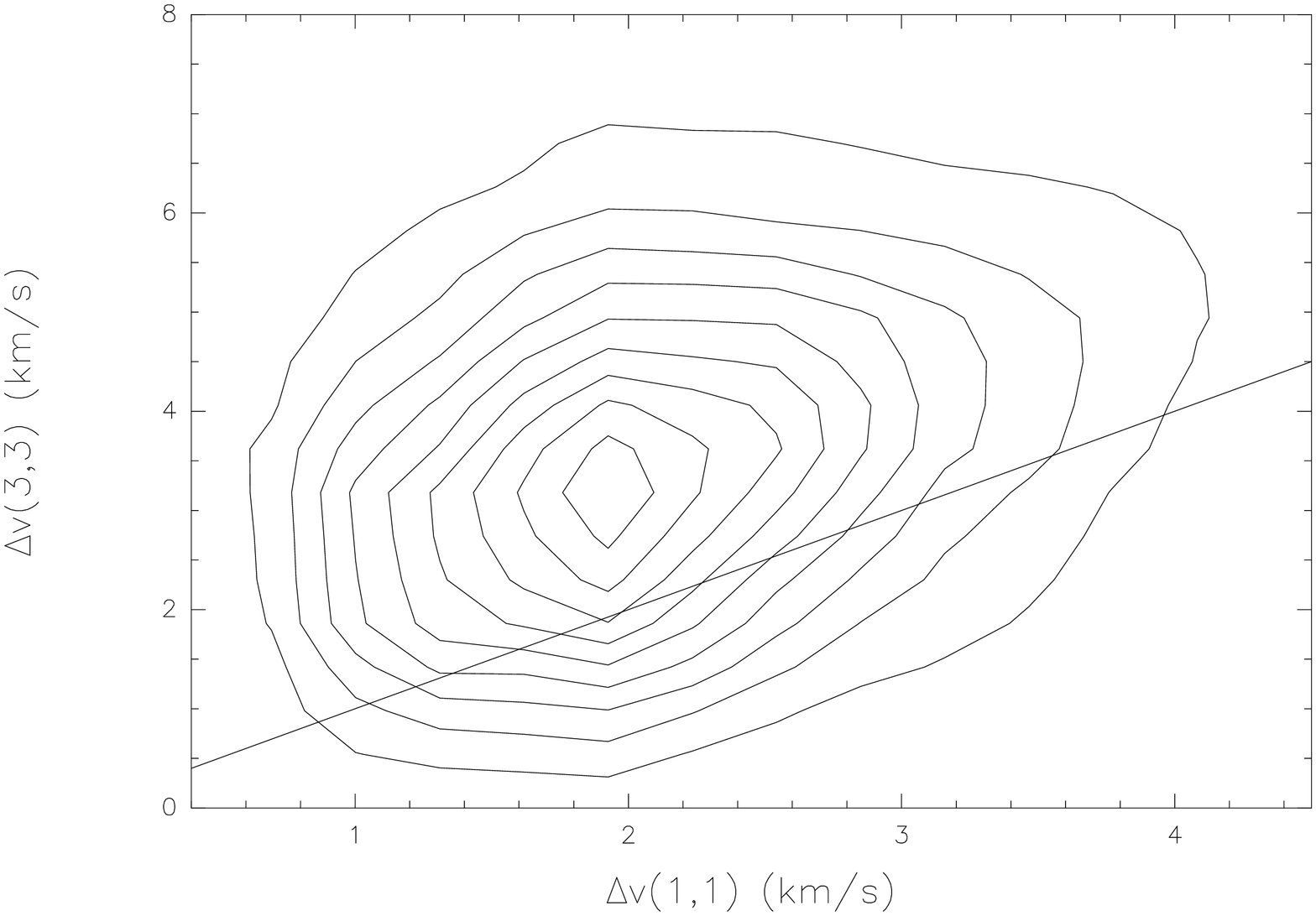}\vspace*{0.5cm}
\includegraphics[angle=-90,width=9.0cm]{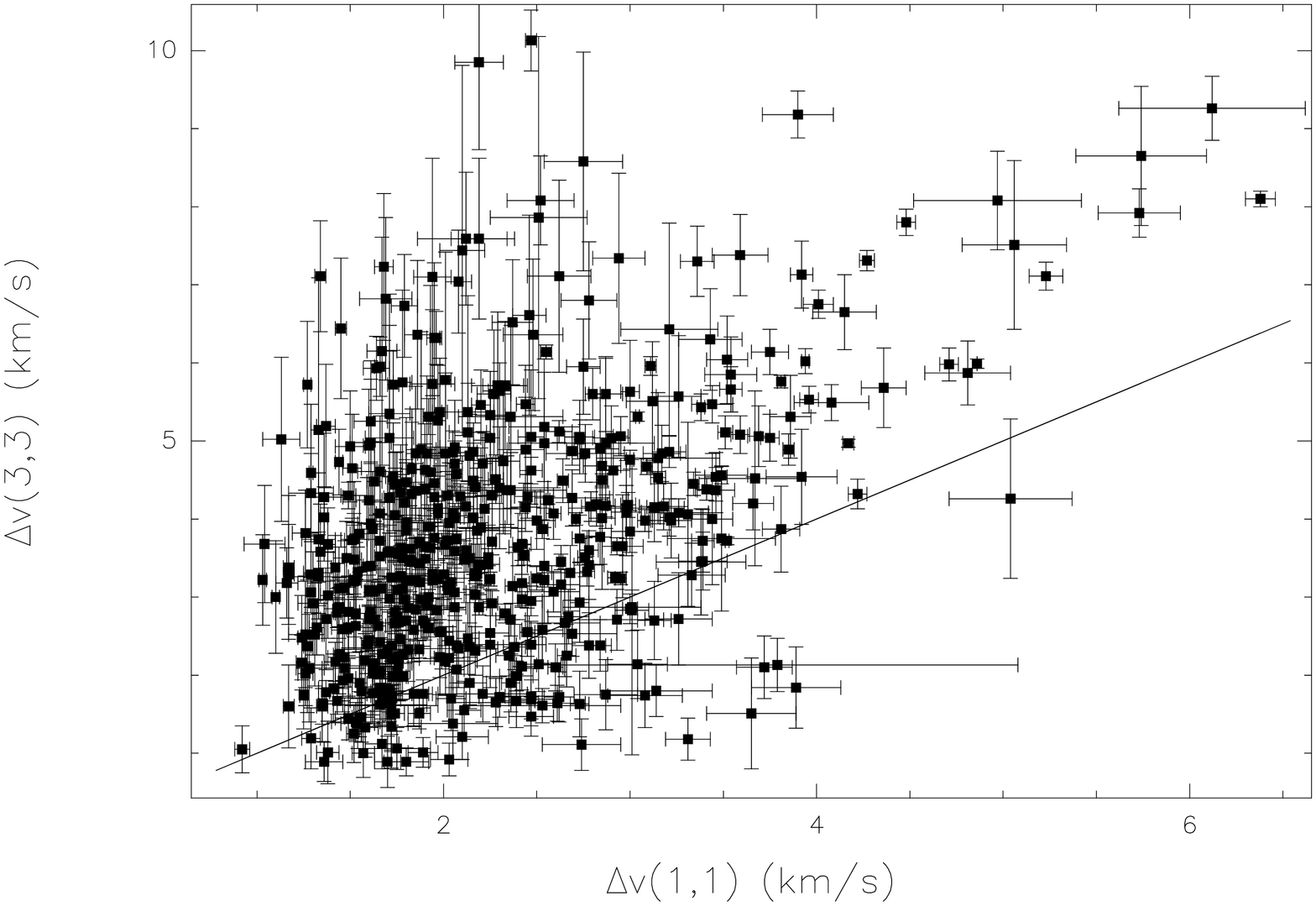}
\caption[correlation of the (1,1) and (3,3) linewidths]{Width of the NH$_3$ (3,3) line plotted against that of the (1,1) line. The straight line shows equal widths. The contour plot is displayed in the top panel with a binning of 0.7 km/s for the range of (1,1) linewidths and of 1 km/s for that of (3,3) linewidths.}\label{dv11-dv33-atlasgal}
\end{figure}

\begin{figure}[!h]
\centering
\includegraphics[angle=-90,width=8.8cm]{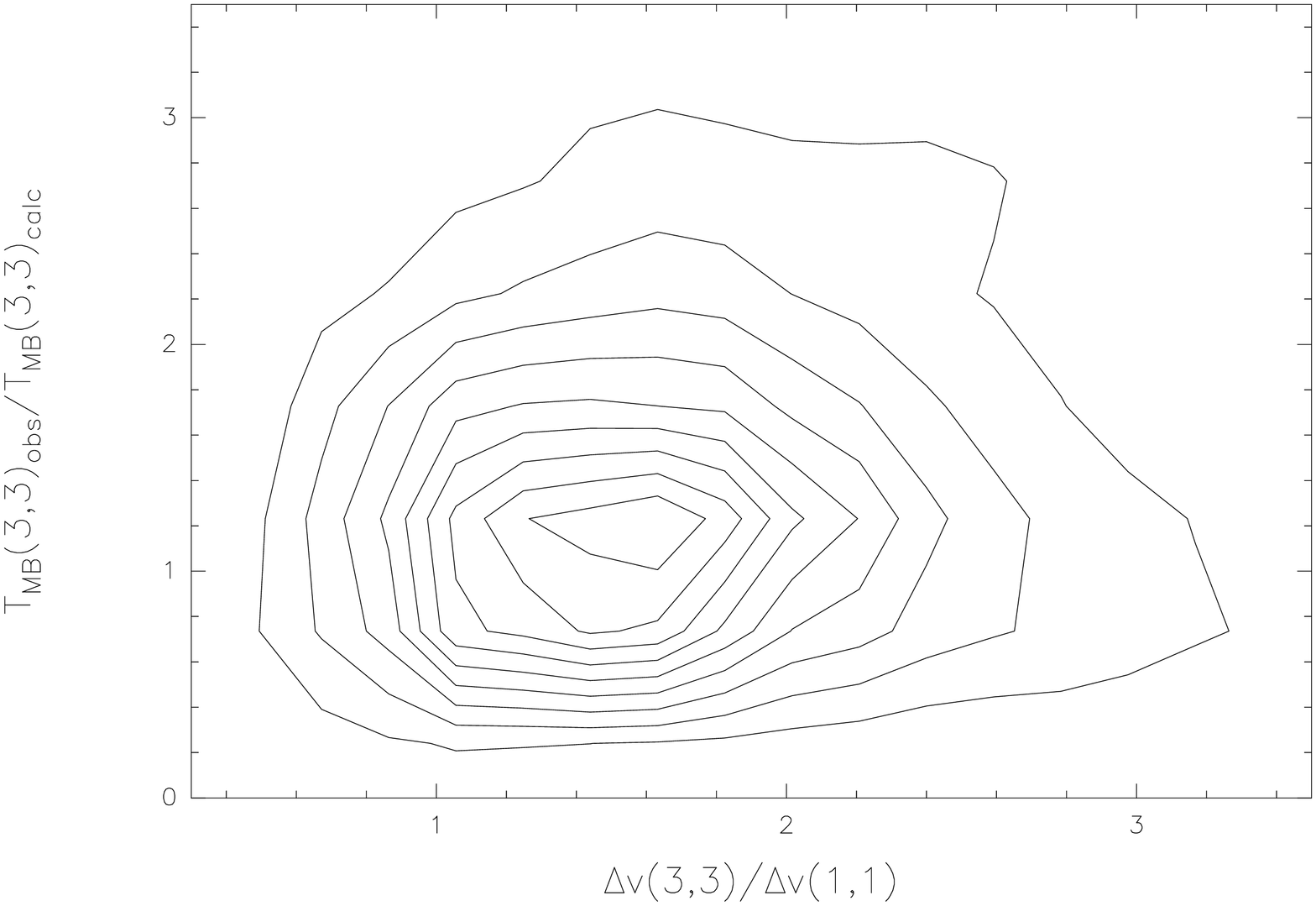}\vspace*{0.5cm}
\includegraphics[angle=-90,width=8.8cm]{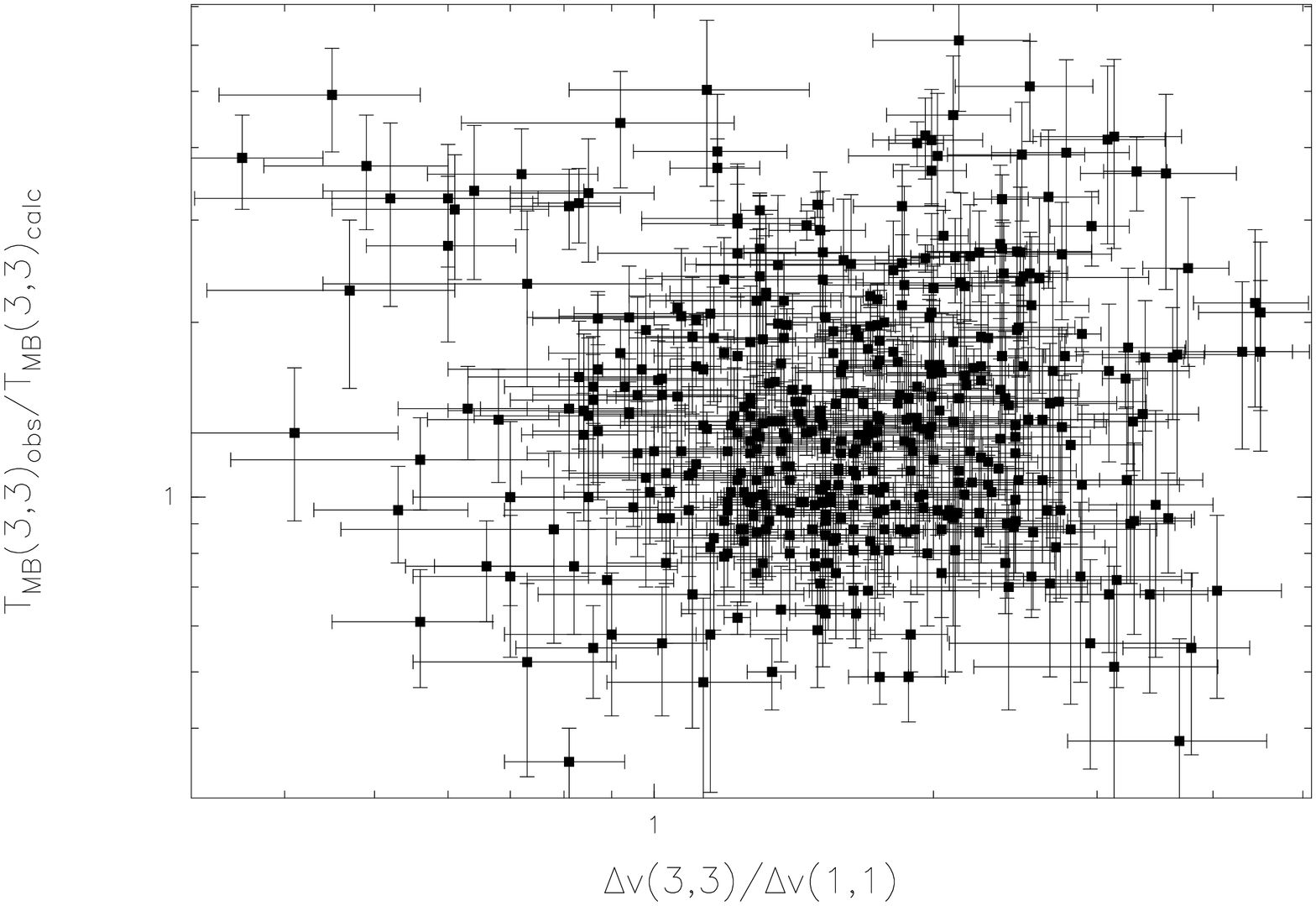}\\
\includegraphics[angle=-90,width=9.5cm]{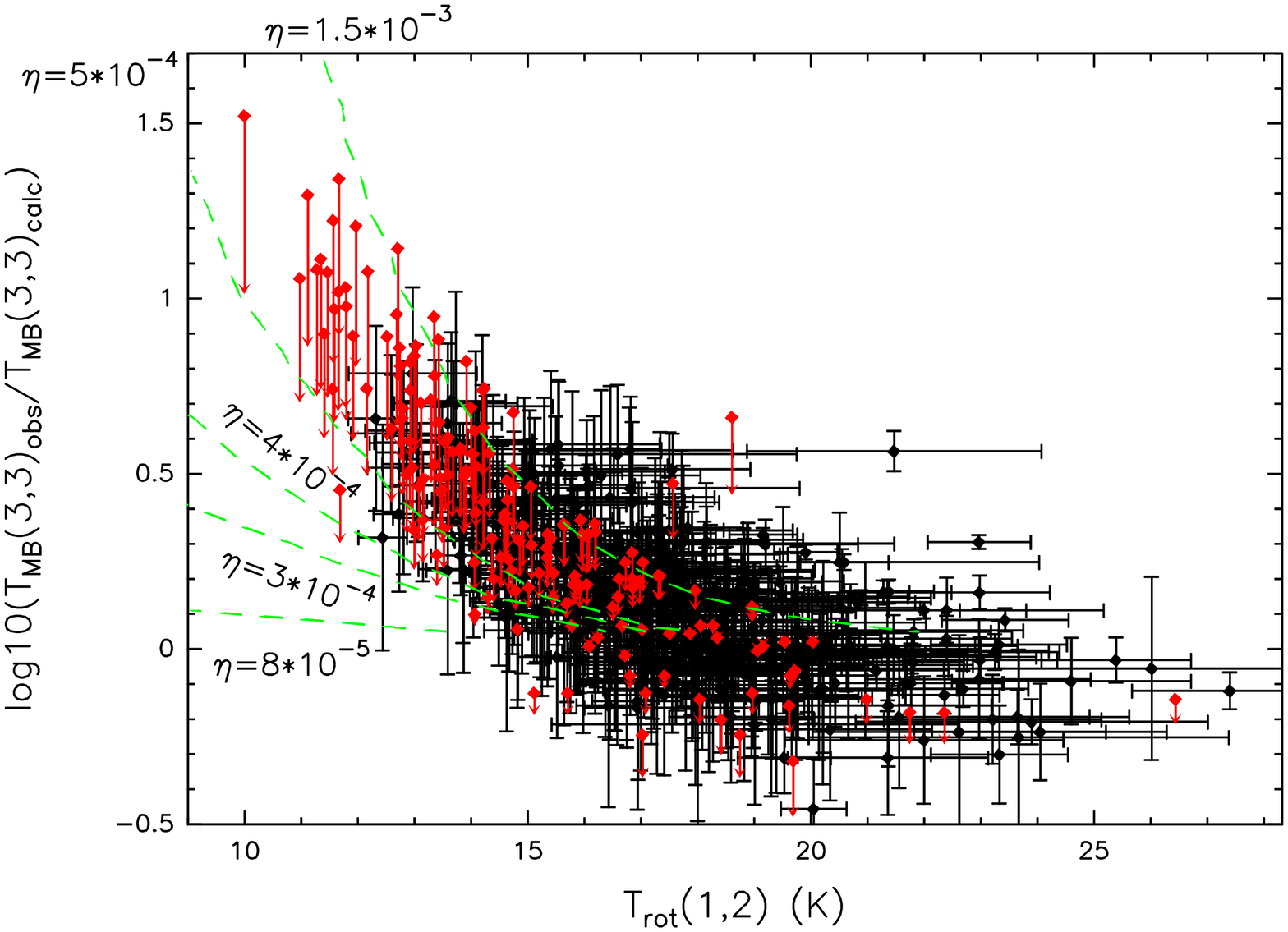}
\caption[correlation of the (1,1) and (3,3) linewidths]{Ratio of observed to calculated (3,3) main beam brightness temperatures compared to the ratio of the (3,3) to (1,1) linewidths as a contour plot in the top panel and as a scatter plot in the middle panel. The correlation plot of the rotational temperature and observed to calculated (3,3) main beam brightness temperature ratio is displayed in the lowest panel. Red points show sources that are not detected in the NH$_3$ (3,3) line. Green dashed curves indicate very low hot core beam filling factors between 0.008\% and 0.15\%.}\label{dv-t33-hotcores}
\end{figure}

\begin{figure}[h]
\centering
\includegraphics[angle=-90,width=9.0cm]{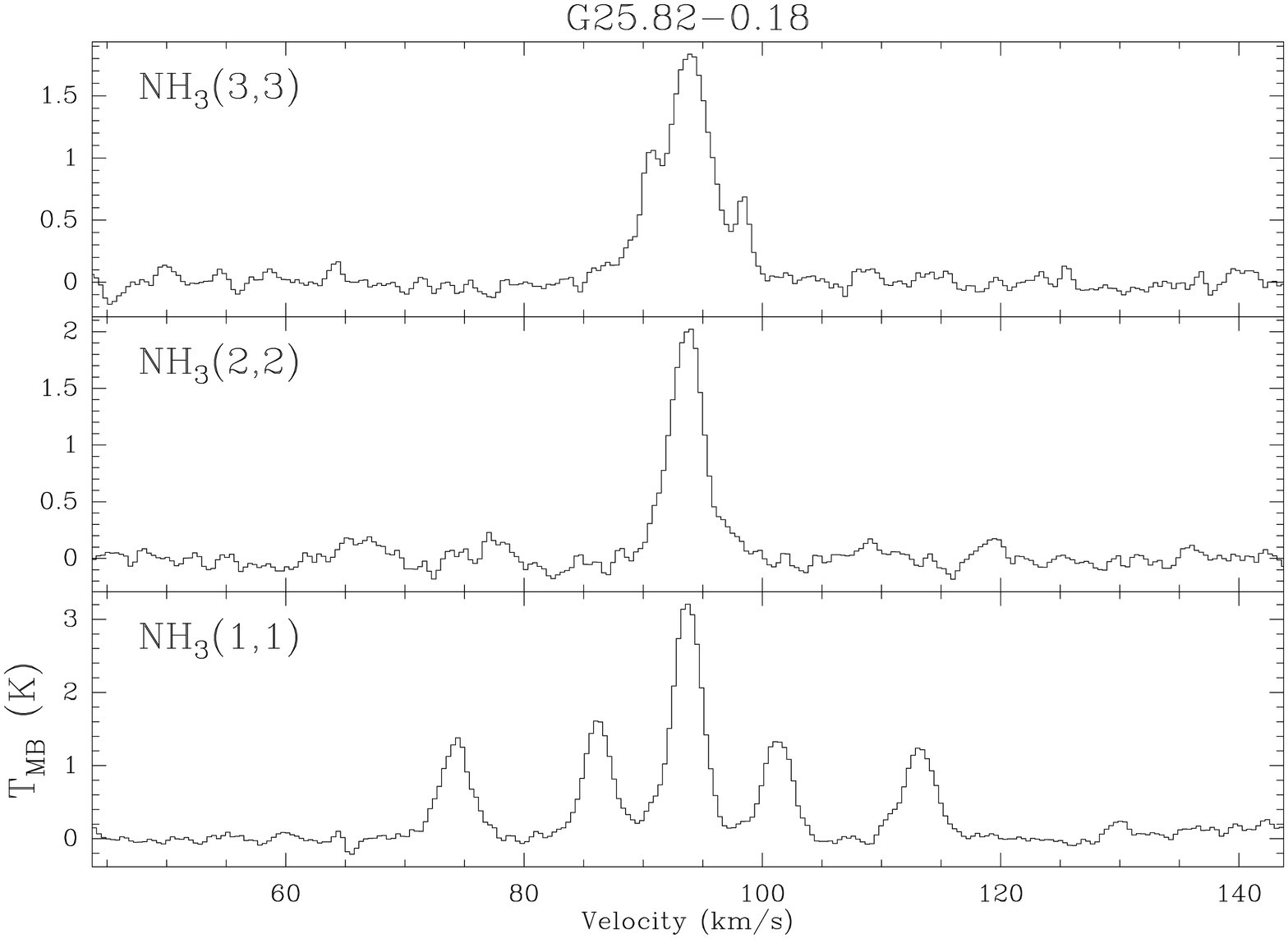}\vspace*{0.5cm}
\includegraphics[angle=-90,width=9.0cm]{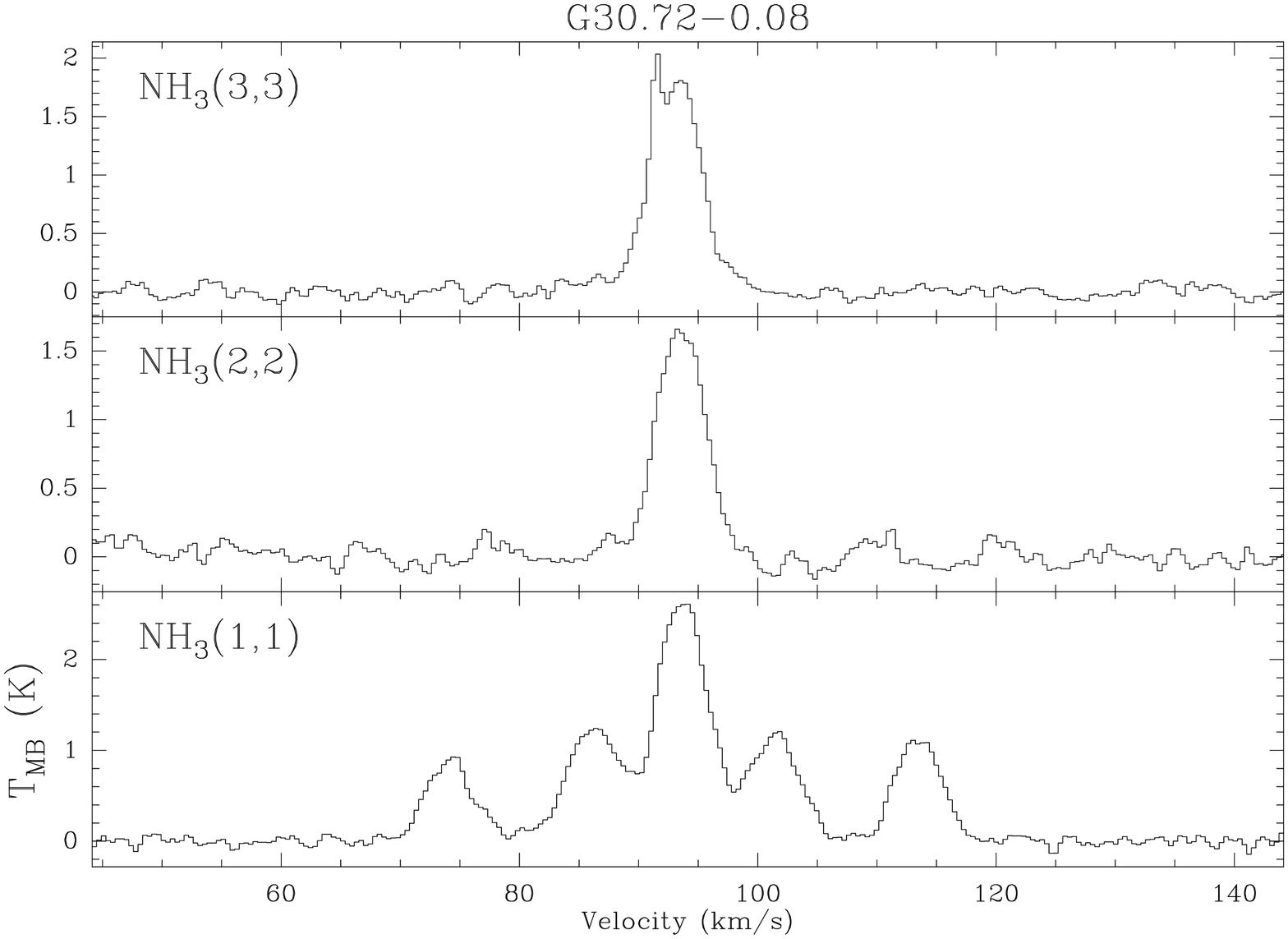}
\caption[maser emission]{NH$_3$ (1,1) to (3,3) lines of two ATLASGAL clumps, G25.82-0.18 at the top and G30.72-0.08 at the bottom: The (3,3) line profile is different from those of the (1,1) and (2,2) transitions, indicating a narrow maser line in addition to the thermal emission.}\label{33maser}
\end{figure}

In Fig. \ref{dv11-dv33-atlasgal} we plot the widths of the ammonia (3,3) line, $\Delta v$(3,3), against the (1,1) line, $\Delta v$(1,1). The straight line shows equal widths. A binning of 0.7 km/s is chosen for $\Delta v$(1,1) and of 1 km/s for $\Delta v$(3,3) for the contour plot. While the (1,1) linewidths lie between 1 and 6 km~s$^{-1}$, the widths of the (3,3) line range from 1 to 10 km~s$^{-1}$ (cf. Table \ref{parline22_line33-atlasgal}) and, with a peak at 3.5 km~s$^{-1}$, are broader than those of the (1,1) inversion transition for most clumps. Because the excitation of the (3,3) inversion transition requires a high temperature, the (3,3) emitting gas is likely heated by embedded already formed or forming stars that generate the turbulence observed in the increased linewidth.\\
Furthermore, we investigated whether the observed NH$_3$ (3,3) temperatures are consistent with the temperatures found from the (1,1) and (2,2) lines or whether they probe an additional, embedded hot core component of the clumps.
We calculated the expected envelope (3,3) main beam brightness temperature from equations of radiative transfer under LTE conditions using the results from Sects. \ref{rotational temperature} and \ref{column density} and compared it to the measured $T_{\mbox{\tiny MB}}$(3,3).\\ 
Sources towards which the NH$_3$ (3,3) line is not detected have an average (1,1) linewidth of 1.8 km~s$^{-1}$, which is slightly below the mean value of the whole ATLASGAL sample of 2 km~s$^{-1}$. Their average rotational temperature is 15.3 K, which is also lower than the mean of 16.5 K.\\
Assuming LTE (and thus $T_{\mbox{\tiny kin}}$=$T_{\mbox{\tiny ex}}$), the (3,3) main beam brightness temperature is
\begin{eqnarray}
T_{\mbox{\tiny MB}}(3,3)_{\mbox{\tiny calc}}=\eta \cdot T_{\mbox{\tiny kin}}(1-{\rm exp}(-\tau (3,3)))
\end{eqnarray}
assuming the same beam filling factor as for the (1,1) line (see Eq. \ref{beamfactor}) and the optical depth $\tau (3,3)$ of the NH$_3$(3,3) line given by 
\begin{eqnarray}\label{ncol-tau}
\tau (3,3) = \alpha \times \left( 1-{\rm exp} \left( {-\frac{ h\nu (3,3)}{ kT_{\mbox{\tiny kin}}}}\right) \right),
\end{eqnarray}
where $\nu (3,3)$ is the frequency of the (3,3) inversion transition (23.87 GHz), $k$ the Boltzmann constant and $T_{\mbox{\tiny kin}}$ the measured kinetic temperatures. 
$\alpha$ is defined as
\begin{eqnarray}
 \alpha = \frac{c^3}{8\pi\nu (3,3)^3}\frac{2\sqrt{{\rm ln}2}}{\sqrt{\pi}}\frac{1}{\Delta v(1,1)}N(3,3) A_{ul}(3,3). \nonumber
\end{eqnarray}
Here, $c$ denotes the speed of light, $\Delta v$(1,1) the observed (1,1) linewidth, $N(3,3)$ the (3,3) level population number and $A_{ul}(3,3)$ the Einstein $A$ coefficient of the (3,3) transition. From the Boltzmann distribution of energy levels we determined the (3,3) level population number using derived column densities $N_{\mbox{\tiny tot}}$ (cf. Table \ref{parabgel-atlasgal}) \citep{2004tra..book.....R}
\begin{eqnarray}
  N(3,3) = N_{\mbox{\tiny tot}}/Q,
\end{eqnarray}
where Q is given by
\begin{eqnarray}
 Q = \frac{1}{14} \, {\rm exp} \left(\frac{123.57}{T_{\mbox{\tiny kin}}}\right) + \frac{3}{14} \, {\rm exp} \left(\frac{99.4}{T_{\mbox{\tiny kin}}}\right) + \frac{5}{14} \, {\rm exp} \left(\frac{59.12}{T_{\mbox{\tiny kin}}}\right) + 1.  \nonumber
\end{eqnarray}
The ratio of observed to calculated (3,3) main beam brightness temperature is compared with the ratio of (3,3) to (1,1) linewidths in Fig. \ref{dv-t33-hotcores}. The ranges of both ratios are divided into bins of 0.4 to create the contour plot. We found values between 0.3 and 5 for the linewidth ratio and from 0.3 to 6 for the (3,3) line temperature ratio. The median of $\Delta v$(3,3)/$\Delta v$(1,1) is 1.6 with a dispersion of 0.8 and the median of $T_{\mbox{\tiny MB}}(3,3)_{\mbox{\tiny obs}}$/$T_{\mbox{\tiny MB}}(3,3)_{\mbox{\tiny calc}}$ is 1.2 with a dispersion of 1.1. Many clumps have higher observed (3,3) line temperatures than is expected from the calculation using the (1,1) and (2,2) line temperatures, which can have various reasons:
\begin{enumerate}
 \item Non-thermal excitation can overpopulate or invert the upper level of the (3,3) doublet. Different examples of these maser lines have already been observed: \cite{1994ApJ...428L..33M} detected $^{14}$NH$_3$ (3,3) maser emission towards the DR 21(OH) star forming region. \cite{1994ApJ...429L..85H} discovered the NH$_3$ (5,5) maser line with interferometric observations towards the H II region complex G9.62+0.19. Some ATLASGAL clumps have (3,3) line profiles, which are different from those of the (1,1) and (2,2) lines and indicate maser emission; two examples are shown in Fig. \ref{33maser}: The (3,3) line of G25.82$-$0.18 shows two narrow components at lower and higher velocities than $v(3,3)$ in addition to the thermal emission. This clump also harbours CH$_3$OH and water masers \citep{2005A&A...434..613S}. In the NH$_3$ (3,3) spectrum of G30.72$-$0.08 we see an additional peak at a slightly lower velocity than that of the (3,3) thermal emission. This clump is located in the W43 high-mass star forming region and is revealed as a compact cloud fragment detected at 1.3 mm and 350 $\mu$m by \cite{2003ApJ...582..277M}. However, they did not find an association of this source with an OH, H$_2$O or CH$_3$OH maser, but instead a bright compact H II region with a flux of $\sim$ 1 Jy at 3.5 cm. Recent Herschel observations show that it is centred on a 1 pc radius IRDC seen in absorption at 70 $\mu$m and bright at wavelengths longer than 160 $\mu$m \citep{2010A&A...518L..90B}. However, because we did not find strong maser lines, but only weak narrow components in the (3,3) spectra of a small subsample of ATLASGAL clumps, it is unlikely that population inversion of the (3,3) transition explains the trend of increased observed $T_{\mbox{\tiny MB}}(3,3)$ values compared to calculations.  
\item Since radiative and collisional transitions let the spin orientations remain constant, transitions between ortho- and para-NH$_3$ are not allowed. There are processes allowing interconversion between both species proposed by \cite{1969ApJ...157L..13C}, but they are very slow ($\sim 10^6$ yr). The distribution between ortho- and para states is a strong function of temperature \citep{2002PASJ...54..195T}: The abundance ratio will be about 1 if the formation of NH$_3$ occurs in high-temperature regions with more than 40 K, but it will rise if NH$_3$ is produced at low temperatures, because the lowest level (0,0) is an ortho state. Our average rotational temperature is 19 K, for which \cite{2002PASJ...54..195T} calculated an ortho-to-para abundance ratio of 2. This ratio would result in an increased population of the (3,3) level and a decrease of observed to calculated (3,3) line temperature ratio.
\item High $T_{\mbox{\tiny MB}}(3,3)_{\mbox{\tiny obs}}$ values might originate from a hot embedded source, which is revealed by broad (3,3) linewidths leading to a high ratio of $\Delta v$(3,3) to $\Delta v$(1,1). We can exclude any calibration issue, because the NH$_3$ (1,1) to (3,3) lines are observed simultaneously and a relative calibration error would be the same for all sources, but a wide range of $T_{\mbox{\tiny MB}}(3,3)_{\mbox{\tiny obs}}$/$T_{\mbox{\tiny MB}}(3,3)_{\mbox{\tiny calc}}$ is found. Since only few typical maser line profiles are observed and the ortho-to-para ratio would only lead to variations of about a factor two and does not explain the increase in linewidth, we favour the hot core explanation. To quantify this effect, we computed the resulting rotational and (3,3) temperatures when adding a hot core with filling factor $\eta_{\mbox{\tiny hc}}$ and temperature of 100 K. In Fig. \ref{dv-t33-hotcores} the resulting increase in the (3,3) line temperatures is shown for various filling factors. The effect is diminished for higher rotational temperatures, since then an observable (3,3) line is produced already by the large-scale cold clump alone. Sources that are not detected in the NH$_3$ (3,3) line are plotted as upper limits (corresponding to three times the rms noise in T$_{\mbox{\tiny MB}}$) in red. 
\end{enumerate}

\section{Comparison with the 870 $\mu$m dust continuum}
\label{dust continuum}
The ATLASGAL data were reduced and calibrated as described in \cite{2009arXiv0903.1369S}. For the sources of our sample, selected as described in Sect. \ref{select}, we integrated the 870 $\mu$m dust continuum flux within an aperture with a diameter of 40$\arcsec$, which corresponds to the beam of the Effelsberg telescope at the ammonia inversion frequencies of $\sim 24$ GHz. The integrated dust continuum fluxes S$_{\mbox{\tiny 870 $\mu$m}}^{40 \arcsec}$ are given in Table \ref{par870mikrom-atlasgal}.
\subsection{Submillimeter flux and NH$_3$}
\label{continuum flux}
\begin{figure}[h]
\centering
\includegraphics[angle=-90,width=9.0cm]{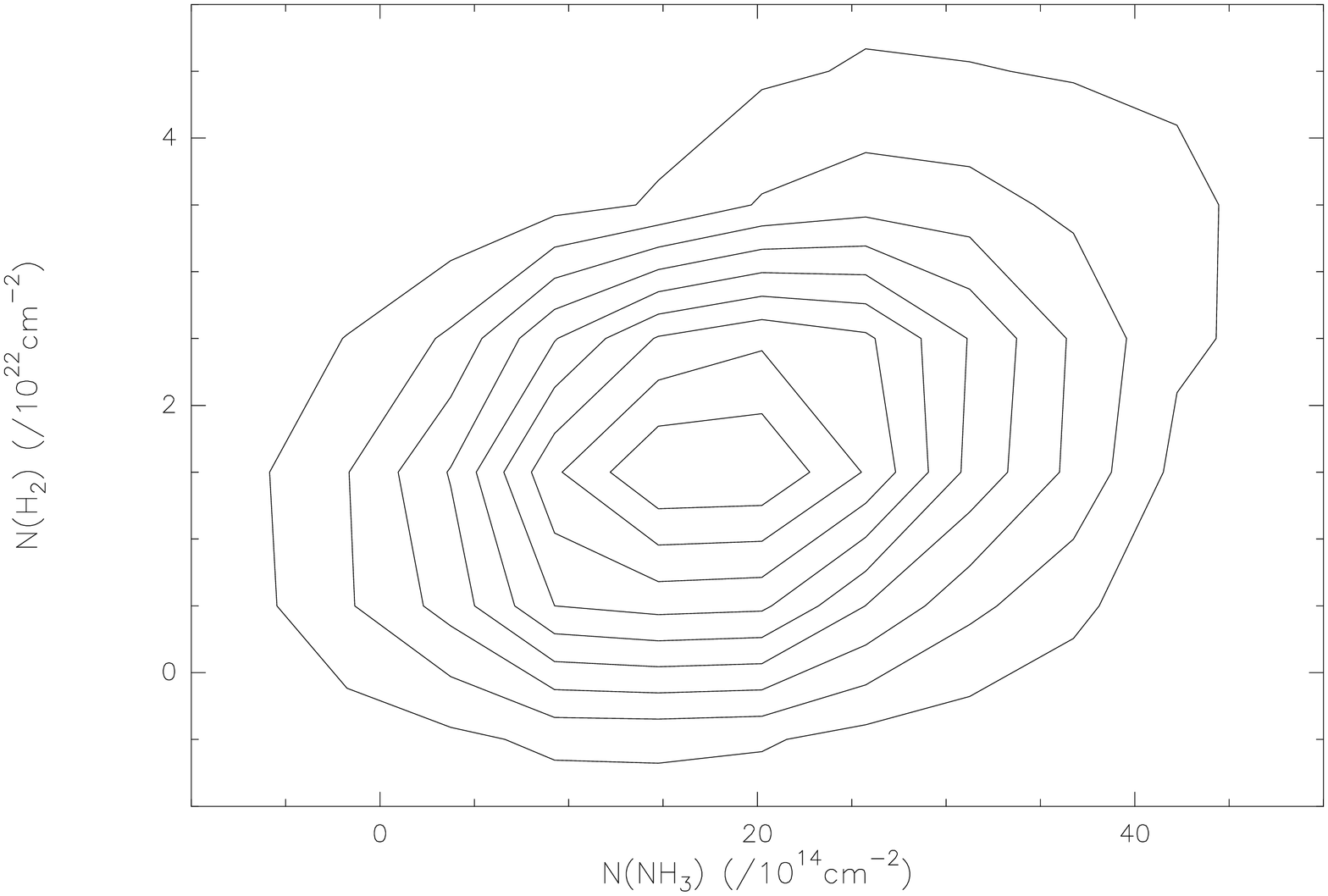}\vspace*{0.5cm}
\includegraphics[angle=-90,width=9.0cm]{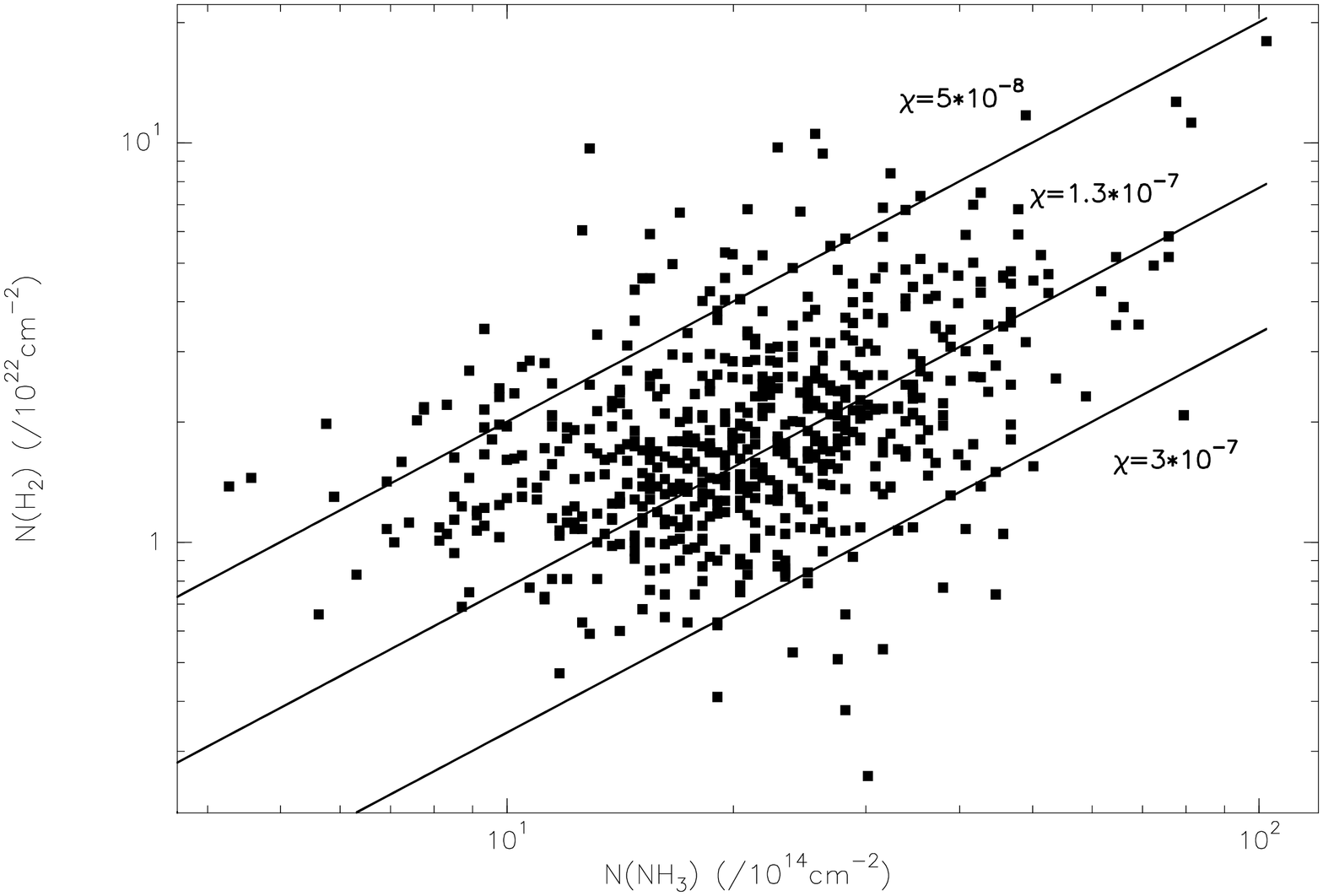}
\caption[comparison of the integrated 870 $\mu$m dust continuum flux and the logarithm of the ammonia column density]{Correlation plot of the ammonia column density and H$_2$ column density. Straight lines indicate a high ammonia abundance $\chi_{\mbox{\tiny NH}_3} > 3 \times 10^{-7}$, an average value of $\sim 1.2 \times 10^{-7}$ and a low NH$_3$ abundance of $5 \times 10^{-8}$. As the contour plot illustration the same correlation is shown in the upper panel, the range of NH$_3$ column densities is divided into bins of 10$^{15}$ cm$^{-2}$ and that of H$_2$ column densities into bins of 10$^{22}$ cm$^{-2}$.}\label{nh3-mm-atlasgal}
\end{figure}

The integrated submillimeter dust continuum flux ranges between 0.6 Jy and 47 Jy and in this section it is compared with the ammonia column density. Because the 870 $\mu$m continuum emission is optically thin, the submm flux is proportional to the dust column density at a given temperature. Assuming a constant gas-to-dust mass ratio, the dust column density is proportional to the H$_2$ density ($N_{\mbox{\tiny H}_2}$), which is related to the ammonia column density ($N_{\mbox{\tiny NH}_3}$) and abundance ($\chi_{\mbox{\tiny NH}_3}$) via the relation $N_{\mbox{\tiny H}_2}$ = $N_{\mbox{\tiny NH}_3}$/$\chi_{\mbox{\tiny NH}_3}$. Hence, assuming a constant NH$_3$ abundance, a correlation between the integrated 870 $\mu$m flux and the ammonia column density is expected, which is confirmed by this study. However, a few clumps exhibit high NH$_3$ column densities and low submm fluxes, which can be explained by an increased ammonia abundance or an overestimation of the temperature. To investigate the two effects we compared the NH$_3$ column density with the H$_2$ column density, computed using the NH$_3$ kinetic temperatures.\\ 
Models of dust grains with thick ice mantles give an absorption coefficient, $\kappa$, of 1.85 cm$^2$/g at 870 $\mu$m at a gas density n(H) = 10$^6$ cm$^{-3}$ \citep{1994A&A...291..943O}. We estimated the H$_2$ column density per the relation \citep{2008A&A...487..993K}
\begin{eqnarray}\label{H2-density}
N({\rm H_2}) = \frac{2.03 \times 10^{16}S_{870\mu m}^{40 \arcsec} \lambda^3 \left ({\rm exp}\left (\frac{1.44 \cdot 10^4}{T_{\mbox{\tiny d}} \lambda}\right) -1 \right)}{\kappa(\lambda) \theta^2}\frac{Z_{\odot}}{Z},
\end{eqnarray}
where $S_{870\mu m}^{40 \arcsec}$ is the 870 $\mu$m flux density in Jy integrated within 40$\arcsec$, $\lambda$ the wavelength in $\mu$m and $T_{\mbox{\tiny d}}$ the dust temperature with $T_{\mbox{\tiny d}}$ = $T_{\mbox{\tiny kin}}$ under the assumption of equal gas and dust temperatures. $\theta$ is the 100-m telescope beamwidth of 40$\arcsec$ and Z/Z$_{\odot}$ the ratio of metallicity to the solar metallicity, assuming Z/Z$_{\odot}$ = 1. We found that most H$_2$ column densities lie between $2.6 \times 10^{21}$ and $1.8 \times 10^{23}$ cm$^{-2}$ with a peak at $1.7 \times 10^{22}$ cm$^{-2}$. For each clump we can now use the ammonia column density $N_{\mbox{\tiny NH}_3}$ to derive the NH$_3$ abundance via $\chi_{\mbox{\tiny NH}_3}$=$N_{\mbox{\tiny NH}_3}$/$N_{\mbox{\tiny H}_2}$, which lies in the range from $7 \times 10^{-9}$ to $1.2 \times 10^{-6}$ with an average of $\sim 1.2 \times 10^{-7}$. The H$_2$ column densities, NH$_3$ abundances and the integrated dust continuum fluxes $S_{\mbox{\tiny 870 $\mu$m}}^{40 \arcsec}$ are given in Table \ref{par870mikrom-atlasgal}. NH$_3$ column densities are plotted against H$_2$ column densities in Fig. \ref{nh3-mm-atlasgal}. They show a better correlation than NH$_3$ column densities and the 870 $\mu$m fluxes, which are affected by variations of the temperature of the sources. Since the kinetic temperature was used in the calculation of the H$_2$ column density, Fig. \ref{nh3-mm-atlasgal} shows the remaining variation in abundance, which can give us a better understanding of the chemical history of the clumps. The scatter, given by the ratio of the rms to the mean value, improved from 0.039 for $S_{\mbox{\tiny 870 $\mu$m}}^{40 \arcsec}$/$N_{\mbox{\tiny NH}_3}$ to 0.0271 for $N_{\mbox{\tiny H}_2}$/$N_{\mbox{\tiny NH}_3}$. Deviations from the trend in Fig. \ref{nh3-mm-atlasgal} are due to a higher ammonia abundance, clumps with $\chi_{\mbox{\tiny NH}_3} > 3 \times 10^{-7}$ are below the lower straight line, or a decreased abundance, values lower than 5 $\times 10^{-8}$ are above the upper straight line. We note that the abundances are upper limits since they are rather derived from the ratio of a source and a beam-averaged column density, hence affected by the filling factor of the cores. With the derived beam filling factors (see Fig. \ref{beamfactor-histo}) the abundances would on average be higher by about a factor ten. With this uncertainty, we find abundances in the range of 5 $\times 10^{-9}$ to $3 \times 10^{-7}$. \cite{2011ApJ...741..110D} found NH$_3$ abundances between $2 \times 10^{-9}$ and $5 \times 10^{-7}$, which is consistent with our results. Moreover, \cite{2006A&A...450..569P} obtained NH$_3$ abundances from 10$^{-8}$ to 10$^{-7}$ for a sample of IRDCs, their $\chi_{\mbox{\tiny NH}_3}$ range is narrower compared to our values and similar to our higher abundances. The average of our derived NH$_3$ abundances agrees with those predicted for low-mass pre-protostellar cores by chemical models \citep{1997ApJ...486..316B}. They show that ammonia is more abundant in cold dense cores than other molecules such as CO, which are depleted from the gas phase at high densities.\\

\subsection{Virial masses and gas masses}
\label{virial mass}
\begin{figure}[h]
\centering
\includegraphics[angle=-90,width=9.0cm]{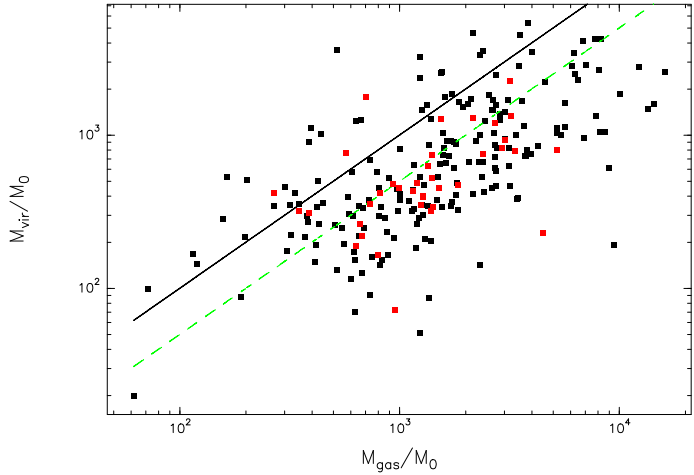}
\caption[dependence of the integrated 870 $\mu$m dust continuum flux on the (1,1) linewidth and the rotational temperature]{Dynamical masses plotted against gas masses for our subsamples, which are associated with GRS clouds (black points) or located at tangential points (red points). The solid black line denotes equal gas and dynamical masses, while the dashed green fit shows $M_{\mbox{\tiny gas}} = 2 M_{\mbox{\tiny dyn}}$. Note that the gas masses are uncertain by a factor 2 due to uncertain dust emissivity.}\label{mass-virialmass}
\end{figure}

We can estimate the masses of the clumps using parameters derived from the 870 $\mu$m dust continuum and from NH$_3$ lines. The gas mass of a source obtained from the dust is 
\citep{2008A&A...487..993K}
\begin{eqnarray}
\label{gas mass}
 M_{\mbox{\tiny gas}} = 1.2\times 10^{-14}\frac{S_{870\mu m}^{40 \arcsec} \lambda^3 d^2 \left({{\rm exp}} \left(\frac{1.44\times 10^4}{T_{\mbox{\tiny d}} \lambda}\right)-1 \right)}{\kappa (\lambda)} \frac{Z_{\odot}}{Z},
\end{eqnarray}
with the distance to the source, $d$, in pc; for the other parameters see equation \ref{H2-density}. The masses can be determined only for the clumps for which we have a preliminary distance estimation: 277 clumps are associated with clouds from the GRS survey with known distances \citep{2009ApJ...699.1153R}, and 71 clumps, which are located at the tangential points, have similar near and far distances, marked in Table \ref{par870mikrom-atlasgal}. To determine distances of the clumps studied by \cite{2009ApJ...699.1153R}, we followed their choice of near/far distance and used the ammonia velocity together with the rotation curve of the Milky Way given by \cite{2009ApJ...700..137R}. The mass can also be calculated using the NH$_3$ (1,1) linewidth under the assumption of virial equilibrium \citep{2004tra..book.....R}
\begin{eqnarray} 
\label{virmass}
 M_{\mbox{\tiny vir}}=250 \Delta v(1,1)_{\mbox{\tiny corr}}^2 \times R,
\end{eqnarray}
with the linewidth $\Delta v(1,1)_{\mbox{\tiny corr}}$, corrected for the resolution of the spectrometer $\Delta v(1,1)_{\mbox{\tiny corr}} = \sqrt{\Delta v(1,1)^2-(0.7{\rm km~s^{-1}})^2}$. The radius $R$ in pc is obtained from Gaussian fits of the 870 $\mu$m continuum flux using the Miriad task sfind \citep{1995ASPC...77..433S} at the distance of the source. Equation \ref{virmass} is derived from the virial theorem by neglecting the magnetic energy and the virial mass is equal to the dynamical mass. The gas and dynamical masses are compared in Fig. \ref{mass-virialmass}, the black points in the scatter plot indicate sources associated with GRS clouds and the red points clumps at the tangential points. Gas masses range from 60 to 1.6 $\times 10^4$ M$_{\odot}$ and virial masses from 20 to 5.5 $\times 10^3$ M$_{\odot}$. The straight solid black line shows equal gas and dynamical masses. Most data lie below that relation, indicating that for most sources their masses exceed their dynamical masses. Since magnetic fields can contribute to prevent the collapse of molecular clouds \citep{2005fds..book.....S}, the magnetic energy has to be taken into account in the virial theorem. We did not measure the magnetic field and assumed equipartition between magnetic and kinetic energy. Then, virialisation is obtained for $M_{\mbox{\tiny gas}}$=$M_{\mbox{\tiny vir}}$=2$M_{\mbox{\tiny dyn}}$. This relation is shown by the dashed green line, but most sources are still located below it. We calculated the virial parameter \citep{1992ApJ...395..140B}
\begin{eqnarray}
 \alpha = \frac{M_{\mbox{\tiny vir}}}{M_{\mbox{\tiny gas}}}
\end{eqnarray}
which is $\sim$ 1 for clumps supported against gravitational collapse. When we use $M_{\mbox{\tiny vir}}$ = $M_{\mbox{\tiny dyn}}$ from equation \ref{virmass}, sources with narrow linewidths have virial parameters on average much smaller (mean $\alpha = 0.21$) than broad linewidth sources with a mean, $\alpha$, of 0.45. With higher density tracers such as H$^{13}$CO$^+$ and C$^{34}$S on average broader linewidths are found, leading to higher virial mass estimates. This will be discussed in Wienen et al. (in prep.). Including again the magnetic energy into the virial theorem gives a factor 2 resulting in $\alpha \sim$ 1 for clumps with broad linewidth. Therefore, these sources are in virial equilibrium. In contrast, sources with small $\Delta v$(1,1) still have a low virial parameter of 0.42.\\
\cite{2010MNRAS.402.2682H}, who observed NH$_3$ lines of high-mass star forming regions with the Parkes telescope, found that their sample is mostly in virial equilibrium. However, they included many sources that have associated infrared sources, UCHII regions, and high-mass protostars, but only a few cores in an early evolutionary phase without any hint at ongoing massive star formation. \cite{2010MNRAS.402.2682H} found a higher average NH$_3$ (1,1) linewidth of 2.9 km~s$^{-1}$ in contrast to our mean $\Delta v$(1,1) of 2 km~s$^{-1}$, which results in larger virial masses than we found.\\
\cite{2010ApJ...717.1157D} obtained an average virial parameter of $\sim$ 1 from their NH$_3$ observations of high-mass star forming sources, which form a similar sample as ours, only at a near distance of about 2 kpc compared to a typical near distance of our sources of $\sim$ 4 kpc. However, \cite{2010ApJ...717.1157D} calculated gas and virial masses between 20 and about 1000 M$_{\odot}$ and therefore probed much smaller masses compared to the ATLASGAL clumps, whose gas masses range up to 1.6$\times10^4$ M$_{\odot}$.\\
\cite{2011ApJ...741..110D} found virial parameters about a factor 2 larger than our study. While the linewidths and dust masses in both studies are comparable, \cite{2011ApJ...741..110D} used for their virial mass computation an ''effective radius'' that on average is about twice as large as the half maximum radii that we used from ATLASGAL, which accounts for this difference.\\
We estimated the magnetic field strength, which is necessary to prevent gravitational collapse, using $B = 2.5 \times N({\rm H_2}) \, \, \mu {\rm G}$ \citep{1993prpl.conf..327M}, where $N({\rm H_2})$ is the hydrogen column density in units of 10$^{21}$ cm$^{-2}$. For our values of the hydrogen column density between $2.3 \times 10^{21}$ and $1.8 \times 10^{23}$ cm$^{-2}$, we obtain magnetic fields between $\sim$ 5 and 450 $\mu$G.\\
Estimates of the gas mass and gas column density as calculated in Eq. \ref{H2-density} can vary strongly. The large uncertainty of Eq. \ref{gas mass} is discussed in \cite{2002ApJ...566..945B} and \cite{2007A&A...476.1243M}, who gave a factor 2 of uncertainty in the mass estimates due to uncertain dust emissivity. In addition, \cite{2012ApJ...751...28M} discussed evidence for changes of the dust opacity in different environments and also found observationally a factor 2 of variation but excluding high-density regions. For denser regions, direct observations of dust opacity changes are difficult. \cite{2011A&A...532A..43O} presented new calculations of dust opacities in dense regions as a function of grain growth and found for moderate time scales ($< 10^7$) a factor 2 increase in the dust opacity with time. The effects of the systematic errors of the linewidths that enter Figs. \ref{nh3-mm-atlasgal} and \ref{mass-virialmass} increase with linewidths, but are much smaller than the uncertainties in the dust opacity. As discussed already in this section, the main uncertainty here is how representative the measured ammonia linewidths are for the dense material of the dust clumps. As we will discuss in more detail in Wienen et al. (2012b, in prep.), observations of higher density probes towards ATLASGAL sources show on average broader linewidths, which in turn lead to higher virial mass estimates.

\section{Correlation of NH$_3$ and $^{13}$CO (1-0) emission}
\label{13CO}
While ammonia is a high-density tracer \citep[$\sim$ 10$^4$ cm$^{-3}$;][]{1986A&A...157..207U}, the $^{13}$CO molecule can be used to probe low-density regions \citep[$\sim 10^3$ cm$^{-3}$;][]{1997ApJ...482..245U}, which are surrounding the dense cores. In this section, the ammonia observations are compared with the $^{13}$CO J = 1 $\rightarrow$ 0 line emission measure at 110.2 GHz in the course of the Boston University-Five College Radio Astronomy Observatory Galactic Ring Survey (GRS) \citep{2006ApJS..163..145J}. It has a greater sensitivity of $\sim$ 0.13 K, a higher spectral resolution of 0.2 km~s$^{-1}$, comparable or better angular resolution (46$\arcsec$) and sampling (22$\arcsec$) than previous molecular line surveys of the inner Galaxy. $^{13}$CO is used instead of the commonly observed $^{12}$CO J = 1 $\rightarrow$ 0 emission, because $^{13}$CO is much less abundant, and consequently possesses an optically thinner transition with narrower linewidths and its analysis can determine the properties of the clumps more reliably. An area of 75.4 deg$^2$ is covered by the whole survey with Galactic longitudes of l = 18$^{\circ}$ - 55.7$^{\circ}$, limited data are available for 14$^{\circ} < l < 18^{\circ}$, and latitudes of $\arrowvert b \arrowvert < 1^{\circ}$. The velocity coverage ranges from -5 to 135 km~s$^{-1}$ up to a Galactic longitude of 40$^{\circ}$ and from -5 to 85 km~s$^{-1}$ for l $> 40^{\circ}$. We extracted $^{13}$CO (1-0) lines from GRS intensity images, which are integrated over the velocity \citep{2006ApJS..163..145J}. The CLASS software was used to derive the $^{13}$CO LSR velocity $v_{\mbox{\tiny $^{13}$CO}}$ of sources detected by GRS and ATLASGAL, their linewidths $\Delta v_{\mbox{\tiny $^{13}$CO}}$ and integrated temperatures T$_{\mbox{\tiny MB, $^{13}$CO}}$, each with errors. The results are given in Table \ref{parlineco-atlasgal}, NH$_3$ and $^{13}$CO line parameters are compared in correlation plots.\\
\subsection{Comparison of ammonia and $^{13}$CO lines}
\begin{figure*}
\centering
\includegraphics[angle=-90,width=18.0cm]{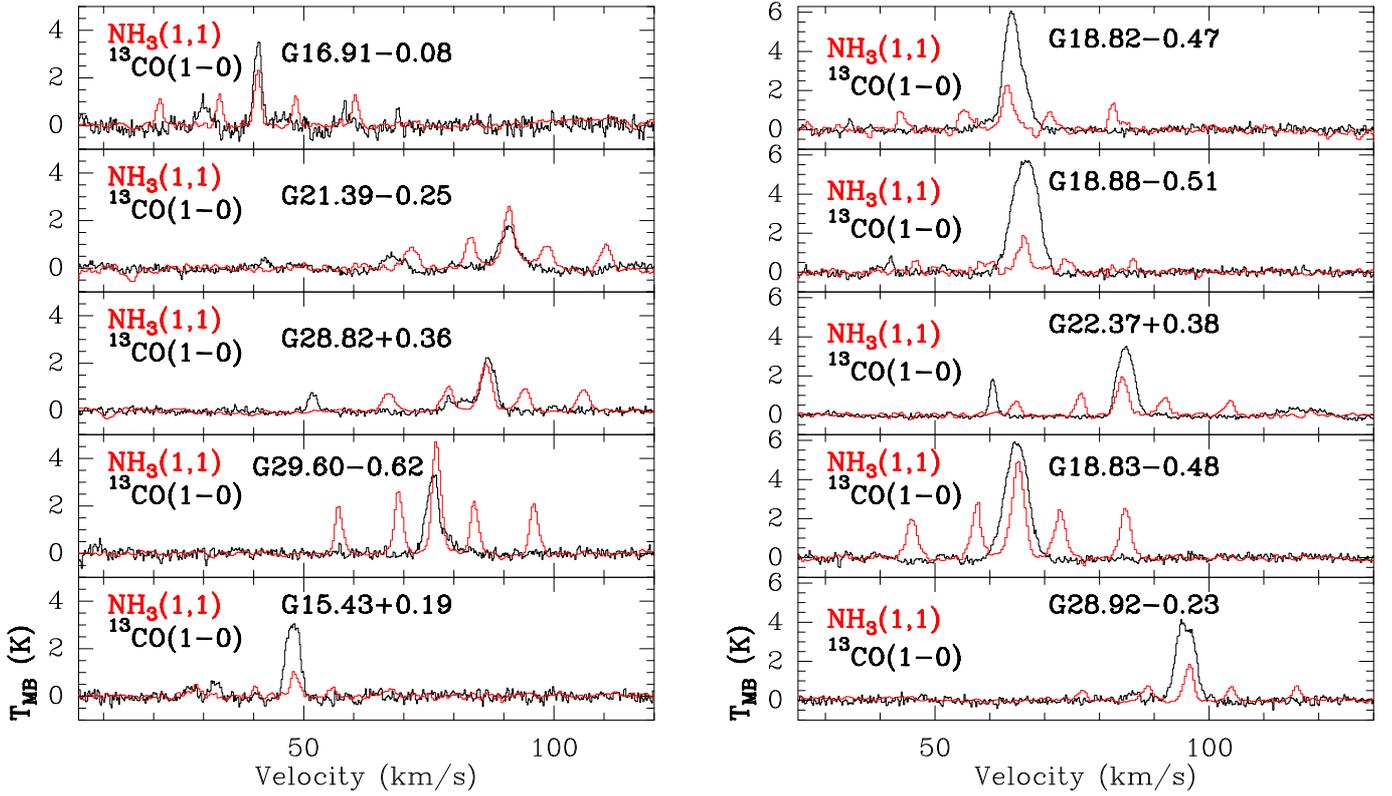}
\caption[NH$_3$ and $^{13}$CO spectra]{NH$_3$(1,1) inversion transition in red overlaid on the $^{13}$CO(1-0) line in black of a few sources. Some clumps exhibit similar widths of the main line of the NH$_3$ (1,1) inversion transitions and $^{13}$CO lines, while most $^{13}$CO linewidths are broader than those of the NH$_3$ (1,1) transitions.}\label{co-nh3sources}
\end{figure*}

We extracted the $^{13}$CO (1-0) line from GRS intensity images at the position of the ammonia observations, fitted a Gaussian to the $^{13}$CO line profile and related the $^{13}$CO line with the velocity closest to the NH$_3$ velocity to the observed ATLASGAL source. Within the ammonia sample we found 517 ATLASGAL clumps associated with $^{13}$CO emission. The spectra of $^{13}$CO and NH$_3$ (1,1) lines are overlaid, some examples of strong sources are shown in Fig. \ref{co-nh3sources}. The $^{13}$CO linewidths are similar to the widths of the NH$_3$ (1,1) inversion transitions for a few sources such as G16.91$-$0.08, G21.39$-$0.25, G28.82+0.36, G29.60$-$0.62, while most clumps possess broader $^{13}$CO than NH$_3$ (1,1) linewidths, e.g. G15.43+0.19, G18.82$-$0.47, G18.88$-$0.51, G22.37+0.38, G18.83$-$0.48, and G28.92$-$0.23. While most ATLASGAL clump lines of sight show only one NH$_3$ (1,1) velocity component, the $^{13}$CO (1-0) often shows multiple components (see e.g. G16.91$-$0.08, G21.39$-$0.25, G28.82+0.36, G15.43+0.19, and G22.37+0.38 in Fig. \ref{co-nh3sources}). Concerning those clumps, we assumed the $^{13}$CO line with the velocity closest to the NH$_3$ velocity to be associated with the ATLASGAL source, which has been demonstrated to be a reasonable choice by \cite{2007ApJ...668.1042K} for low-mass dense cores.

\subsection{Clump-to-cloud motion}
\begin{figure}[!h]
\centering
\includegraphics[angle=-90,width=9.0cm]{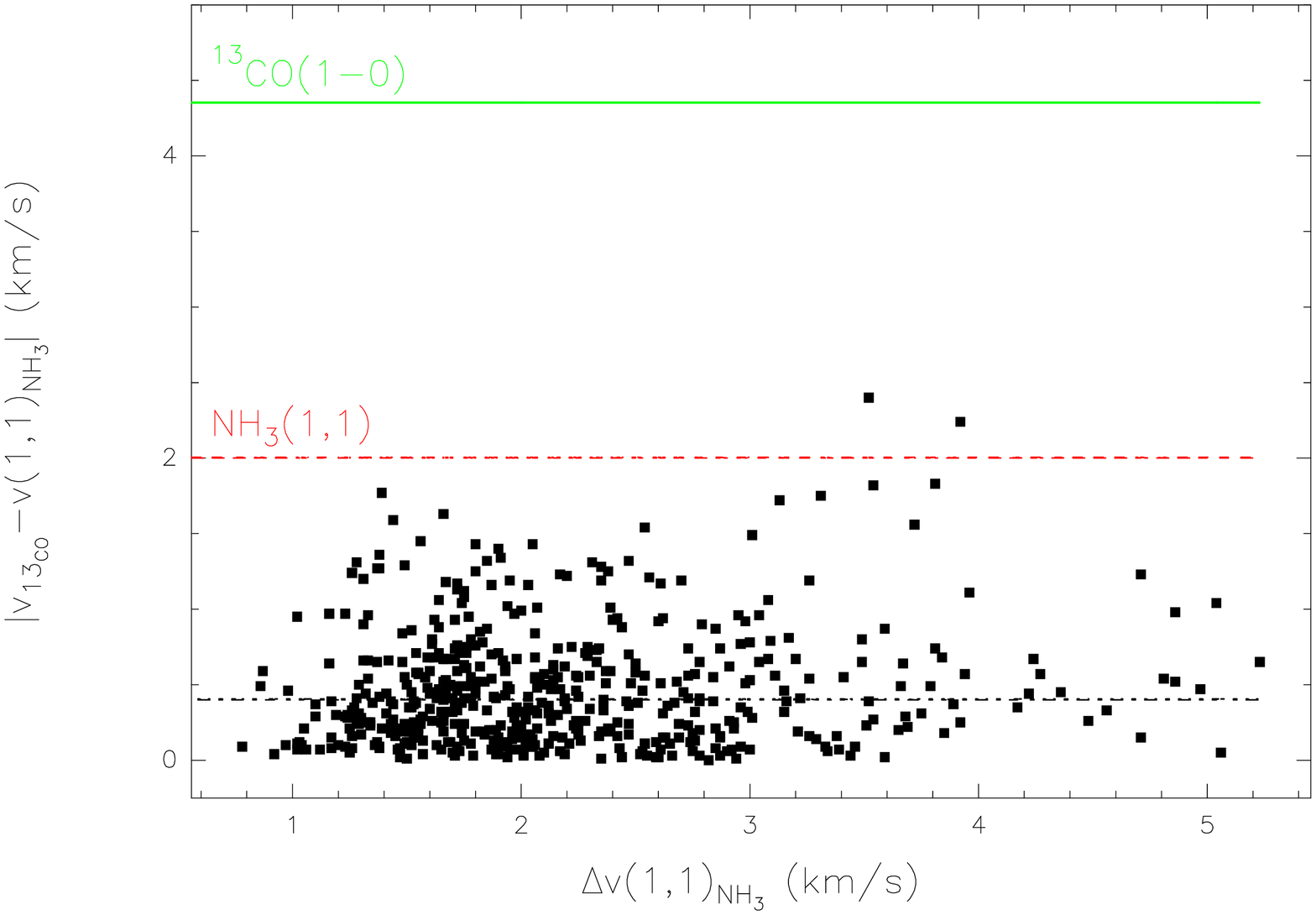}\vspace*{0.5cm}
\includegraphics[angle=-90,width=9.0cm]{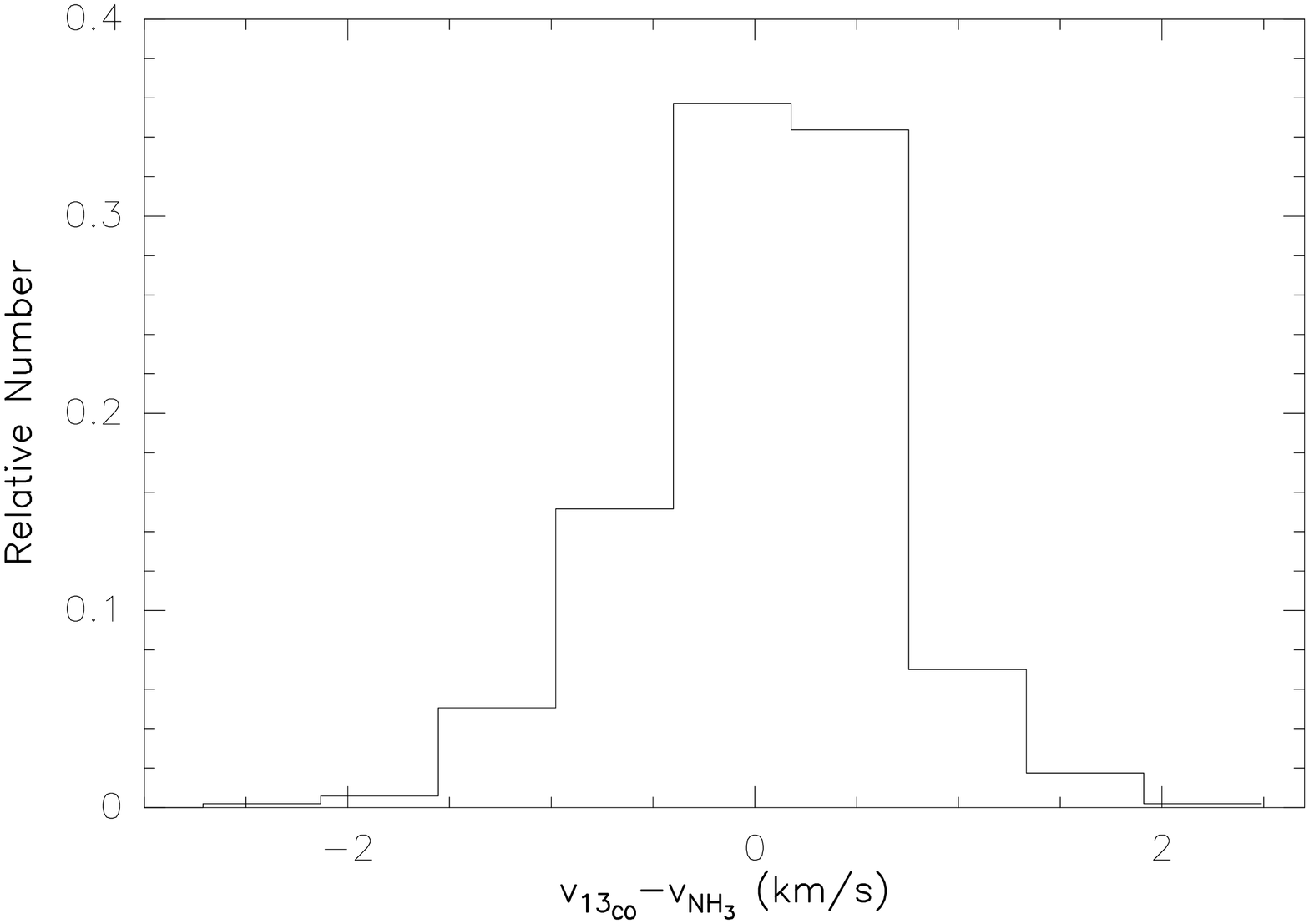}
\caption[comparison of $^{13}$CO and NH$_3$ velocities]{Absolute value of the difference in line-centre velocities of $^{13}$CO (1-0) and NH$_3$ lines detected in the ATLASGAL sources plotted against the ammonia linewidth in the top panel. The differences of most clumps are similar to the typical NH$_3$ linewidth, which is indicated by the dashed red line, and smaller than the width of the $^{13}$CO transition labelled by the solid green line, the rms of the absolute value of the difference in $^{13}$CO and NH$_3$ (1,1) velocities is shown by the dashed-dotted line. The bottom panel shows the relative number distribution of the ATLASGAL clumps with the velocity differences of both molecules. The $^{13}$CO and NH$_3$ velocities of most sources differ by more than the sound speed at 0.23 km~s$^{-1}$.}\label{diff-vnh3vco-atlasgal}
\end{figure}

\begin{figure}[h]
\centering
\includegraphics[angle=-90,width=9.0cm]{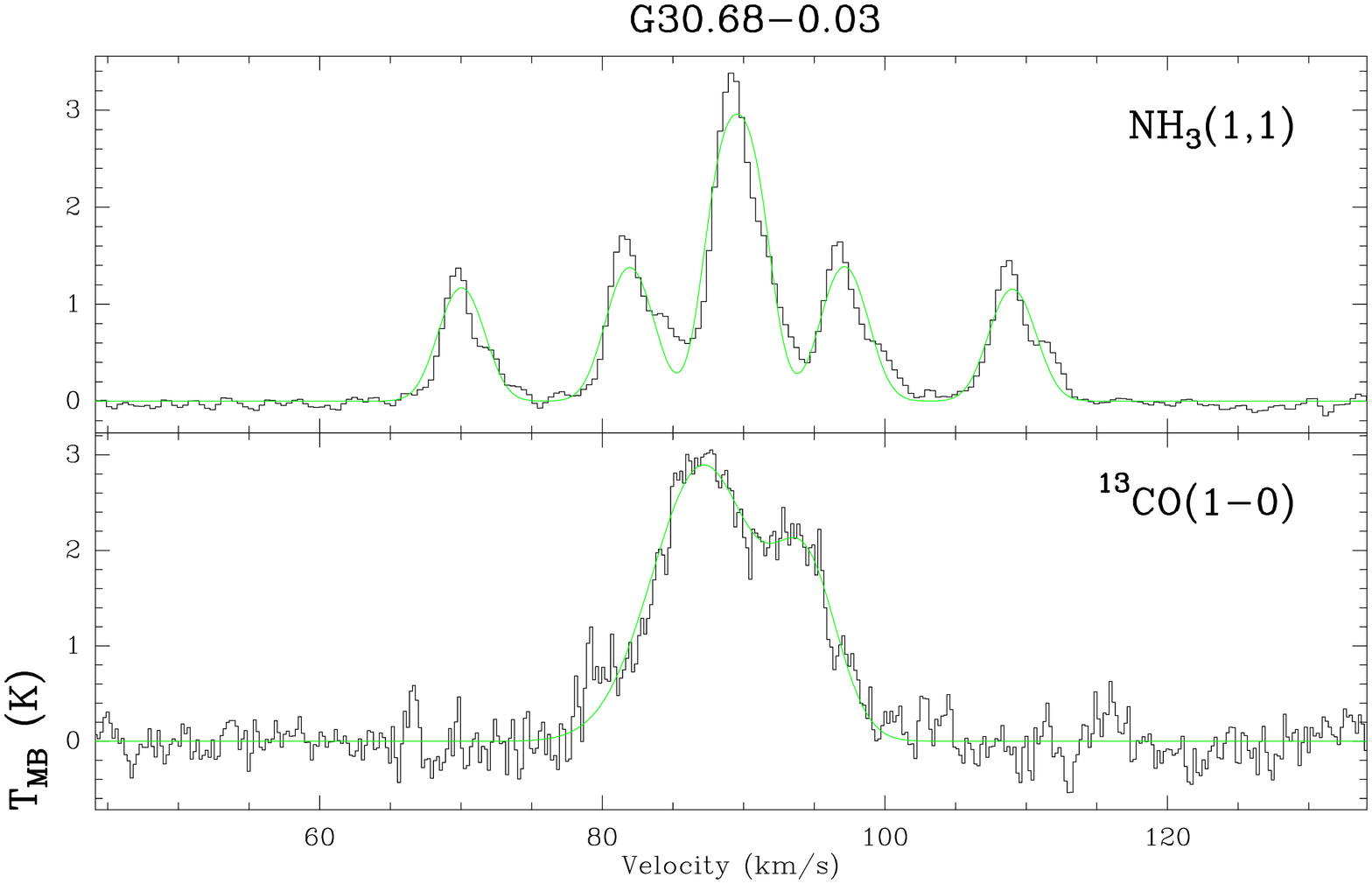}\vspace*{0.5cm}
\includegraphics[angle=-90,width=9.0cm]{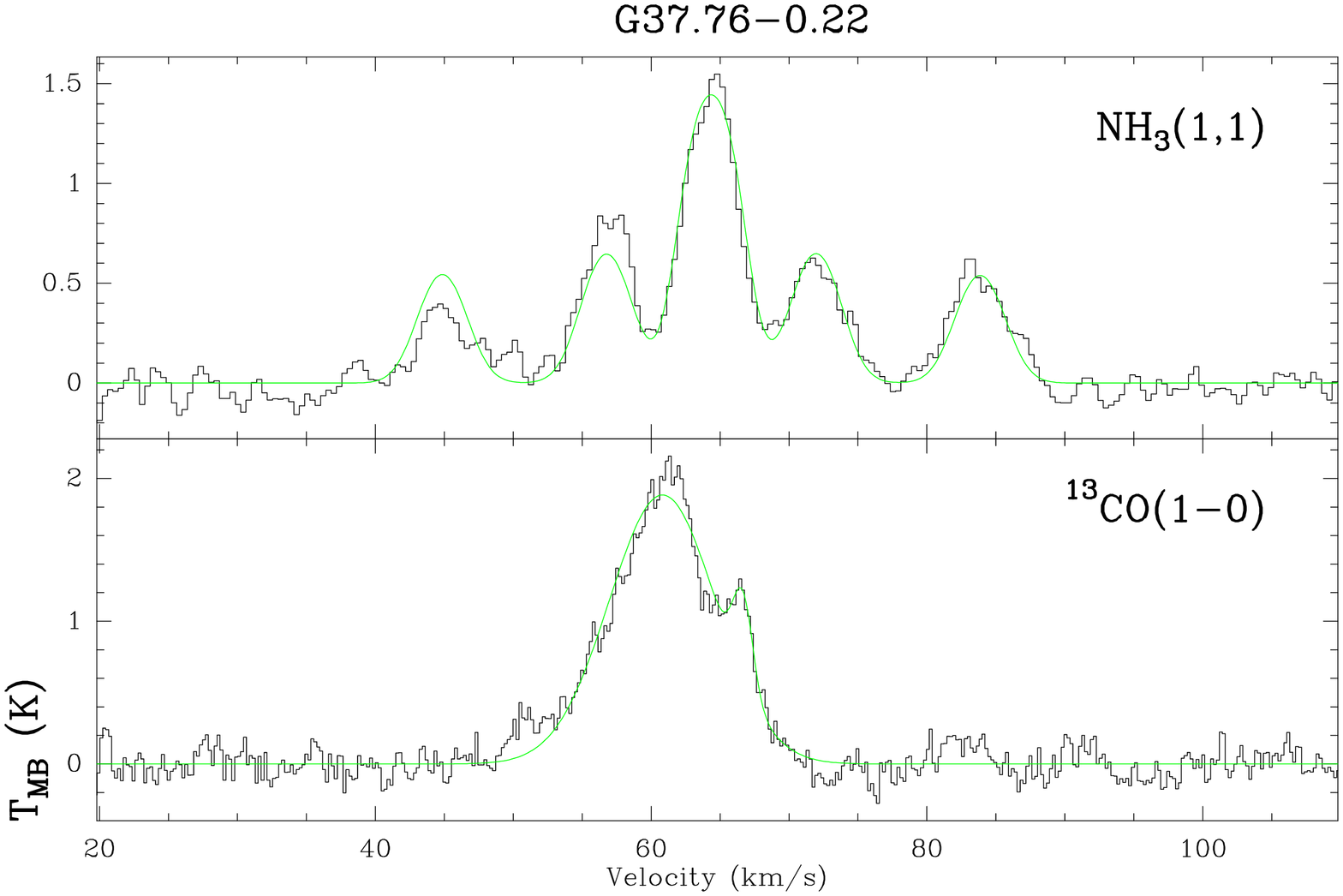}
\caption[spectra of G23.01-0.41 and G18.83-0.30]{NH$_3$(1,1) and $^{13}$CO(1-0) lines of G30.68$-$0.03 are plotted in the upper panel, the lower one shows these transitions of the clump G37.76$-$0.22. The Gaussian fit of the spectra is shown.}\label{two-velocity-spectra}
\end{figure}

While it is not yet understood how high-mass stars form, two main theoretical models are currently under consideration: (1) in \textit{the turbulent core model} massive star forming clouds are supported against gravitational collapse mainly by supersonic turbulence \citep{2003ApJ...585..850M, 2004RvMP...76..125M}. (2) According to recent simulations of star formation in clusters \citep[e.g.][]{1997MNRAS.285..201B, 2001MNRAS.323..785B}, the accretion rate and in turn the mass of the forming star are strongly affected by the density of the molecular cloud through which it is moving. In addition, material is supposed to diffuse into the potential well of the cloud at the centre of a cluster, where high-mass stars are preferentially forming. This \textit{theory of competitive accretion} also agrees with observations, which show that the most massive stars of a cluster are located at its centre \citep{2010A&A...524A..18B}. According to \cite{2004ApJ...614..194W}, this moving accretion is only possible if the protostellar core is moving at appreciable velocity with respect to its surrounding envelope, which will be associated with a significant difference in the line-centre velocity of high- and low-density tracers. Previous studies of low-mass star forming regions have analysed the motions of dense cores with respect to their surrounding envelope, using N$_2$H$^+$ as high-density tracer and $^{13}$CO as well as C$^{18}$O as probes of less dense gas \citep[e.g.][]{2004ApJ...614..194W, 2007ApJ...655..958W, 2007ApJ...668.1042K}. While \cite{2004ApJ...614..194W} investigated mostly single isolated cores, clustered regions of dense cores such as NGC1333 and the Perseus molecular cloud were also studied by \cite{2007ApJ...655..958W} and \cite{2007ApJ...668.1042K}. However, they did not find significant motions of the cores relative to their lower density surroundings. In this section we will compare their results to conditions prevailing in high-mass star forming regions.\\
Although in the more distant high-mass star forming regions we sample clumps on larger scales than the aforementioned studies, it is still interesting to investigate motions of clouds relative to their host clouds. It is especially useful that the NH$_3$ and $^{13}$CO lines, good tracers for the former and the latter, respectively, were measured with similar beamwidths, 40$\arcsec$ and 46$\arcsec$, respectively. The absolute value of the difference in $^{13}$CO and NH$_3$ (1,1) line-centre velocities is plotted against the linewidth of the ammonia (1,1) inversion transition in the top panel of Fig. \ref{diff-vnh3vco-atlasgal}. If ammonia clumps showed significant motions with respect to their envelopes, the differences in line-centre velocities should be comparable to the width of the $^{13}$CO lines \citep{2004ApJ...614..194W}. Deviations in the range of NH$_3$ linewidths would be associated with random motions within the ammonia clumps. The dashed red line in the top panel of Fig. \ref{diff-vnh3vco-atlasgal} at 2 km~s$^{-1}$ indicates the average widths of NH$_3$ (1,1) lines and the solid green line at $\sim 4.5$ km~s$^{-1}$ illustrates the average $^{13}$CO (1-0) linewidths. The rms of the difference in both velocities is 0.4 km~s$^{-1}$, which is shown by the dashed dotted black line. Because the distribution of almost all clumps is narrower than the average NH$_3$ linewidth and also narrower than the average width of the $^{13}$CO (1-0) transition, random motions of the ammonia clumps are expected and hint at turbulence, which could prevail in the sources. Hence, these low relative velocities between the ammonia clump and $^{13}$CO cloud put strong constraints on the role of turbulence in the clump formation. There are only two sources whose absolute value of NH$_3$ (1,1) and $^{13}$CO velocity difference is slightly higher than the NH$_3$ linewidth: G30.68$-$0.03 with a deviation in both velocities of -2.4 km~s$^{-1}$ has a lower $^{13}$CO velocity, while G37.76$-$0.22 exhibits a lower NH$_3$ velocity with a difference to $^{13}$CO of 2.24 km~s$^{-1}$. Fig. \ref{two-velocity-spectra} illustrates that the $^{13}$CO lines of both clumps might show either self-absorption or each $^{13}$CO spectrum could be emitted by two clumps, that lie on the same line of sight to the observer, leading to higher uncertainties in the $^{13}$CO velocities. Concerning low-mass star formation, the line-centre velocity differences of N$_2$H$^+$ and $^{13}$CO or C$^{18}$O in cores observed by \cite{2004ApJ...614..194W, 2007ApJ...655..958W} are also similar to their N$_2$H$^+$ linewidths, which is consistent with the analysis of the ATLASGAL sources. This result indicates that the velocity deviations of both molecular probes in low-mass sources are much lower than in massive ones, because low-mass cores have a much more narrow average N$_2$H$^+$ and $^{13}$CO linewidth of $\sim 0.38$ km~s$^{-1}$ and $\sim 1.18$ km~s$^{-1}$ \citep{2007ApJ...655..958W} compared to the broader ammonia and $^{13}$CO lines of high-mass cores. This also agrees with and is additionally supported by the analysis of \cite{2007ApJ...668.1042K}, who measured differences in N$_2$H$^+$ and C$^{18}$O velocities similar to the N$_2$H$^+$ thermal linewidths of 0.065 km~s$^{-1}$ for most dense cores in the Perseus molecular cloud assuming a temperature of 15 K.\\
The distribution of differences in line-centre velocities of $^{13}$CO and NH$_3$ in the bottom panel of Fig. \ref{diff-vnh3vco-atlasgal} illustrates that both, higher and lower $^{13}$CO velocities than ammonia velocities are present. The differences can be compared to the sound speed of the ambient medium, which is 0.23 km~s$^{-1}$ assuming gas that consists of 90\% H$_2$ and 10\% He with a mean molecular weight of 2.33 at a temperature of 15 K. Most clumps show differences larger than the sound speed, only 32\% of them have subsonic motions. A high percentage of sources shows a deviation in clump-to-cloud velocities between 1 km~s$^{-1}$ and -0.7 km~s$^{-1}$, some range to 2.2 km~s$^{-1}$ and -2.9 km~s$^{-1}$. This result can be compared to the difference in N$_2$H$^+$ and C$^{18}$O line-centre velocities of low-mass, starless and protostellar cores observed in the Perseus molecular cloud by \cite{2007ApJ...668.1042K}. Although the motion of protostellar cores with respect to the envelope could be influenced by turbulence generated by heating or outflows, most of them as well as the majority of the starless cores show relative velocities of both molecules slower than the sound speed, while the remaining sources have only slightly higher differences. This agrees with the analysis of \cite{2004ApJ...614..194W}, who measured a core-to-envelope velocity exceeding the sound speed in only 3\% of their sample, which included mostly isolated low-mass cores. Hence, those comparisons also illustrate that the ATLASGAL clumps as a high-mass star forming region sample show much more distinct velocity differences between low- and high-density tracers than low-mass cores.

\subsection{$^{13}$CO and ammonia linewidths}
\begin{figure}[h]
\centering
\includegraphics[angle=-90,width=9.0cm]{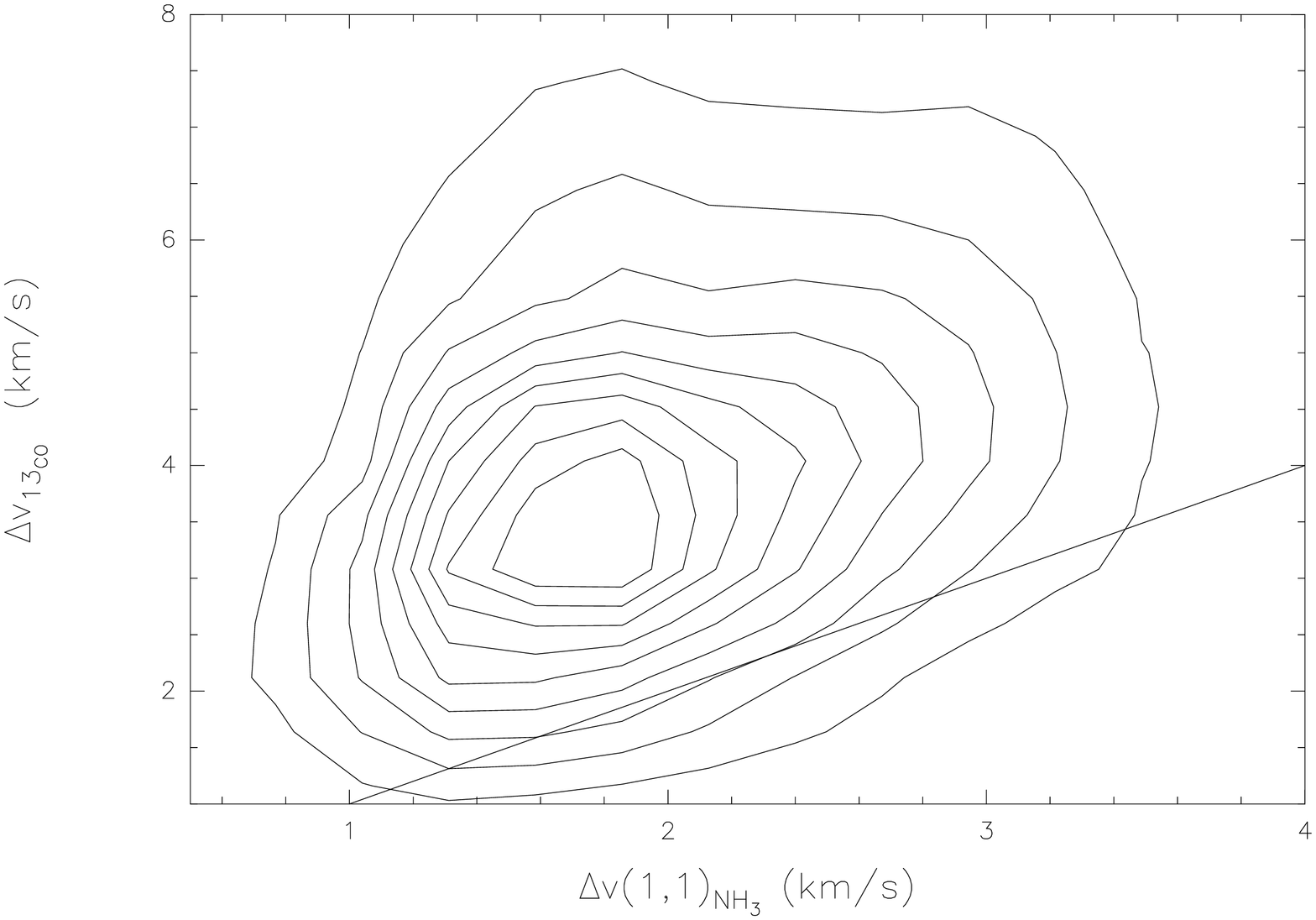}\vspace*{0.5cm}
\includegraphics[angle=-90,width=9.0cm]{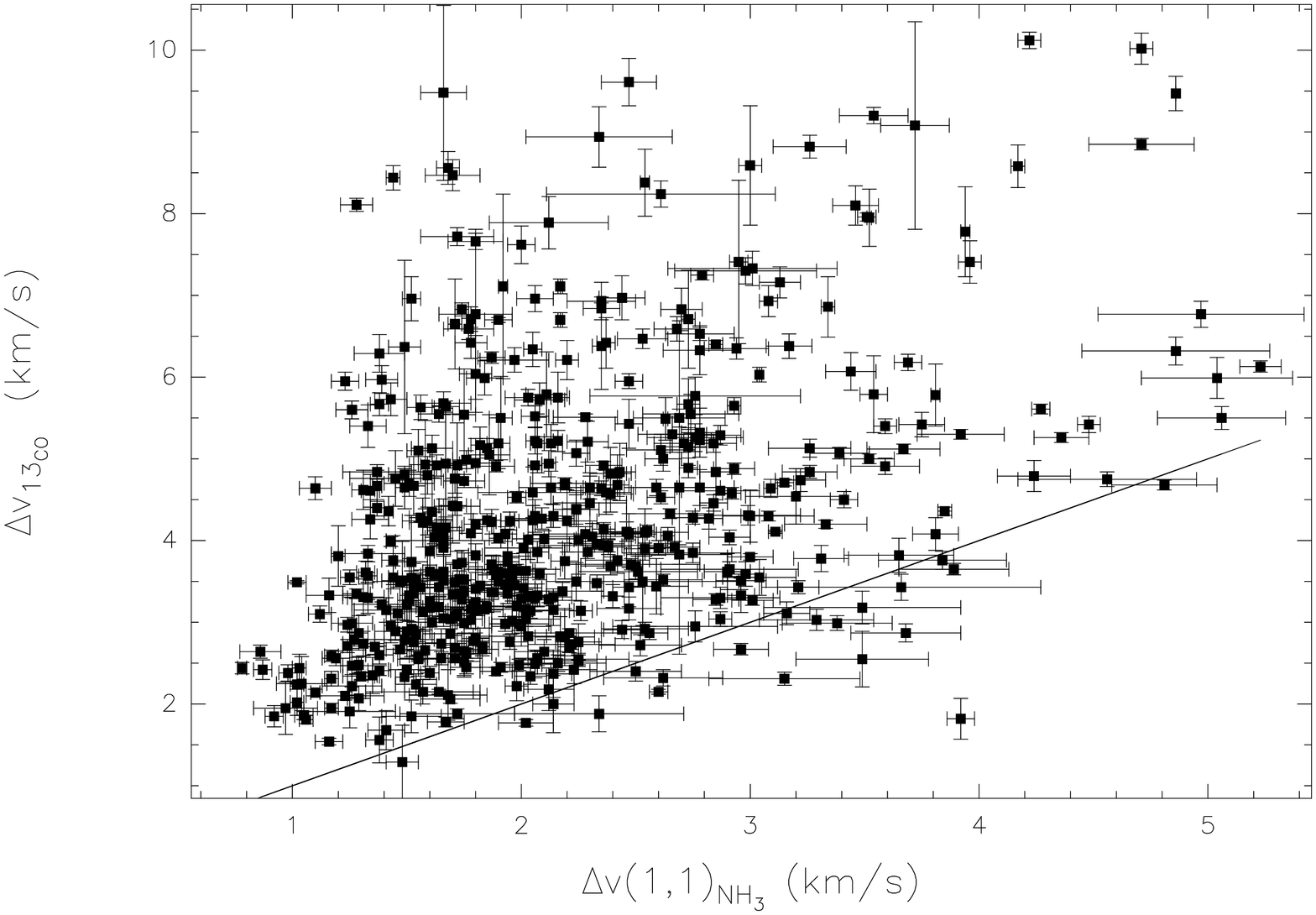}
\caption[comparison of $^{13}$CO and NH$_3$ linewidth]{Widths of the $^{13}$CO (1-0) transition compared to those of the NH$_3$ (1,1) line as contour plot in the upper panel and as scatter plot in the lower panel. The straight line corresponds to equal widths. The binning used for the NH$_3$ (1,1) linewidth is 0.4 km/s and for the $^{13}$CO linewidth is 1 km/s.}\label{dv11-dvco-atlasgal}
\end{figure}

The width of the $^{13}$CO (1-0) transition is plotted against that of the NH$_3$ (1,1) inversion line in the lower panel of Fig. \ref{dv11-dvco-atlasgal}, the straight line represents equal widths. For the contour plot in the upper panel of Fig. \ref{dv11-dvco-atlasgal} a binning of 0.4 km/s is chosen for the NH$_3$ (1,1) linewidth and of 1 km/s for the $^{13}$CO linewidth. The $^{13}$CO line, exhibiting widths from 1 km~s$^{-1}$ to 10 km~s$^{-1}$ with a peak at 3.3 km~s$^{-1}$, is mostly broader than the ammonia line, which may indicate that the lines arise from different parts of the clumps. It is expected that $^{13}$CO traces a larger scale than ammonia, although the beamwidth used for the GRS of 46$\arcsec$ is only slightly broader than that of the Effelsberg telescope. $^{13}$CO traces densities of $\sim 10^3$ cm$^{-3}$, but is frozen out at densities of 10$^5$ cm$^{-3}$, where NH$_3$ is still existing in the gas phase. Hence, ammonia probes the massive, denser inner parts of a clump, while $^{13}$CO is also sensitive to the outer less dense regions of the cloud. Turbulence within large structures can cause broader $^{13}$CO than NH$_3$ linewidths found in small-scale regions of the clumps.\\ 
Another explanation of the difference in linewidths could be that multiple $^{13}$CO emitting clumps may lie on the same line of sight to the observer, since $^{13}$CO traces low densities and its abundance of 10$^{-6}$ \citep{2008ApJ...686..384D} is higher compared to the ammonia molecule with $7 \times 10^{-9}$ to $1.2 \times 10^{-6}$ (cf. Sect. \ref{continuum flux}). Fig. \ref{co-nh3sources} also shows that some ATLASGAL source lines of sight have more than one $^{13}$CO component. Those different clumps may also have narrow linewidths, which would add up to the observed increased $^{13}$CO widths due to velocity dispersion within the beam.

\section{Temperature and linewidth distributions in different evolutionary stages}
\label{discussion}
\begin{figure}[h]
\centering
\includegraphics[angle=-90,width=9.0cm]{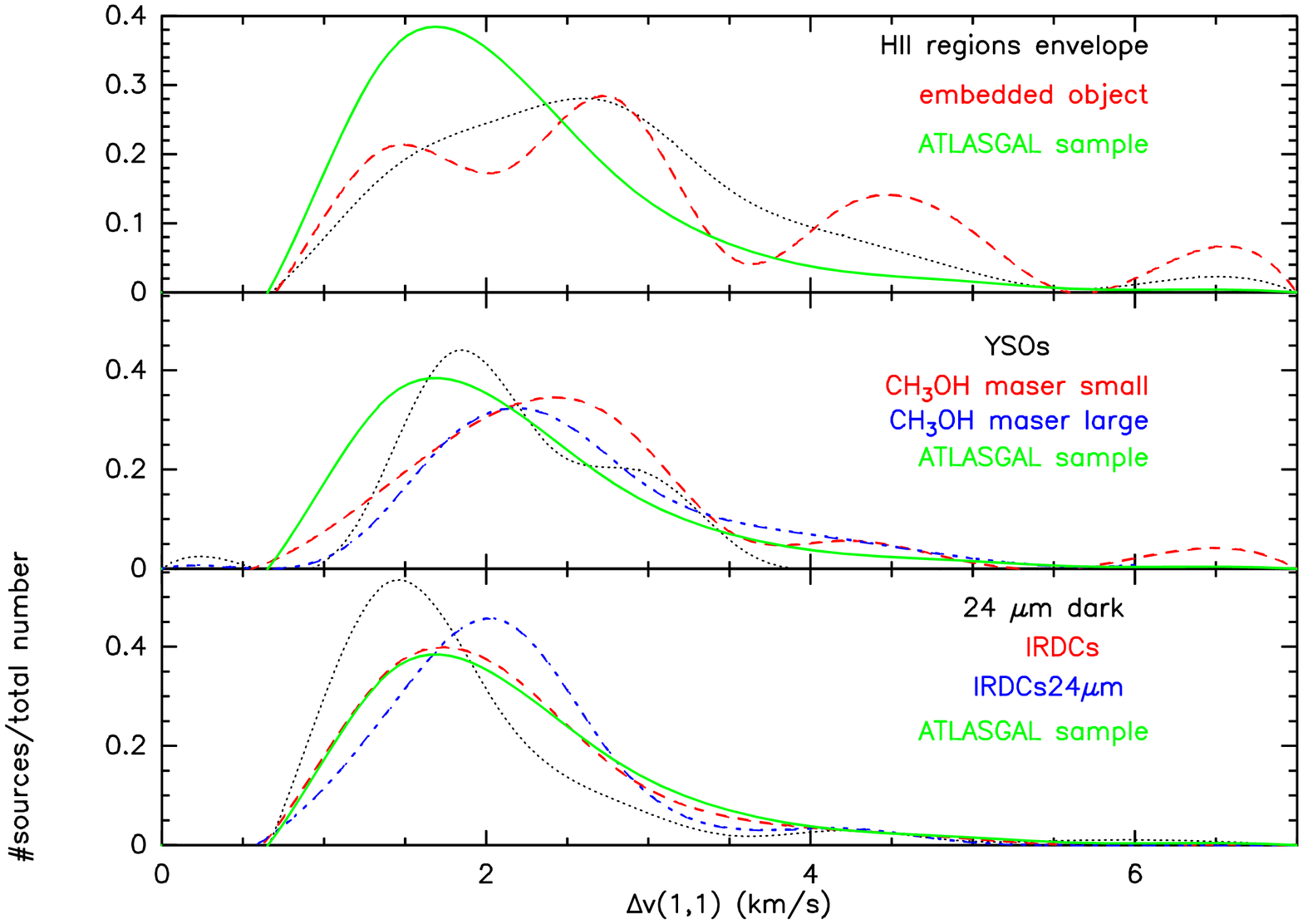}\vspace*{0.5cm}
\includegraphics[angle=-90,width=9.0cm]{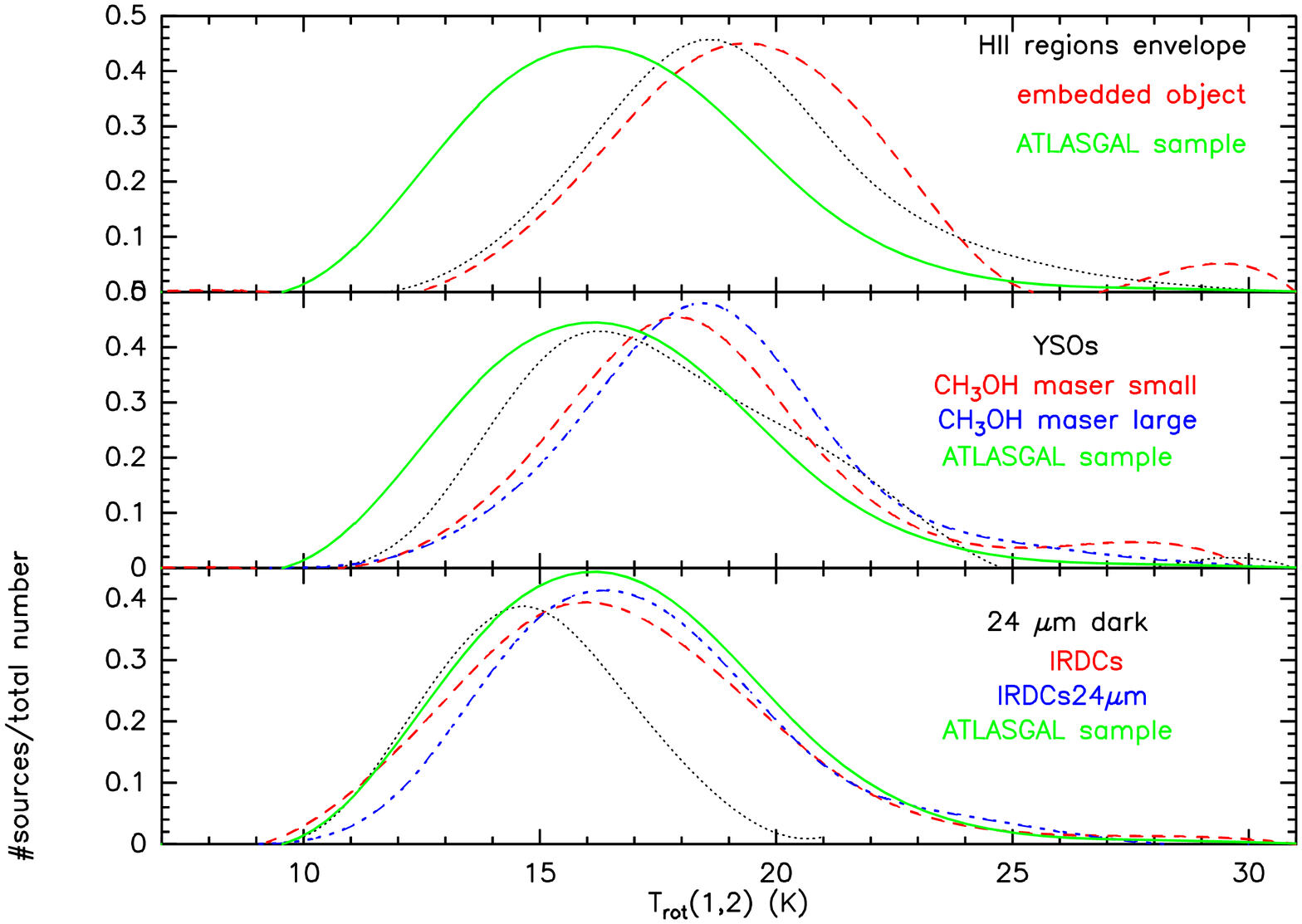}
\caption[histogram of the (1,1) linewidth and the rotational temperature]{Relative number distribution of ATLASGAL subsamples in different evolutionary phases of high-mass star formation showing the NH$_3$ (1,1) linewidth in the upper part and the rotational temperature in the lower part. The whole ATLASGAL sample is shown as solid green curve in each panel, while the other subsamples are indicated by different patterns and colours: \textit{Lower panel:} dotted black curve: ATLASGAL clumps not detected at 24 $\mu$m, dashed red: IRDCs from \cite{2009A&A...505..405P}, dashed-dotted blue: an IRDC subsample emitting at 24 $\mu$m. \textit{Middle panel:} dotted black curve: young stellar objects (YSOs) from the RMS survey \citep{2005IAUS..227..370H}, dashed red: CH$_3$OH masers from \cite{2011ApJ...730...55P}, \cite{2002A&A...392..277S}, and \cite{2010MNRAS.404.1029C} found within a small search radius (see Sect. \ref{discussion}), dashed-dotted blue: same CH$_3$OH maser samples found in a large radius (see Sect. \ref{discussion}). \textit{Upper panel:} UCHII regions from \cite{2005IAUS..227..370H}, dotted black curve: associations with the envelope in a large search radius (see Sect. \ref{discussion}), dashed red curve: associations with embedded object in a small search radius (see Sect. \ref{discussion}).}\label{dv11-trot-histoatlasgal}
\end{figure}

\begin{figure}[h]
\centering
\includegraphics[angle=-90,width=7.5cm]{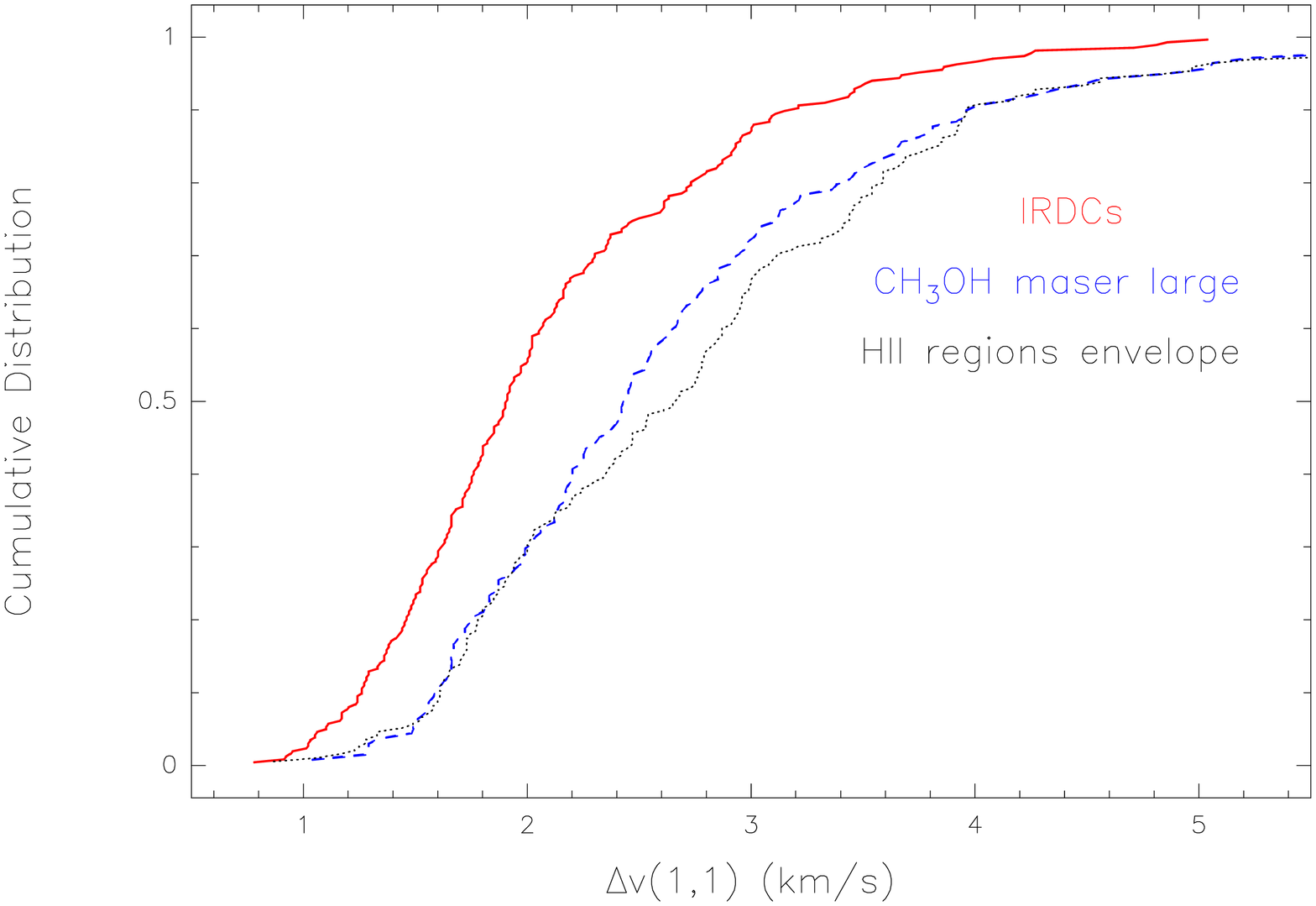}\vspace*{0.5cm}
\includegraphics[angle=-90,width=7.5cm]{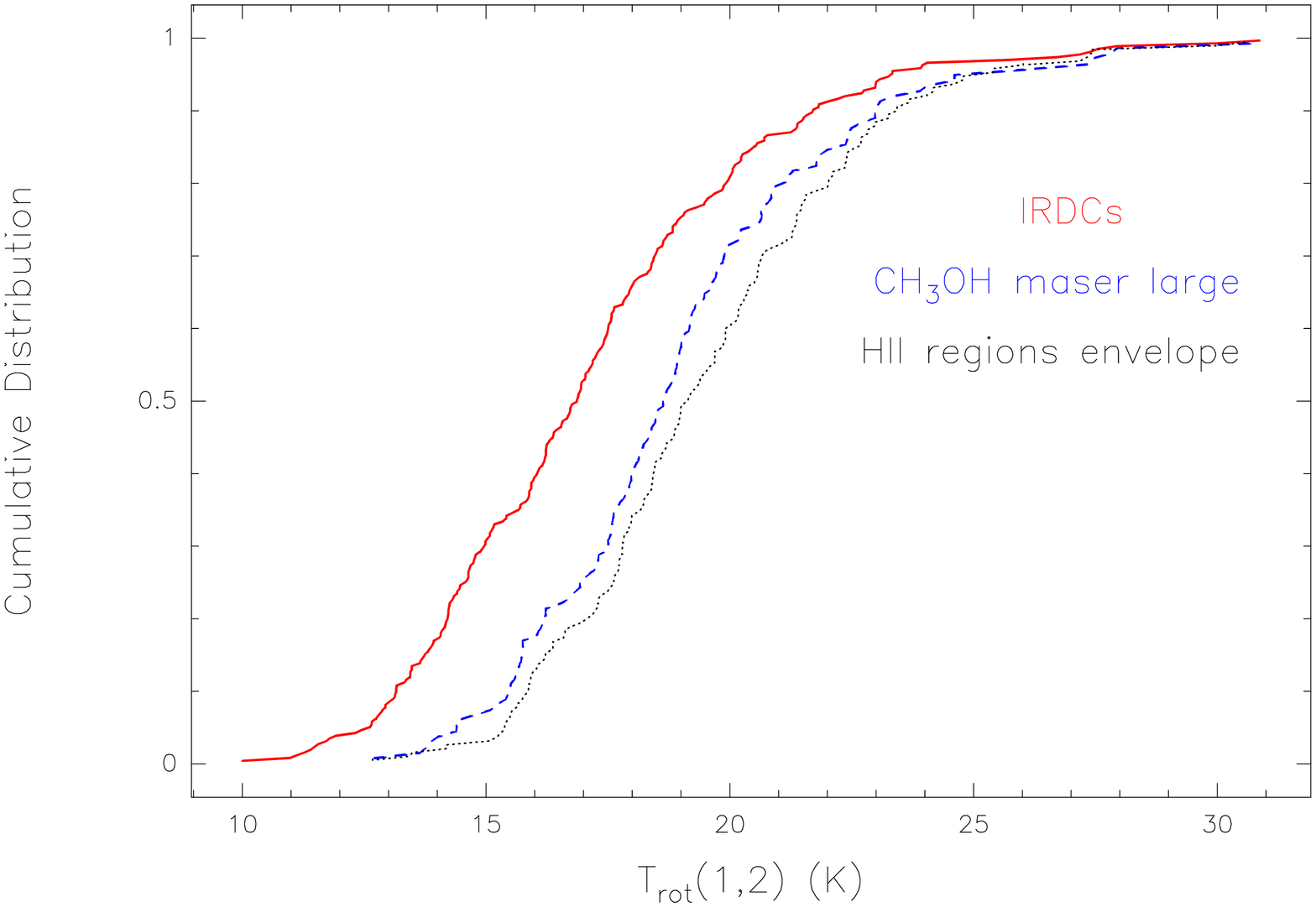}
\caption[cumulative distributions of the (1,1) linewidth and the rotational temperature]{Cumulative distribution functions illustrating the NH$_3$ (1,1) linewidth (upper panel) and rotational temperature (lower panel) for the three largest subsamples in Table \ref{sample}. The distribution of IRDCs from \cite{2009A&A...505..405P} is shown as a solid red line, the CH$_3$OH sample found within the large search radius is plotted as a dashed blue curve and the H II regions envelope as a dotted black curve.}\label{dv11-trot-distribution}
\end{figure}

To investigate trends of NH$_3$ line parameters with evolutionary phases, we divided most clumps into subsamples by searching for their associations with other samples that cover the main evolutionary stages of high-mass star formation.
We used search radii dependent on the type of sample. Search radius, number of associated sources, mean and error of the NH$_3$ (1,1) linewidth, and the rotational temperature of the resulting ATLASGAL subsamples are given in Table \ref{sample}. While a few sources do not match any sample, some clumps show characteristics of more than one phase of high-mass star formation and are therefore associated with different subsamples. Similar studies using smaller samples have already analysed the association of sources traced by NH$_3$ with different evolutionary stages of massive star formation \citep{2005ApJ...634L..57S,2006A&A...450..569P}. Fig. \ref{dv11-trot-histoatlasgal} shows the number distribution with the (1,1) linewidth and the rotational temperature of the different ATLASGAL subsamples for clarity as curves and not histograms, although the data sampling is not as good as suggested.\\ 
The lowest panel compares the whole ATLASGAL sample as a solid curve in green with a subsample of ATLASGAL clumps not detected at 24 $\mu$m, denoted as ''24 $\mu$m dark'' as a dotted curve in black, IRDCs as a dashed red curve from \cite{2009A&A...505..405P} and a subsample of those IRDCs emitting at 24 $\mu$m as a dashed-dotted curve in blue. As search radius for associations of IRDCs with ATLASGAL sources we used the equivalent radius given by \cite{2009A&A...505..405P}, which agrees with the radius of a disk, whose area is the same as that of the IRDC. An average of the search radii used is given in Table \ref{sample}. The 24 $\mu$m dark ATLASGAL clumps are the coldest sources with the narrowest linewidths, suggesting that they are in a very early phase without any hint of star formation. The distribution of the IRDCs and the ATLASGAL sample is similar in linewidth and rotational temperature. Although the IRDCs that emit at 24 $\mu$m show signs of star formation activity, they have only sligthly broader linewidths than the average of ATLASGAL sources and similar rotational temperatures. The 24 $\mu$m emitting star embedded in those IRDCs is a local heating component of little influence over the volume of the whole clump and thus has no significant influence on the source properties. Examples of the NH$_3$ (1,1) to (3,3) lines of IRDCs (G18.09$-$0.30, G10.67$-$0.22) are displayed in Fig. \ref{nh3lines}.\\
We plot the distribution of young stellar objects (YSOs as dotted black curve) from the RMS survey \citep{2005IAUS..227..370H}, CH$_3$OH masers from \cite{2011ApJ...730...55P}, \cite{2002A&A...392..277S}, and \cite{2010MNRAS.404.1029C} (dashed and dashed-dotted curves) and the ATLASGAL sample as a solid curve in green in the middle panel. The YSOs are MSX sources \citep{2001AJ....121.2819P}, which emit at infrared wavelengths and are therefore in a more evolved phase than the IRDCs. We searched for them within the ATLASGAL beam in addition to the MSX positional uncertainty. The embedded infrared sources have broader (1,1) linewidths than the mean value of ATLASGAL clumps. This hints at star formation activity in the YSOs, which results in higher rotational temperatures than found for the ATLASGAL sample on average (Fig. \ref{dv11-trot-histoatlasgal}) and at turbulence, e.g. due to molecular outflows. Concerning the methanol masers, we used two search radii for each of the three samples: Sources found within a small radius, consisting of the pointing error of the ATLASGAL survey of 5$\arcsec$ and those of the telescopes used to observe the CH$_3$OH samples, might be directly related to the maser (CH$_3$OH maser small as a dashed red curve). ATLASGAL clumps located in the same complex as the methanol maser source can be revealed within the large radius that includes the Effelsberg beamwidth in addition to the pointing errors (CH$_3$OH maser large as dashed-dotted blue curve). The range of small and large search radii is given in Table \ref{sample}. The CH$_3$OH samples found within both search radii have a similar range of $\Delta v$(1,1) and $T_{\mbox{\tiny rot}}$, although they have broader linewidths and higher $T_{\mbox{\tiny rot}}$ than the ATLASGAL sample on average. An ammonia spectrum of a YSO (G27.02+0.20) is shown in Fig. \ref{nh3lines}.\\
The upper plot compares the ATLASGAL sample (the solid green curve) with UCHII regions from \cite{2005IAUS..227..370H}. We again used a small and large search radius as described above to distinguish between the embedded object as a dashed red curve, and the envelope of the H II region (the dotted black curve). Because they already contain a protostar heating up the molecular cloud, they are at a more evolved stage of high-mass star formation than the IRDCs and YSOs and are therefore the warmest sources with the broadest linewidths. Since the embedded object traces the central part of the molecular cloud, it covers slightly higher temperatures and broader linewidths than the envelope. In contrast, the envelope includes cold regions as well, which results in a decrease of $T_{\mbox{\tiny rot}}$ and $\Delta v$(1,1) by averaging over the large search radius. Fig. \ref{nh3lines} illustrates an example of NH$_3$ lines of an embedded UCHIIR (G12.89+0.49).\\
To investigate if the differences in physical properties given in Table \ref{sample} are statistically significant, we performed a Kolmogorov-Smirnov (KS) test using the three largest subsamples from different evolutionary phases. The comparison of the NH$_3$ (1,1) linewidth and rotational temperature distributions of the IRDCs from \cite{2009A&A...505..405P} and the CH$_3$OH sample found within the large search radius using the KS test showed that both are measurably different (probability $>$ 99.9\%). In addition, the KS test used for the (1,1) linewidths and rotational temperatures of the IRDCs with those of the H II regions envelope from \cite{2005IAUS..227..370H} reveals that those are significantly different as well. Cumulative distribution plots (Fig. \ref{dv11-trot-distribution}) show the NH$_3$ (1,1) linewidth and rotational temperature distribution for the different subsamples. They illustrate a trend of broader linewidths and higher temperatures for sources with CH$_3$OH masers (dashed curve in blue) and with H II regions (dotted black line) compared to IRDCs (solid red curve). The distributions of methanol masers and H II regions are similar. Our KS test results for the NH$_3$ (1,1) linewidths and rotational temperatures do not contradict that the CH$_3$OH sample and the H II regions envelope sample are drawn from the same parent population.

\vspace*{0.5cm}

\vspace*{0.5cm}

\section{Conclusions}
\label{conclusion}
The (1,1) to (3,3) ammonia inversion transitions of 898 ATLASGAL sources, located between 5$^{\circ}$ and $60^{\circ}$ in Galactic longitude and $\pm 1.5^{\circ}$ in Galactic latitude, were observed with the Effelsberg telescope. They build the largest available ammonia sample of high-mass star forming clumps. Our results are summarised in this section.\\
We analysed the correlation of different ammonia line parameters and compared them to the submm dust continuum and $^{13}$CO emission.\\
The NH$_3$ (1,1) line's LSR velocity of most clumps lies between -10 km~s$^{-1}$ and 120 km~s$^{-1}$, most clumps show a strong correlation with CO emission \citep{2001ApJ...547..792D}, probing the larger giant molecular clouds. The velocities yield near/far kinematical distances, that were calculated for 750 unknown sources using the revised rotation parameters of the Milky Way presented by \cite{2009ApJ...700..137R}. Galactocentric distances show an enhancement of ATLASGAL sources at 4.5 and 6 kpc, tracing the Scutum arm and Sagittarius arm, which agrees with previous results for other tracers of massive star formation \citep{2009ApJ...690..706A}.\\
Most sources exhibit NH$_3$ (1,1) linewidths between 0.7 km~s$^{-1}$ and 3.5 km~s$^{-1}$, some range up to 7 km~s$^{-1}$, which is much broader than the thermal NH$_3$ linewidth and common linewidths observed in cores of low-mass \citep{1999ApJS..125..161J}. The rotational temperature between the (1,1) and (2,2) inversion levels was derived and was used to calculate the kinetic temperature. Most ATLASGAL clumps have rotational temperatures between 10 K and 25 K, some exhibit up to 30 K, while their average kinetic temperatures range from 12 K to 35 K, only few up to 40 K. The increasing rotational temperature and the broader widths of the (3,3) line, which 
probes higher temperature, hint at star formation activity inside the clumps. Indeed, many clumps show a stronger than expected intensity of the NH$_3$ (3,3) line as estimated for a homogeneous, cold clump with a temperature derived from the (1,1) and (2,2) lines. This results probably from an embedded hot core component with a low beam filling factor and high temperature.   \\
We used the kinetic temperature and submm dust continuum flux to determine the H$_2$ density. Its correlation with the NH$_3$ column density leads to ammonia abundances relative to H$_2$ mostly from 5$\times$10$^{-8}$ to 3$\times$10$^{-7}$.\\
For ATLASGAL clumps located at tangential points and associated with clouds from the GRS survey, preliminary distance estimates were given and gas masses were calculated, that lie in the range from 60 to 10$^4$ M$_{\odot}$. Comparing them with the virial masses yields a virial parameter of $\sim$ 1 for sources with broad linewidths, which are supported against gravitational collapse. In contrast, clumps with narrow linewidths have a virial parameter of only $\sim$ 0.5.\\
The comparison of the NH$_3$ lines as high-density probe with the GRS $^{13}$CO emission as low-density envelope tracer yields broader $^{13}$CO than ammonia linewidths. This can result from turbulence within large structures probed by $^{13}$CO.
We also analysed the relative motion of the clump and its cloud using NH$_3$, which represents the dense core material and $^{13}$CO to trace the more diffuse molecular cloud gas. Since our comparison yields small differences between NH$_3$ and $^{13}$CO velocities, less than the average $^{13}$CO linewidth, random motions of the ammonia clumps are expected and hint at turbulence. In contrast to low-mass cores, which show mainly subsonic core-to-envelope motions \citep{2004ApJ...614..194W, 2007ApJ...655..958W, 2007ApJ...668.1042K}, the relative motions of our sample mostly exceed the sound speed.\\
Recent NH$_3$ (1,1) to (3,3) line observations of northern BGPS sources were carried out by \cite{2011ApJ...741..110D} at the Robert F. Byrd Green Bank Telescope (GBT) with a lower FWHM of 31$\arcsec$ compared to the Effelsberg beamwidth. Our results agree well with those of \cite{2011ApJ...741..110D}, yielding similar ranges of kinetic temperature, NH$_3$ column density, H$_2$ column density, thus NH$_3$ abundance, and gas masses.\\
Association of the ATLASGAL sources with other high-mass star forming samples led to subsamples in various evolutionary stages, whose NH$_3$ line parameters were compared. Clumps with low temperatures, narrow linewidths, and high column densities may be IRDCs, which are assumed to be the cold precursors to hot cores and are consequently in an early evolutionary phase of high-mass star formation \citep{2006ApJ...641..389R}. Moreover, few ATLASGAL sources show higher temperatures, greater linewidths and lower column densities than the IRDCs. Those properties hint at objects in a more evolved stage of massive star formation, such as YSOs that contain a luminous infrared source or clumps with embedded UCHII regions.\\
Because these NH$_3$ observations and analyses can also build a valuable database for the selection of different phases of massive star formation and the determined velocities are important especially for the so far unknown ATLASGAL sources, we extended the NH$_3$ survey to the south using the Parkes telescope. The comparison of first and fourth quadrant properties and the distinction between near and far distances will be published in a forthcoming paper (Wienen et al. in prep.).\\

\begin{acknowledgements}
We thank the anonymous referee for very useful comments which improved the paper. LB acknowledges support from CONICYT projects FONDAP 15010003 and Basal PFB-06. M. Wienen was supported for this research through a stipend from the International Max Planck Research School (IMPRS) for Astronomy and Astrophysics at the Universities of Bonn and Cologne. 
\end{acknowledgements}

\begin{appendix}

\section{Pointing and calibration}
In this section, we describe the pointing and calibration of the NH$_3$ data obtained with the Effelsberg telescope.\\
Continuum drift scans were taken on nearby pointing sources such as NGC7027, 1741-038, G10.62, and NRAO530 every hour to correct for pointing offsets. NGC7027 was also used to calibrate the temperature scale. The focal point was determined by measuring NGC7027 at the beginning of each observational period, before and after sunrise and sunset.\\
The data taken with the Effelsberg telescope are given in noise tube units and must be converted into main beam brightness temperature during the calibration procedure. Because many sources could only be observed at low elevations between 15$^\circ$ and 25$^\circ$ because of their southern position, several flux calibrators at different elevations had to be observed. These are the continuum sources, whose flux density was measured during the pointing. In addition, the atmospheric opacity has to be included in the calibration procedure owing to the low elevations of the sources. The primary flux calibration was performed using NGC7027, since its flux density is precisely known. The other pointing sources can be used as secondary calibrators, which were observed over the entire elevation range and had to be adapted to NGC7027 as relative calibration.\\
The sources were calibrated by the total calibration factor $T_{\mbox{\tiny cal,tot}}$, which is given by a constant factor $T_{\mbox{\tiny cal}}$ belonging to the primary flux calibration divided by a gain elevation curve for each observation day according to
\begin{eqnarray}
T_{\mbox{\tiny cal,tot}}=\frac{T_{\mbox{\tiny cal}}}{(a+b\cdot El+c\cdot El^2)\cdot {\rm exp}(-\tau\cdot amass)}.
\end{eqnarray}
The term y1$ = a+b\cdot El + c\cdot El^2$ with the elevation $El$ describes the distortion of the telescope caused by its weight, when it is inclined, as well as thermal conditions. y2$=e^{-\tau\cdot amass}$ with the zenith opacity $\tau$ and the factor \textit{airmass}, $amass=1/{\rm sin}(El)$, takes the weakening of the radiation penetrating the atmosphere into account.\\
During the primary flux calibration, the factor $T_{\mbox{\tiny cal}}$ was determined using continuum scans of NGC7027, which possesses a flux density $S_\lambda \approx 5.6$ Jy at 1.3 cm \citep{1994A&A...284..331O}. The main beam brightness temperature $T_{\mbox{\tiny MB}}$ of this calibrator for a given beamwidth $\theta$ and wavelength $\lambda$ is given by \citep{2004tra..book.....R}
\begin{eqnarray}
(T_{\mbox{\tiny MB}}/{\rm K})=\frac{(S_\lambda/{\rm Jy})\cdot (\lambda/{\rm cm})^2}{(\theta/{\rm arcmin})^2\cdot 2.65}.
\end{eqnarray}
With a beamwidth of $\theta=40''$ and wavelength of $\lambda=1.25$ cm, $T_{\mbox{\tiny MB}}$ of NGC7027 is 7.5 K. We used only one calibration factor for the NH$_3$ (1,1) to (3,3) inversion transitions, because the variation of the noise diode temperature\footnote{see http://www.mpifr-bonn.mpg.de/div/effelsberg/calibration/1.3cmpf.html} is between 4 and 6\% and of the flux of NGC7027 \citep{1994A&A...284..331O} is only $\sim$ 0.1\%, both are therefore smaller than the average calibration uncertainty of 10\%.\\
The opacity is given by the water vapour radiometer, although it could not be used in May 2008, because it did not work during this month. Hence, the opacity of the sources observed in that time was obtained by plotting their system temperatures against elevation and fitting them. The system temperature is related to the elevation according to
\begin{eqnarray}
T_{\mbox{\tiny sys}}=T_0+T_{\mbox{\tiny atm}}(1- {\rm exp}(-\tau \cdot amass)),
\end{eqnarray}
with the system temperature $T_{\mbox{\tiny sys}}$ for an atmospheric temperature $T_{\mbox{\tiny atm}}$, the zenith opacity $\tau$ and the temperature $T_0$, which contains the temperature of the receiver and the ground. G15.66$-$0.50 was observed every day with the Effelsberg telescope. Hence, the $T_{\mbox{\tiny MB}}$ variation of its NH$_3$(1,1) line could be used to estimate the rms uncertainty of the calibration, which was determined to be $\sim 10\%$, which is small considering the low elevations of the sources up to 25$^\circ$. But this does not influence important NH$_3$ line parameters such as the optical depth or the rotational and kinetic temperature of the clumps, because their determination does not require calibration. The optical depth was obtained by the ratio of the hyperfine components of one inversion transition. The rotational and thus kinetic temperature were calculated by the intensity ratio of the ammonia (1,1) and (2,2) lines, which can be observed simultaneously. However, the data needed to be calibrated to estimate e.g. the source-averaged column density.

\end{appendix}

\newpage

\begin{table*}
\begin{minipage}{\textwidth}
\caption{NH$_3$(1,1) line parameters. Errors are given in parentheses. The full table is available at CDS.}              
\label{parline11-atlasgal}      
\centering                                      
\begin{tabular}{l l l l l l l}          
\hline\hline                        
 & RA\tablefootmark{1} & Dec\tablefootmark{1} & $\tau$(1,1) & $v$(1,1) & $\Delta v$(1,1) & T$_{\mbox{\tiny MB}}$(1,1) \\ 
Name  & (J2000) &  (J2000) &  & (km~s$^{-1}$) &  (km~s$^{-1}$) & (K) \\                    
\hline                                  
 G5.39+0.19 & 17 57 12.9150 & -24 12 28.150 & 1.71 $(\pm$0.29) & 10.58 $(\pm$0.03) & 1.42 $(\pm$0.06) & 1.7 $(\pm$0.11) \\
G5.35+0.10 & 17 57 29.2160 & -24 16 57.670 & 2.11 $(\pm$0.31) & 11.21 $(\pm$0.03) & 1.48 $(\pm$0.07) & 1.73 $(\pm$0.13) \\
G5.64+0.24 & 17 57 34.6840 & -23 58 01.950 & 2.2 $(\pm$0.13) & 7.94 $(\pm$0.01) & 1.61 $(\pm$0.03) & 4.22 $(\pm$0.12) \\
G5.62-0.08 & 17 58 44.6080 & -24 08 39.020 & 3.14 $(\pm$0.14) & -25.74 $(\pm$0.02) & 2.43 $(\pm$0.03) & 3.24 $(\pm$0.11) \\
G6.22-0.05 & 17 59 55.2530 & -23 36 10.150 & 3.13 $(\pm$0.63) & 19 $(\pm$0.09) & 2.78 $(\pm$0.15) & 0.62 $(\pm$0.09) \\
G6.19-0.12 & 18 00 07.4100 & -23 39 51.830 & 1.5 $(\pm$0.41) & 10.88 $(\pm$0.06) & 1.99 $(\pm$0.14) & 0.99 $(\pm$0.12) \\
G5.89-0.29 & 18 00 07.5280 & -24 00 29.480 & 2.09 $(\pm$0.23) & 10.11 $(\pm$0.02) & 1.42 $(\pm$0.04) & 1.88 $(\pm$0.1) \\
G5.89-0.32 & 18 00 14.4580 & -24 01 23.810 & 0.9 $(\pm$0.13) & 10.11 $(\pm$0.02) & 1.74 $(\pm$0.05) & 2.6 $(\pm$0.09) \\
G6.25-0.12 & 18 00 16.1890 & -23 37 09.950 & 1.71 $(\pm$0.25) & 14.48 $(\pm$0.06) & 3.31 $(\pm$0.11) & 1.02 $(\pm$0.08) \\
G5.83-0.40 & 18 00 24.8380 & -24 06 50.090 & 0.91 $(\pm$0.21) & 7.64 $(\pm$0.03) & 1.67 $(\pm$0.07) & 1.54 $(\pm$0.08) \\
G5.89-0.39 & 18 00 30.3380 & -24 04 00.209 & 0.69 $(\pm$0.09) & 9.18 $(\pm$0.03) & 3.96 $(\pm$0.06) & 4.05 $(\pm$0.13) \\
G5.90-0.43 & 18 00 40.3520 & -24 04 18.700 & 1.8 $(\pm$0.06) & 6.64 $(\pm$0.01) & 2.53 $(\pm$0.03) & 3.8 $(\pm$0.09) \\
G5.83-0.51 & 18 00 51.2040 & -24 10 23.290 & 1.99 $(\pm$0.12) & 16.22 $(\pm$0.01) & 1.74 $(\pm$0.03) & 2.8 $(\pm$0.08) \\
G6.19-0.36 & 18 01 01.7140 & -23 47 05.130 & 2.11 $(\pm$0.12) & -33.81 $(\pm$0.02) & 2.2 $(\pm$0.04) & 3.72 $(\pm$0.12) \\
G5.91-0.54 & 18 01 08.2520 & -24 07 14.800 & 2.09 $(\pm$0.22) & 14.79 $(\pm$0.02) & 1.37 $(\pm$0.05) & 2.32 $(\pm$0.11) \\

\hline                                             
\end{tabular}
\tablefoot{
\tablefoottext{1}{Units of right ascension are hours, minutes, and seconds, and units of declination are degrees, arcminutes, and arcseconds.}
}
\end{minipage}
\end{table*}

\begin{table*}
\begin{minipage}{\textwidth}
\caption{NH$_3$(2,2) and (3,3) line parameters. Errors are given in parentheses. The full table is available at CDS.}              
\label{parline22_line33-atlasgal}      
\centering                                      
\begin{tabular}{l l l l l l l}          
\hline\hline                        
 & $v$(2,2) & $\Delta v$(2,2) & T$_{\mbox{\tiny MB}}$(2,2) & $v$(3,3) & $\Delta v$(3,3) & T$_{\mbox{\tiny MB}}$(3,3) \\ 
Name  & (km~s$^{-1}$) &  (km~s$^{-1}$) &  (K) & (km~s$^{-1}$) &  (km~s$^{-1}$) & (K) \\                    
\hline                                  
 G5.39+0.19 & 10.6 $(\pm$0.13) & 1.53 $(\pm$0.38) & 0.47 $(\pm$0.12) & - & - & $<$ 0.09  \\
G5.35+0.10 & 10.85 $(\pm$0.17) & 1.8 $(\pm$0.35) & 0.38 $(\pm$0.11) & - & - & $<$ 0.08  \\
G5.64+0.24 & 7.95 $(\pm$0.02) & 1.89 $(\pm$0.06) & 2.88 $(\pm$0.12) & 8.29 $(\pm$0.09) & 2.81 $(\pm$0.2) & 0.82 $(\pm$0.09) \\
G5.62-0.08 & -25.57 $(\pm$0.05) & 2.93 $(\pm$0.11) & 1.93 $(\pm$0.14) & -25.47 $(\pm$0.05) & 3.54 $(\pm$0.13) & 1.85 $(\pm$0.11) \\
G6.22-0.05 & - & - & $<$ 0.1  & - & - & $<$ 0.08  \\
G6.19-0.12 & 10.63 $(\pm$0.26) & 2.91 $(\pm$0.62) & 0.3 $(\pm$0.1) & - & - & $<$ 0.1  \\
G5.89-0.29 & 10.13 $(\pm$0.06) & 1.65 $(\pm$0.16) & 0.85 $(\pm$0.1) & - & - & $<$ 0.1  \\
G5.89-0.32 & 10.09 $(\pm$0.06) & 1.87 $(\pm$0.14) & 1.21 $(\pm$0.13) & 10.04 $(\pm$0.06) & 1.51 $(\pm$0.23) & 1.01 $(\pm$0.1) \\
G6.25-0.12 & 13.97 $(\pm$0.2) & 3.02 $(\pm$0.42) & 0.44 $(\pm$0.12) & 14.35 $(\pm$0.39) & 4.06 $(\pm$1.18) & 0.31 $(\pm$0.1) \\
G5.83-0.40 & 7.77 $(\pm$0.11) & 2.48 $(\pm$0.3) & 0.67 $(\pm$0.11) & - & - & $<$ 0.09  \\
G5.89-0.39 & 9.28 $(\pm$0.04) & 4.58 $(\pm$0.09) & 3.37 $(\pm$0.15) & - & - & - \\
G5.90-0.43 & 6.58 $(\pm$0.04) & 3.29 $(\pm$0.08) & 2.92 $(\pm$0.14) & 6.6 $(\pm$0.05) & 3.87 $(\pm$0.12) & 1.76 $(\pm$0.09) \\
G5.83-0.51 & 16.16 $(\pm$0.06) & 2.35 $(\pm$0.14) & 1.13 $(\pm$0.1) & - & - & $<$ 0.09  \\
G6.19-0.36 & -33.84 $(\pm$0.04) & 2.83 $(\pm$0.1) & 2.71 $(\pm$0.14) & -33.98 $(\pm$0.04) & 3.41 $(\pm$0.1) & 1.82 $(\pm$0.09) \\
G5.91-0.54 & 14.9 $(\pm$0.05) & 1.59 $(\pm$0.13) & 1.32 $(\pm$0.13) & 14.43 $(\pm$0.33) & 5.19 $(\pm$0.79) & 0.29 $(\pm$0.1) \\

\hline                   
\end{tabular}                              
\end{minipage}
\end{table*}

\begin{table*}
\begin{minipage}{\textwidth}
\caption{Parameters derived from the NH$_3$(1,1) to (3,3) inversion transitions. Errors are given in parentheses. The full table is available at CDS.}              
\label{parabgel-atlasgal}      
\centering                                      
\begin{tabular}{l l l l }          
\hline\hline                        
 & T$_{{\mbox{\tiny rot}}}$ & T$_{\mbox{\tiny kin}}$ & log$_{10}$(N$_{\mbox{NH}_3}$)   \\ 
Name  & (K) &  (K) &   (cm$^{-2}$)  \\                    
\hline                                  
 G5.39+0.19 & 13.12 $(\pm$1.27) & 14.39 $(\pm$1.67) & 15.12 $(\pm$0.08) \\
G5.35+0.10 & 11.66 $(\pm$1.15) & 12.52 $(\pm$1.43) & 15.23 $(\pm$0.07) \\
G5.64+0.24 & 19.53 $(\pm$0.82) & 23.94 $(\pm$1.39) & 15.34 $(\pm$0.03) \\
G5.62-0.08 & 16.08 $(\pm$0.81) & 18.5 $(\pm$1.19) & 15.63 $(\pm$0.02) \\
G6.22-0.05 & - & - & - \\
G6.19-0.12 & 13.92 $(\pm$1.99) & 15.45 $(\pm$2.7) & 15.21 $(\pm$0.12) \\
G5.89-0.29 & 15.45 $(\pm$1.02) & 17.58 $(\pm$1.47) & 15.21 $(\pm$0.05) \\
G5.89-0.32 & 18.16 $(\pm$1.08) & 21.67 $(\pm$1.73) & 14.97 $(\pm$0.07) \\
G6.25-0.12 & 15.79 $(\pm$2.13) & 18.07 $(\pm$3.09) & 15.5 $(\pm$0.07) \\
G5.83-0.40 & 17.51 $(\pm$1.53) & 20.66 $(\pm$2.38) & 14.95 $(\pm$0.1) \\
G5.89-0.39 & 27.33 $(\pm$1.33) & 39.65 $(\pm$3.21) & 15.36 $(\pm$0.06) \\
G5.90-0.43 & 22.69 $(\pm$1.15) & 29.63 $(\pm$2.22) & 15.5 $(\pm$0.03) \\
G5.83-0.51 & 14.82 $(\pm$0.64) & 16.68 $(\pm$0.89) & 15.28 $(\pm$0.03) \\
G6.19-0.36 & 20.85 $(\pm$1.17) & 26.22 $(\pm$2.08) & 15.48 $(\pm$0.03) \\
G5.91-0.54 & 17.44 $(\pm$1.22) & 20.55 $(\pm$1.89) & 15.22 $(\pm$0.05) \\

\hline                                             
\end{tabular}
\end{minipage}
\end{table*}

\begin{table*}
\begin{minipage}{\textwidth}
\caption{Parameters from the 870 $\mu$m dust continuum and NH$_3$ lines. The full table is available at CDS.}              
\label{par870mikrom-atlasgal}      
\centering                                      
\begin{tabular}{l l l l l l l}          
\hline\hline                        
 &  & N$_{\mbox{H}_2}$ & $\chi$ &  S$_{\mbox{\tiny 870 $\mu$m}}^{40 \arcsec}$ & Distance (near) & Distance (far) \\ 
Name & $\eta$ & (10$^{22}$ cm$^{-2}$) & (10$^{-7}$) &  (Jy) & (kpc) &  (kpc)  \\                    
\hline                                  
 G5.39+0.19 & 0.18 & 1.18 & 1.12 & 1.21 & 2.8 & 13.9 \\
G5.35+0.10 & 0.2 & 1.79 & 0.95 & 1.44 & 2.9 & 13.8 \\
G5.64+0.24 & 0.22 & 3.79 & 0.58 & 8.42 & 2.2 & 14.5 \\
G5.62-0.08 & 0.21 & 2.59 & 1.65 & 3.96 & - & - \\
G6.22-0.05 & - & - & - & 2.74 & 3.7 & 13.0 \\
G6.19-0.12 & 0.1 & 1.32 & 1.23  & 1.52 & 2.7 & 14.1 \\
G5.89-0.29 & 0.14 & 2.1 & 0.77 & 2.98 & 2.5 & 14.2 \\
G5.89-0.32 & 0.23 & 2.15 & 0.43 & 4.15 & 2.6 & 14.1 \\
G6.25-0.12 & 0.08 & 2.16 & 1.46 & 3.19 & 3.1 & 13.6 \\
G5.83-0.40 & 0.14 & 1.45 & 0.61 & 2.62 & 2.1 & 14.6 \\
G5.89-0.39 & 0.22 & 9.74 & 0.24 & 41.64 & 2.4 & 14.4 \\
G5.90-0.43 & 0.17 & 6.88 & 0.46 & 20.37 & 1.9 & 14.8 \\
G5.83-0.51 & 0.23 & 3.8 & 0.5 & 4.96 & 3.4 & 13.3 \\
G6.19-0.36 & 0.18 & 3.73 & 0.81 & 9.38 & - & - \\
G5.91-0.54 & 0.15 & 1.37 & 1.21 & 2.45 & 3.3 & 13.5 \\

\hline                                             
\end{tabular}
\tablefoot{
Columns are beam filling factor, H$_2$ column density, NH$_3$ abundance, 870 $\mu$m dust continuum flux integrated over 40$\arcsec$, near kinematic distance, far kinematic distance. * are sources with known distances (cf. Sect. \ref{virial mass}.)
}
\end{minipage}
\end{table*}

\begin{table*}
\begin{minipage}{\textwidth}
\caption{Line parameters derived from $^{13}$CO (1-0) transition. Errors are given in parentheses. The full table is available at CDS.}              
\label{parlineco-atlasgal}      
\centering                                      
\begin{tabular}{l l l l l l}          
\hline\hline                        
 & $v_{\mbox{\tiny $^{13}$CO}}$ & $\Delta v_{\mbox{\tiny $^{13}$CO}}$ & T$_{\mbox{\tiny MB, $^{13}$CO}}$  \\ 
Name & (km~s$^{-1}$) & (km~s$^{-1}$) & (K)   \\                    
\hline                                  
 G14.62+0.33 & 26.73 $(\pm$0.03) & 2.56 $(\pm$0.09) & 3.26 $(\pm$0.17) \\
G14.63+0.31 & 26.03 $(\pm$0.04) & 2.34 $(\pm$0.1) & 3.65 $(\pm$0.16) \\
G14.92+0.07 & 25.93 $(\pm$0.05) & 3.35 $(\pm$0.12) & 3.76 $(\pm$0.26) \\
G15.01+0.01 & 25.38 $(\pm$0.13) & 4.35 $(\pm$0.42) & 2.65 $(\pm$0.15) \\
G15.43+0.19 & 47.85 $(\pm$0.03) & 3.13 $(\pm$0.06) & 3.31 $(\pm$0.17) \\
G14.99-0.12 & 49.46 $(\pm$0.06) & 3.15 $(\pm$0.19) & 1.67 $(\pm$0.18) \\
G15.46-0.00 & 66.92 $(\pm$0.06) & 3.3 $(\pm$0.13) & 1.66 $(\pm$0.12) \\
G14.89-0.40 & 61.64 $(\pm$0.13) & 2.65 $(\pm$0.32) & 2.74 $(\pm$0.66) \\
G15.20-0.44 & 20.86 $(\pm$0.03) & 2.38 $(\pm$0.08) & 3.69 $(\pm$0.13) \\
G15.53-0.41 & 39.27 $(\pm$0.07) & 4.36 $(\pm$0.18) & 1.97 $(\pm$0.19) \\
G15.56-0.46 & 15.24 $(\pm$0.03) & 1.78 $(\pm$0.06) & 3.92 $(\pm$0.28) \\
G16.36-0.07 & 47.25 $(\pm$0.05) & 2.72 $(\pm$0.13) & 5.75 $(\pm$0.19) \\
G16.58-0.05 & 59.41 $(\pm$0.02) & 3.27 $(\pm$0.06) & 6.53 $(\pm$0.17) \\
G16.83+0.08 & 62.43 $(\pm$0.03) & 2.48 $(\pm$0.07) & 2.35 $(\pm$0.15) \\
G16.58-0.08 & 40.85 $(\pm$0.17) & 3.74 $(\pm$0.57) & 2.32 $(\pm$0.13) \\

\hline                                             
\end{tabular}
\end{minipage}
\end{table*}

\begin{table*}[htbp]
\caption[]{Properties of ATLASGAL subsamples belonging to different evolutionary stages of high-mass star formation.}
\label{sample}
\centering
\begin{tabular}{c c c c c c c}
\hline\hline
Sample & Search radius ($\arcsec$) & Number & T$_{\mbox{\tiny rot}}$ (K) & $\sigma$(T$_{\mbox{\tiny rot}}$) & $\Delta v$(1,1) (km~s$^{-1}$) & $\sigma$($\Delta v$(1,1)) \\ \hline 
24$\mu$m dark  &   40   &  98  &  15.25 &  0.21  & 1.85 & 0.09 \\
IRDCs   &   32.27   &  264 & 16.31  & 0.19 & 1.96  & 0.04 \\
IRDCs24$\mu$m  &  32.27 & 117   & 16.89  & 0.27  &  2.08  & 0.06 \\
YSOs  &   40.05  &  57   & 17.31 &  0.35 & 2.14 & 0.08 \\
CH$_3$OH maser small   & 5 - 25  & 61 & 19.23 & 1.61 &  2.77 & 0.1 \\
CH$_3$OH maser large  &  40 - 47   &   137 &  18.97 &  1.65 & 2.65 & 0.1 \\
HII regions embedded object  &  5   &  57   & 18.98  &  0.37  & 2.79 & 0.18 \\
HII regions envelope  &  40.3  &  194  & 18.63 &  0.2  & 2.63 & 0.08 \\ \hline 
\end{tabular}
\end{table*}

\bibliography{paperslibrary}

\begin{thebibliography}{93}
\expandafter\ifx\csname natexlab\endcsname\relax\def\natexlab#1{#1}\fi

\bibitem[{{Anderson} \& {Bania}(2009)}]{2009ApJ...690..706A}
{Anderson}, L.~D. \& {Bania}, T.~M. 2009, \apj, 690, 706

\bibitem[{{Andr\'e} {et~al.}(2000){Andr\'e}, {Ward-Thompson}, \&
  {Barsony}}]{2000prpl.conf...59A}
{Andr\'e}, P., {Ward-Thompson}, D., \& {Barsony}, M. 2000, Protostars and
  Planets IV, 59

\bibitem[{{Bally} {et~al.}(2010){Bally}, {Anderson}, {Battersby}, {Calzoletti},
  {Digiorgio}, {Faustini}, {Ginsburg}, {Li}, {Nguyen-Luong}, {Molinari},
  {Motte}, {Pestalozzi}, {Plume}, {Rodon}, {Schilke}, {Schlingman},
  {Schneider-Bontemps}, {Shirley}, {Stringfellow}, {Testi}, {Traficante},
  {Veneziani}, \& {Zavagno}}]{2010A&A...518L..90B}
{Bally}, J., {Anderson}, L.~D., {Battersby}, C., {et~al.} 2010, \aap, 518, L90+

\bibitem[{{Becker} {et~al.}(1994){Becker}, {White}, {Helfand}, \&
  {Zoonematkermani}}]{1994ApJS...91..347B}
{Becker}, R.~H., {White}, R.~L., {Helfand}, D.~J., \& {Zoonematkermani}, S.
  1994, \apjs, 91, 347

\bibitem[{{Benjamin} {et~al.}(2003){Benjamin}, {Churchwell}, {Babler}, {Bania},
  {Clemens}, {Cohen}, {Dickey}, {Indebetouw}, {Jackson}, {Kobulnicky},
  {Lazarian}, {Marston}, {Mathis}, {Meade}, {Seager}, {Stolovy}, {Watson},
  {Whitney}, {Wolff}, \& {Wolfire}}]{2003PASP..115..953B}
{Benjamin}, R.~A., {Churchwell}, E., {Babler}, B.~L., {et~al.} 2003, \pasp,
  115, 953

\bibitem[{{Bergin} \& {Langer}(1997)}]{1997ApJ...486..316B}
{Bergin}, E.~A. \& {Langer}, W.~D. 1997, \apj, 486, 316

\bibitem[{{Bertoldi} \& {McKee}(1992)}]{1992ApJ...395..140B}
{Bertoldi}, F. \& {McKee}, C.~F. 1992, \apj, 395, 140

\bibitem[{{Beuther} {et~al.}(2007){Beuther}, {Churchwell}, {McKee}, \&
  {Tan}}]{2007prpl.conf..165B}
{Beuther}, H., {Churchwell}, E.~B., {McKee}, C.~F., \& {Tan}, J.~C. 2007,
  Protostars and Planets V, 165

\bibitem[{{Beuther} {et~al.}(2002){Beuther}, {Schilke}, {Menten}, {Motte},
  {Sridharan}, \& {Wyrowski}}]{2002ApJ...566..945B}
{Beuther}, H., {Schilke}, P., {Menten}, K.~M., {et~al.} 2002, \apj, 566, 945

\bibitem[{{Bonnell} {et~al.}(1997){Bonnell}, {Bate}, {Clarke}, \&
  {Pringle}}]{1997MNRAS.285..201B}
{Bonnell}, I.~A., {Bate}, M.~R., {Clarke}, C.~J., \& {Pringle}, J.~E. 1997,
  \mnras, 285, 201

\bibitem[{{Bonnell} {et~al.}(2001){Bonnell}, {Bate}, {Clarke}, \&
  {Pringle}}]{2001MNRAS.323..785B}
{Bonnell}, I.~A., {Bate}, M.~R., {Clarke}, C.~J., \& {Pringle}, J.~E. 2001,
  \mnras, 323, 785

\bibitem[{{Bontemps} {et~al.}(2010){Bontemps}, {Motte}, {Csengeri}, \&
  {Schneider}}]{2010A&A...524A..18B}
{Bontemps}, S., {Motte}, F., {Csengeri}, T., \& {Schneider}, N. 2010, \aap,
  524, A18+

\bibitem[{{Braz} \& {Epchtein}(1983)}]{1983A&AS...54..167B}
{Braz}, M.~A. \& {Epchtein}, N. 1983, \aaps, 54, 167

\bibitem[{{Bronfman} {et~al.}(2000){Bronfman}, {Casassus}, {May}, \&
  {Nyman}}]{2000A&A...358..521B}
{Bronfman}, L., {Casassus}, S., {May}, J., \& {Nyman}, L.-{\AA}. 2000, \aap,
  358, 521

\bibitem[{{Carey} {et~al.}(1998){Carey}, {Clark}, {Egan}, {Price}, {Shipman},
  \& {Kuchar}}]{1998ApJ...508..721C}
{Carey}, S.~J., {Clark}, F.~O., {Egan}, M.~P., {et~al.} 1998, \apj, 508, 721

\bibitem[{{Carey} {et~al.}(2000){Carey}, {Feldman}, {Redman}, {Egan},
  {MacLeod}, \& {Price}}]{2000ApJ...543L.157C}
{Carey}, S.~J., {Feldman}, P.~A., {Redman}, R.~O., {et~al.} 2000, \apjl, 543,
  L157

\bibitem[{{Carey} {et~al.}(2009){Carey}, {Noriega-Crespo}, {Mizuno}, {Shenoy},
  {Paladini}, {Kraemer}, {Price}, {Flagey}, {Ryan}, {Ingalls}, {Kuchar},
  {Pinheiro Gon{\c c}alves}, {Indebetouw}, {Billot}, {Marleau}, {Padgett},
  {Rebull}, {Bressert}, {Ali}, {Molinari}, {Martin}, {Berriman}, {Boulanger},
  {Latter}, {Miville-Deschenes}, {Shipman}, \& {Testi}}]{2009PASP..121...76C}
{Carey}, S.~J., {Noriega-Crespo}, A., {Mizuno}, D.~R., {et~al.} 2009, \pasp,
  121, 76

\bibitem[{{Caswell} {et~al.}(2010){Caswell}, {Fuller}, {Green}, {Avison},
  {Breen}, {Brooks}, {Burton}, {Chrysostomou}, {Cox}, {Diamond}, {Ellingsen},
  {Gray}, {Hoare}, {Masheder}, {McClure-Griffiths}, {Pestalozzi}, {Phillips},
  {Quinn}, {Thompson}, {Voronkov}, {Walsh}, {Ward-Thompson}, {Wong-McSweeney},
  {Yates}, \& {Cohen}}]{2010MNRAS.404.1029C}
{Caswell}, J.~L., {Fuller}, G.~A., {Green}, J.~A., {et~al.} 2010, \mnras, 404,
  1029

\bibitem[{{Cesaroni} {et~al.}(1992){Cesaroni}, {Walmsley}, \&
  {Churchwell}}]{1992A&A...256..618C}
{Cesaroni}, R., {Walmsley}, C.~M., \& {Churchwell}, E. 1992, \aap, 256, 618

\bibitem[{{Cheung} {et~al.}(1969){Cheung}, {Rank}, {Townes}, {Knowles}, \&
  {Sullivan}}]{1969ApJ...157L..13C}
{Cheung}, A.~C., {Rank}, D.~M., {Townes}, C.~H., {Knowles}, S.~H., \&
  {Sullivan}, III, W.~T. 1969, \apjl, 157, L13+

\bibitem[{{Churchwell} {et~al.}(1990){Churchwell}, {Walmsley}, \&
  {Cesaroni}}]{1990A&AS...83..119C}
{Churchwell}, E., {Walmsley}, C.~M., \& {Cesaroni}, R. 1990, \aaps, 83, 119

\bibitem[{{Csengeri} {et~al.}(2011){Csengeri}, {Bontemps}, {Schneider}, \&
  {Motte}}]{2011IAUS..270...53C}
{Csengeri}, T., {Bontemps}, S., {Schneider}, N., \& {Motte}, F. 2011, in IAU
  Symposium, Vol. 270, IAU Symposium, ed. {J.~Alves, B.~G.~Elmegreen,
  J.~M.~Girart, \& V.~Trimble}, 53--56

\bibitem[{{Dame} {et~al.}(2001){Dame}, {Hartmann}, \&
  {Thaddeus}}]{2001ApJ...547..792D}
{Dame}, T.~M., {Hartmann}, D., \& {Thaddeus}, P. 2001, \apj, 547, 792

\bibitem[{{Devine} {et~al.}(2011){Devine}, {Chandler}, {Brogan}, {Churchwell},
  {Indebetouw}, {Shirley}, \& {Borg}}]{2011ApJ...733...44D}
{Devine}, K.~E., {Chandler}, C.~J., {Brogan}, C., {et~al.} 2011, \apj, 733, 44

\bibitem[{{Du} \& {Yang}(2008)}]{2008ApJ...686..384D}
{Du}, F. \& {Yang}, J. 2008, \apj, 686, 384

\bibitem[{{Dunham} {et~al.}(2011){Dunham}, {Rosolowsky}, {Evans}, {Cyganowski},
  \& {Urquhart}}]{2011ApJ...741..110D}
{Dunham}, M.~K., {Rosolowsky}, E., {Evans}, II, N.~J., {Cyganowski}, C., \&
  {Urquhart}, J.~S. 2011, \apj, 741, 110

\bibitem[{{Dunham} {et~al.}(2010){Dunham}, {Rosolowsky}, {Evans}, {Cyganowski},
  {Aguirre}, {Bally}, {Battersby}, {Bradley}, {Dowell}, {Drosback}, {Ginsburg},
  {Glenn}, {Harvey}, {Merello}, {Schlingman}, {Shirley}, {Stringfellow},
  {Walawender}, \& {Williams}}]{2010ApJ...717.1157D}
{Dunham}, M.~K., {Rosolowsky}, E., {Evans}, II, N.~J., {et~al.} 2010, \apj,
  717, 1157

\bibitem[{{Egan} {et~al.}(1998){Egan}, {Shipman}, {Price}, {Carey}, {Clark}, \&
  {Cohen}}]{1998ApJ...494L.199E}
{Egan}, M.~P., {Shipman}, R.~F., {Price}, S.~D., {et~al.} 1998, \apjl, 494,
  L199+

\bibitem[{{Hill} {et~al.}(2010){Hill}, {Longmore}, {Pinte}, {Cunningham},
  {Burton}, \& {Minier}}]{2010MNRAS.402.2682H}
{Hill}, T., {Longmore}, S.~N., {Pinte}, C., {et~al.} 2010, \mnras, 402, 2682

\bibitem[{{Ho} \& {Townes}(1983)}]{1983ARA&A..21..239H}
{Ho}, P.~T.~P. \& {Townes}, C.~H. 1983, \araa, 21, 239

\bibitem[{{Hoare} {et~al.}(2005){Hoare}, {Lumsden}, {Oudmaijer}, {Urquhart},
  {Busfield}, {Sheret}, {Clarke}, {Moore}, {Allsopp}, {Burton}, {Purcell},
  {Jiang}, \& {Wang}}]{2005IAUS..227..370H}
{Hoare}, M.~G., {Lumsden}, S.~L., {Oudmaijer}, R.~D., {et~al.} 2005, in IAU
  Symposium, Vol. 227, Massive Star Birth: A Crossroads of Astrophysics, ed.
  {R.~Cesaroni, M.~Felli, E.~Churchwell, \& M.~Walmsley}, 370--375

\bibitem[{{Hofner} {et~al.}(1994){Hofner}, {Kurtz}, {Churchwell}, {Walmsley},
  \& {Cesaroni}}]{1994ApJ...429L..85H}
{Hofner}, P., {Kurtz}, S., {Churchwell}, E., {Walmsley}, C.~M., \& {Cesaroni},
  R. 1994, \apjl, 429, L85

\bibitem[{{Jackson} {et~al.}(2008){Jackson}, {Finn}, {Rathborne}, {Chambers},
  \& {Simon}}]{2008ApJ...680..349J}
{Jackson}, J.~M., {Finn}, S.~C., {Rathborne}, J.~M., {Chambers}, E.~T., \&
  {Simon}, R. 2008, \apj, 680, 349

\bibitem[{{Jackson} {et~al.}(2006){Jackson}, {Rathborne}, {Shah}, {Simon},
  {Bania}, {Clemens}, {Chambers}, {Johnson}, {Dormody}, {Lavoie}, \&
  {Heyer}}]{2006ApJS..163..145J}
{Jackson}, J.~M., {Rathborne}, J.~M., {Shah}, R.~Y., {et~al.} 2006, \apjs, 163,
  145

\bibitem[{{Jijina} {et~al.}(1999){Jijina}, {Myers}, \&
  {Adams}}]{1999ApJS..125..161J}
{Jijina}, J., {Myers}, P.~C., \& {Adams}, F.~C. 1999, \apjs, 125, 161

\bibitem[{{Kauffmann} {et~al.}(2008){Kauffmann}, {Bertoldi}, {Bourke}, {Evans},
  \& {Lee}}]{2008A&A...487..993K}
{Kauffmann}, J., {Bertoldi}, F., {Bourke}, T.~L., {Evans}, II, N.~J., \& {Lee},
  C.~W. 2008, \aap, 487, 993

\bibitem[{{Kirk} {et~al.}(2007){Kirk}, {Johnstone}, \&
  {Tafalla}}]{2007ApJ...668.1042K}
{Kirk}, H., {Johnstone}, D., \& {Tafalla}, M. 2007, \apj, 668, 1042

\bibitem[{{Kroupa} {et~al.}(2011){Kroupa}, {Weidner}, {Pflamm-Altenburg},
  {Thies}, {Dabringhausen}, {Marks}, \& {Maschberger}}]{2011arXiv1112.3340K}
{Kroupa}, P., {Weidner}, C., {Pflamm-Altenburg}, J., {et~al.} 2011, ArXiv
  e-prints

\bibitem[{{Lekht}(2000)}]{2000A&AS..141..185L}
{Lekht}, E.~E. 2000, \aaps, 141, 185

\bibitem[{{Mac Low} \& {Klessen}(2004)}]{2004RvMP...76..125M}
{Mac Low}, M.-M. \& {Klessen}, R.~S. 2004, Reviews of Modern Physics, 76, 125

\bibitem[{{Mangum} \& {Wootten}(1994)}]{1994ApJ...428L..33M}
{Mangum}, J.~G. \& {Wootten}, A. 1994, \apjl, 428, L33

\bibitem[{{Martin} {et~al.}(2012){Martin}, {Roy}, {Bontemps},
  {Miville-Desch{\^e}nes}, {Ade}, {Bock}, {Chapin}, {Devlin}, {Dicker},
  {Griffin}, {Gundersen}, {Halpern}, {Hargrave}, {Hughes}, {Klein}, {Marsden},
  {Mauskopf}, {Netterfield}, {Olmi}, {Patanchon}, {Rex}, {Scott}, {Semisch},
  {Truch}, {Tucker}, {Tucker}, {Viero}, \& {Wiebe}}]{2012ApJ...751...28M}
{Martin}, P.~G., {Roy}, A., {Bontemps}, S., {et~al.} 2012, \apj, 751, 28

\bibitem[{{McKee} \& {Tan}(2003)}]{2003ApJ...585..850M}
{McKee}, C.~F. \& {Tan}, J.~C. 2003, \apj, 585, 850

\bibitem[{{McKee} {et~al.}(1993){McKee}, {Zweibel}, {Goodman}, \&
  {Heiles}}]{1993prpl.conf..327M}
{McKee}, C.~F., {Zweibel}, E.~G., {Goodman}, A.~A., \& {Heiles}, C. 1993, in
  Protostars and Planets III, ed. {E.~H.~Levy \& J.~I.~Lunine}, 327

\bibitem[{{Molinari} {et~al.}(1996){Molinari}, {Brand}, {Cesaroni}, \&
  {Palla}}]{1996A&A...308..573M}
{Molinari}, S., {Brand}, J., {Cesaroni}, R., \& {Palla}, F. 1996, \aap, 308,
  573

\bibitem[{{Molinari} {et~al.}(2010){Molinari}, {Swinyard}, {Bally}, {Barlow},
  {Bernard}, {Martin}, {Moore}, {Noriega-Crespo}, {Plume}, {Testi}, {Zavagno},
  {Abergel}, {Ali}, {Andr{\'e}}, {Baluteau}, {Benedettini}, {Bern{\'e}},
  {Billot}, {Blommaert}, {Bontemps}, {Boulanger}, {Brand}, {Brunt}, {Burton},
  {Campeggio}, {Carey}, {Caselli}, {Cesaroni}, {Cernicharo}, {Chakrabarti},
  {Chrysostomou}, {Codella}, {Cohen}, {Compiegne}, {Davis}, {de Bernardis}, {de
  Gasperis}, {Di Francesco}, {di Giorgio}, {Elia}, {Faustini}, {Fischera},
  {Fukui}, {Fuller}, {Ganga}, {Garcia-Lario}, {Giard}, {Giardino}, {Glenn},
  {Goldsmith}, {Griffin}, {Hoare}, {Huang}, {Jiang}, {Joblin}, {Joncas},
  {Juvela}, {Kirk}, {Lagache}, {Li}, {Lim}, {Lord}, {Lucas}, {Maiolo},
  {Marengo}, {Marshall}, {Masi}, {Massi}, {Matsuura}, {Meny}, {Minier},
  {Miville-Desch{\^e}nes}, {Montier}, {Motte}, {M{\"u}ller}, {Natoli}, {Neves},
  {Olmi}, {Paladini}, {Paradis}, {Pestalozzi}, {Pezzuto}, {Piacentini},
  {Pomar{\`e}s}, {Popescu}, {Reach}, {Richer}, {Ristorcelli}, {Roy}, {Royer},
  {Russeil}, {Saraceno}, {Sauvage}, {Schilke}, {Schneider-Bontemps},
  {Schuller}, {Schultz}, {Shepherd}, {Sibthorpe}, {Smith}, {Smith},
  {Spinoglio}, {Stamatellos}, {Strafella}, {Stringfellow}, {Sturm}, {Taylor},
  {Thompson}, {Tuffs}, {Umana}, {Valenziano}, {Vavrek}, {Viti}, {Waelkens},
  {Ward-Thompson}, {White}, {Wyrowski}, {Yorke}, \&
  {Zhang}}]{2010PASP..122..314M}
{Molinari}, S., {Swinyard}, B., {Bally}, J., {et~al.} 2010, \pasp, 122, 314

\bibitem[{{Molinari} {et~al.}(2002){Molinari}, {Testi}, {Rodr{\'{\i}}guez}, \&
  {Zhang}}]{2002ApJ...570..758M}
{Molinari}, S., {Testi}, L., {Rodr{\'{\i}}guez}, L.~F., \& {Zhang}, Q. 2002,
  \apj, 570, 758

\bibitem[{{Motte} {et~al.}(2007){Motte}, {Bontemps}, {Schilke}, {Schneider},
  {Menten}, \& {Brogui{\`e}re}}]{2007A&A...476.1243M}
{Motte}, F., {Bontemps}, S., {Schilke}, P., {et~al.} 2007, \aap, 476, 1243

\bibitem[{{Motte} {et~al.}(2003){Motte}, {Schilke}, \&
  {Lis}}]{2003ApJ...582..277M}
{Motte}, F., {Schilke}, P., \& {Lis}, D.~C. 2003, \apj, 582, 277

\bibitem[{{Nguyen Luong} {et~al.}(2011{\natexlab{a}}){Nguyen Luong}, {Motte},
  {Hennemann}, {Hill}, {Rygl}, {Schneider}, {Bontemps}, {Men'shchikov},
  {Andr{\'e}}, {Peretto}, {Anderson}, {Arzoumanian}, {Deharveng}, {Didelon},
  {di Francesco}, {Griffin}, {Kirk}, {K{\"o}nyves}, {Martin}, {Maury},
  {Minier}, {Molinari}, {Pestalozzi}, {Pezzuto}, {Reid}, {Roussel}, {Sauvage},
  {Schuller}, {Testi}, {Ward-Thompson}, {White}, \&
  {Zavagno}}]{2011A&A...535A..76N}
{Nguyen Luong}, Q., {Motte}, F., {Hennemann}, M., {et~al.} 2011{\natexlab{a}},
  \aap, 535, A76

\bibitem[{{Nguyen Luong} {et~al.}(2011{\natexlab{b}}){Nguyen Luong}, {Motte},
  {Schuller}, {Schneider}, {Bontemps}, {Schilke}, {Menten}, {Heitsch},
  {Wyrowski}, {Carlhoff}, {Bronfman}, \& {Henning}}]{2011A&A...529A..41N}
{Nguyen Luong}, Q., {Motte}, F., {Schuller}, F., {et~al.} 2011{\natexlab{b}},
  \aap, 529, A41

\bibitem[{{Olmi} {et~al.}(2010){Olmi}, {Araya}, {Chapin}, {Gibb}, {Hofner},
  {Martin}, \& {Poventud}}]{2010ApJ...715.1132O}
{Olmi}, L., {Araya}, E.~D., {Chapin}, E.~L., {et~al.} 2010, \apj, 715, 1132

\bibitem[{{Ormel} {et~al.}(2011){Ormel}, {Min}, {Tielens}, {Dominik}, \&
  {Paszun}}]{2011A&A...532A..43O}
{Ormel}, C.~W., {Min}, M., {Tielens}, A.~G.~G.~M., {Dominik}, C., \& {Paszun},
  D. 2011, \aap, 532, A43

\bibitem[{{Ossenkopf} \& {Henning}(1994)}]{1994A&A...291..943O}
{Ossenkopf}, V. \& {Henning}, T. 1994, \aap, 291, 943

\bibitem[{{Ott} {et~al.}(1994){Ott}, {Witzel}, {Quirrenbach}, {Krichbaum},
  {Standke}, {Schalinski}, \& {Hummel}}]{1994A&A...284..331O}
{Ott}, M., {Witzel}, A., {Quirrenbach}, A., {et~al.} 1994, \aap, 284, 331

\bibitem[{{Pandian} {et~al.}(2011){Pandian}, {Momjian}, {Xu}, {Menten}, \&
  {Goldsmith}}]{2011ApJ...730...55P}
{Pandian}, J.~D., {Momjian}, E., {Xu}, Y., {Menten}, K.~M., \& {Goldsmith},
  P.~F. 2011, \apj, 730, 55

\bibitem[{{Perault} {et~al.}(1996){Perault}, {Omont}, {Simon}, {Seguin},
  {Ojha}, {Blommaert}, {Felli}, {Gilmore}, {Guglielmo}, {Habing}, {Price},
  {Robin}, {de Batz}, {Cesarsky}, {Elbaz}, {Epchtein}, {Fouque}, {Guest},
  {Levine}, {Pollock}, {Prusti}, {Siebenmorgen}, {Testi}, \&
  {Tiphene}}]{1996A&A...315L.165P}
{Perault}, M., {Omont}, A., {Simon}, G., {et~al.} 1996, \aap, 315, L165

\bibitem[{{Peretto} \& {Fuller}(2009)}]{2009A&A...505..405P}
{Peretto}, N. \& {Fuller}, G.~A. 2009, \aap, 505, 405

\bibitem[{{Peretto} \& {Fuller}(2010)}]{2010ApJ...723..555P}
{Peretto}, N. \& {Fuller}, G.~A. 2010, \apj, 723, 555

\bibitem[{{Pillai} {et~al.}(2006){Pillai}, {Wyrowski}, {Carey}, \&
  {Menten}}]{2006A&A...450..569P}
{Pillai}, T., {Wyrowski}, F., {Carey}, S.~J., \& {Menten}, K.~M. 2006, \aap,
  450, 569

\bibitem[{{Price} {et~al.}(2001){Price}, {Egan}, {Carey}, {Mizuno}, \&
  {Kuchar}}]{2001AJ....121.2819P}
{Price}, S.~D., {Egan}, M.~P., {Carey}, S.~J., {Mizuno}, D.~R., \& {Kuchar},
  T.~A. 2001, \aj, 121, 2819

\bibitem[{{Ragan} {et~al.}(2011){Ragan}, {Bergin}, \&
  {Wilner}}]{2011ApJ...736..163R}
{Ragan}, S.~E., {Bergin}, E.~A., \& {Wilner}, D. 2011, \apj, 736, 163

\bibitem[{{Rathborne} {et~al.}(2005){Rathborne}, {Jackson}, {Chambers},
  {Simon}, {Shipman}, \& {Frieswijk}}]{2005ApJ...630L.181R}
{Rathborne}, J.~M., {Jackson}, J.~M., {Chambers}, E.~T., {et~al.} 2005, \apjl,
  630, L181

\bibitem[{{Rathborne} {et~al.}(2006){Rathborne}, {Jackson}, \&
  {Simon}}]{2006ApJ...641..389R}
{Rathborne}, J.~M., {Jackson}, J.~M., \& {Simon}, R. 2006, \apj, 641, 389

\bibitem[{{Reid} {et~al.}(2009){Reid}, {Menten}, {Zheng}, {Brunthaler},
  {Moscadelli}, {Xu}, {Zhang}, {Sato}, {Honma}, {Hirota}, {Hachisuka}, {Choi},
  {Moellenbrock}, \& {Bartkiewicz}}]{2009ApJ...700..137R}
{Reid}, M.~J., {Menten}, K.~M., {Zheng}, X.~W., {et~al.} 2009, \apj, 700, 137

\bibitem[{{Reifenstein} {et~al.}(1970){Reifenstein}, {Wilson}, {Burke},
  {Mezger}, \& {Altenhoff}}]{1970A&A.....4..357R}
{Reifenstein}, E.~C., {Wilson}, T.~L., {Burke}, B.~F., {Mezger}, P.~G., \&
  {Altenhoff}, W.~J. 1970, \aap, 4, 357

\bibitem[{{Rohlfs} \& {Wilson}(2004)}]{2004tra..book.....R}
{Rohlfs}, K. \& {Wilson}, T.~L. 2004, {Tools of radio astronomy} (Tools of
  radio astronomy, 4th rev.~and enl.~ed., by K.~Rohlfs and T.L.~Wilson.~
  Berlin: Springer, 2004)

\bibitem[{{Roman-Duval} {et~al.}(2009){Roman-Duval}, {Jackson}, {Heyer},
  {Johnson}, {Rathborne}, {Shah}, \& {Simon}}]{2009ApJ...699.1153R}
{Roman-Duval}, J., {Jackson}, J.~M., {Heyer}, M., {et~al.} 2009, \apj, 699,
  1153

\bibitem[{{Russeil} {et~al.}(2010){Russeil}, {Zavagno}, {Motte}, {Schneider},
  {Bontemps}, \& {Walsh}}]{2010A&A...515A..55R}
{Russeil}, D., {Zavagno}, A., {Motte}, F., {et~al.} 2010, \aap, 515, A55+

\bibitem[{{Sault} {et~al.}(1995){Sault}, {Teuben}, \&
  {Wright}}]{1995ASPC...77..433S}
{Sault}, R.~J., {Teuben}, P.~J., \& {Wright}, M.~C.~H. 1995, in Astronomical
  Society of the Pacific Conference Series, Vol.~77, Astronomical Data Analysis
  Software and Systems IV, ed. {R.~A.~Shaw, H.~E.~Payne, \& J.~J.~E.~Hayes},
  433--+

\bibitem[{{Schuller} {et~al.}(2009){Schuller}, {Menten}, {Contreras},
  {Wyrowski}, {Schilke}, {Bronfman}, {Henning}, {Walmsley}, {Beuther},
  {Bontemps}, {Cesaroni}, {Deharveng}, {Garay}, {Herpin}, {Lefloch}, {Linz},
  {Mardones}, {Minier}, {Molinari}, {Motte}, {Nyman}, {Reveret}, {Risacher},
  {Russeil}, {Schneider}, {Testi}, {Troost}, {Vasyunina}, {Wienen}, {Zavagno},
  {Kovacs}, {Kreysa}, {Siringo}, \& {Weiss}}]{2009arXiv0903.1369S}
{Schuller}, F., {Menten}, K.~M., {Contreras}, Y., {et~al.} 2009, ArXiv e-prints

\bibitem[{{Schulz}(2005)}]{2005fds..book.....S}
{Schulz}, N.~S. 2005, {From Dust To Stars Studies of the Formation and Early
  Evolution of Stars}, ed. {Schulz, N.~S.}

\bibitem[{{Simon} {et~al.}(2006{\natexlab{a}}){Simon}, {Jackson}, {Rathborne},
  \& {Chambers}}]{2006ApJ...639..227S}
{Simon}, R., {Jackson}, J.~M., {Rathborne}, J.~M., \& {Chambers}, E.~T.
  2006{\natexlab{a}}, \apj, 639, 227

\bibitem[{{Simon} {et~al.}(2006{\natexlab{b}}){Simon}, {Rathborne}, {Shah},
  {Jackson}, \& {Chambers}}]{2006ApJ...653.1325S}
{Simon}, R., {Rathborne}, J.~M., {Shah}, R.~Y., {Jackson}, J.~M., \&
  {Chambers}, E.~T. 2006{\natexlab{b}}, \apj, 653, 1325

\bibitem[{{Siringo} {et~al.}(2008){Siringo}, {Kreysa}, {Kovacs}, {Schuller},
  {Wei{\ss}}, {Esch}, {Gem{\"u}nd}, {Jethava}, {Lundershausen}, {G{\"u}sten},
  {Menten}, {Beelen}, {Bertoldi}, {Beeman}, {Haller}, \&
  {Colin}}]{2008SPIE.7020E...2S}
{Siringo}, G., {Kreysa}, E., {Kovacs}, A., {et~al.} 2008, in Society of
  Photo-Optical Instrumentation Engineers (SPIE) Conference Series, Vol. 7020,
  Society of Photo-Optical Instrumentation Engineers (SPIE) Conference Series

\bibitem[{{Siringo} {et~al.}(2007){Siringo}, {Weiss}, {Kreysa}, {Schuller},
  {Kovacs}, {Beelen}, {Esch}, {Gem{\"u}nd}, {Jethava}, {Lundershausen},
  {Menten}, {G{\"u}sten}, {Bertoldi}, {De Breuck}, {Nyman}, {Haller}, \&
  {Beeman}}]{2007Msngr.129....2S}
{Siringo}, G., {Weiss}, A., {Kreysa}, E., {et~al.} 2007, The Messenger, 129, 2

\bibitem[{{Skrutskie} {et~al.}(2006){Skrutskie}, {Cutri}, {Stiening},
  {Weinberg}, {Schneider}, {Carpenter}, {Beichman}, {Capps}, {Chester},
  {Elias}, {Huchra}, {Liebert}, {Lonsdale}, {Monet}, {Price}, {Seitzer},
  {Jarrett}, {Kirkpatrick}, {Gizis}, {Howard}, {Evans}, {Fowler}, {Fullmer},
  {Hurt}, {Light}, {Kopan}, {Marsh}, {McCallon}, {Tam}, {Van Dyk}, \&
  {Wheelock}}]{2006AJ....131.1163S}
{Skrutskie}, M.~F., {Cutri}, R.~M., {Stiening}, R., {et~al.} 2006, \aj, 131,
  1163

\bibitem[{{Sridharan} {et~al.}(2005){Sridharan}, {Beuther}, {Saito},
  {Wyrowski}, \& {Schilke}}]{2005ApJ...634L..57S}
{Sridharan}, T.~K., {Beuther}, H., {Saito}, M., {Wyrowski}, F., \& {Schilke},
  P. 2005, \apjl, 634, L57

\bibitem[{{Sridharan} {et~al.}(2002){Sridharan}, {Beuther}, {Schilke},
  {Menten}, \& {Wyrowski}}]{2002ApJ...566..931S}
{Sridharan}, T.~K., {Beuther}, H., {Schilke}, P., {Menten}, K.~M., \&
  {Wyrowski}, F. 2002, \apj, 566, 931

\bibitem[{{Szymczak} {et~al.}(2002){Szymczak}, {Kus}, {Hrynek}, {K{\v e}pa}, \&
  {Pazderski}}]{2002A&A...392..277S}
{Szymczak}, M., {Kus}, A.~J., {Hrynek}, G., {K{\v e}pa}, A., \& {Pazderski}, E.
  2002, \aap, 392, 277

\bibitem[{{Szymczak} {et~al.}(2005){Szymczak}, {Pillai}, \&
  {Menten}}]{2005A&A...434..613S}
{Szymczak}, M., {Pillai}, T., \& {Menten}, K.~M. 2005, \aap, 434, 613

\bibitem[{{Tafalla} {et~al.}(2004){Tafalla}, {Myers}, {Caselli}, \&
  {Walmsley}}]{2004A&A...416..191T}
{Tafalla}, M., {Myers}, P.~C., {Caselli}, P., \& {Walmsley}, C.~M. 2004, \aap,
  416, 191

\bibitem[{{Tafalla} {et~al.}(2002){Tafalla}, {Myers}, {Caselli}, {Walmsley}, \&
  {Comito}}]{2002ApJ...569..815T}
{Tafalla}, M., {Myers}, P.~C., {Caselli}, P., {Walmsley}, C.~M., \& {Comito},
  C. 2002, \apj, 569, 815

\bibitem[{{Takano} {et~al.}(2002){Takano}, {Nakai}, \&
  {Kawaguchi}}]{2002PASJ...54..195T}
{Takano}, S., {Nakai}, N., \& {Kawaguchi}, K. 2002, \pasj, 54, 195

\bibitem[{{Ungerechts} {et~al.}(1997){Ungerechts}, {Bergin}, {Goldsmith},
  {Irvine}, {Schloerb}, \& {Snell}}]{1997ApJ...482..245U}
{Ungerechts}, H., {Bergin}, E.~A., {Goldsmith}, P.~F., {et~al.} 1997, \apj,
  482, 245

\bibitem[{{Ungerechts} {et~al.}(1986){Ungerechts}, {Winnewisser}, \&
  {Walmsley}}]{1986A&A...157..207U}
{Ungerechts}, H., {Winnewisser}, G., \& {Walmsley}, C.~M. 1986, \aap, 157, 207

\bibitem[{{van der Tak} {et~al.}(2007){van der Tak}, {Black}, {Sch{\"o}ier},
  {Jansen}, \& {van Dishoeck}}]{2007A&A...468..627V}
{van der Tak}, F.~F.~S., {Black}, J.~H., {Sch{\"o}ier}, F.~L., {Jansen}, D.~J.,
  \& {van Dishoeck}, E.~F. 2007, \aap, 468, 627

\bibitem[{{Walmsley} \& {Ungerechts}(1983)}]{1983A&A...122..164W}
{Walmsley}, C.~M. \& {Ungerechts}, H. 1983, \aap, 122, 164

\bibitem[{{Walsh} {et~al.}(2004){Walsh}, {Myers}, \&
  {Burton}}]{2004ApJ...614..194W}
{Walsh}, A.~J., {Myers}, P.~C., \& {Burton}, M.~G. 2004, \apj, 614, 194

\bibitem[{{Walsh} {et~al.}(2007){Walsh}, {Myers}, {Di Francesco}, {Mohanty},
  {Bourke}, {Gutermuth}, \& {Wilner}}]{2007ApJ...655..958W}
{Walsh}, A.~J., {Myers}, P.~C., {Di Francesco}, J., {et~al.} 2007, \apj, 655,
  958

\bibitem[{{Wood} \& {Churchwell}(1989{\natexlab{a}})}]{1989ApJ...340..265W}
{Wood}, D.~O.~S. \& {Churchwell}, E. 1989{\natexlab{a}}, \apj, 340, 265

\bibitem[{{Wood} \& {Churchwell}(1989{\natexlab{b}})}]{1989ApJS...69..831W}
{Wood}, D.~O.~S. \& {Churchwell}, E. 1989{\natexlab{b}}, \apjs, 69, 831

\bibitem[{{Zoonematkermani} {et~al.}(1990){Zoonematkermani}, {Helfand},
  {Becker}, {White}, \& {Perley}}]{1990ApJS...74..181Z}
{Zoonematkermani}, S., {Helfand}, D.~J., {Becker}, R.~H., {White}, R.~L., \&
  {Perley}, R.~A. 1990, \apjs, 74, 181

\end{thebibliography}
\bibliographystyle{aa}

\end{document}